\documentclass[aps,twocolumn,superscriptaddress,showpacs]{revtex4}
\usepackage{amsmath}
\usepackage{graphicx}
\usepackage{color}

\newcommand{\be}{\begin{equation}}
\newcommand{\ee}{\end{equation}}
\newcommand{\bea}{\begin{eqnarray}}
\newcommand{\eea}{\end{eqnarray}}
\newcommand{\bse}{\begin{subequations}}
\newcommand{\ese}{\end{subequations}}
\newcommand{\ema}{${\rm EuMn_2As_2}$}
\newcommand{\ekxma}{Eu$_{1-x}$K$_x$Mn$_2$As$_2$}
\newcommand{\ekyma}{${\rm Eu_{0.96}K_{0.04}Mn_2As_2}$}
\newcommand{\ekzma}{${\rm Eu_{0.93}K_{0.07}Mn_2As_2}$}

\begin{document}
\title{Metallic behavior induced by potassium doping\\ of the trigonal antiferromagnetic insulator EuMn$_2$As$_2$}
\author{V. K. Anand}
\email{vivekkranand@gmail.com}
\affiliation {Ames Laboratory and Department of Physics and Astronomy, Iowa State University, Ames, Iowa 50011, USA}
\affiliation{\mbox{Helmholtz-Zentrum Berlin f\"{u}r Materialien und Energie GmbH, Hahn-Meitner Platz 1, D-14109 Berlin, Germany}}
\author{D. C. Johnston}
\email{johnston@ameslab.gov}
\affiliation {Ames Laboratory and Department of Physics and Astronomy, Iowa State University, Ames, Iowa 50011, USA}

\date{\today}

\begin{abstract}

We report magnetic susceptibility $\chi$, isothermal magnetization $M$, heat capacity $C_{\rm p}$ and electrical resistivity $\rho$ measurements on undoped \ema\ and K-doped \ekyma\ and \ekzma\ single crystals with the trigonal ${\rm CaAl_2Si_2}$-type structure as a function of temperature $T$ and magnetic field $H$\@. \ema\ has an insulating ground state with an activation energy of 52~meV and exhibits antiferromagnetic (AFM) ordering of the Eu$^{+2}$ spins $S=7/2$ at $T_{\rm N1} = 15$~K from $C_{\rm p}(T)$ and $\chi(T)$ data with a likely spin-reorientation transition at $T_{\rm N2} = 5.0$~K\@.  The Mn$^{+2}$ $3d^5$ spins-5/2 exhibit AFM ordering at $T_{\rm N}=142$~K from all three types of measurements.  The $M(H)$ isotherm and $\chi(T)$ data indicate that the Eu AFM structure is both noncollinear and noncoplanar.  The AFM structure of the Mn spins is also unclear.  A 4\% substitution of K for Eu in ${\rm Eu_{0.96}K_{0.04}Mn_2As_2}$ is sufficient to induce a metallic ground state.  Evidence is found for a difference in the AFM structure of the Eu moments in the metallic crystals from that of undoped EuMn$_2$As$_2$ versus both $T$ and~$H$\@.  For metallic ${\rm Eu_{0.96}K_{0.04}Mn_2As_2}$ and ${\rm Eu_{0.93}K_{0.07}Mn_2As_2}$, an anomalous S-shape $T$ dependence of $\rho$ related to the Mn magnetism is found.  Upon cooling from 200~K, $\rho$ exhibits a strong negative curvature, reaches maximum positive slope at the Mn $T_{\rm N} \approx 150$~K, and then continues to decrease but more slowly below $T_{\rm N}$\@.  This suggests that dynamic short-range AFM order of the Mn spins {\it above} the Mn $T_{\rm N}$ strongly suppresses the resistivity, contrary to the conventional decrease of $\rho$ that is only observed upon cooling below $T_{\rm N}$ of an antiferromagnet.

\end{abstract}


\maketitle

\section{\label{Intro} Introduction}

The Eu-based 122 pnictides such as ${\rm EuFe_2As_2}$, which present a potential avenue for investigations of the interplay and coexistence of long-range magnetic order and superconductivity, have recently attracted considerable attention in the ongoing research activities on high-$T_{\rm c}$ pnictide superconductors  \cite{Johnston2010, Stewart2011, Scalapino2012, Dagotto2013, Fernandes2014, Hosono2015, Dai2015, Inosov2016, Si2016, Ren2008, Jeevan2008, Anupam2011, Jiang2009b, Guguchia2011, Ren2009a, Jeevan2011, Miclea2009, Paramanik2013, Kurita2011}. The substitutions at Eu, Fe or As sites as well as the application of pressure have been found to suppress the antiferromagnetic (AFM)  spin-density-wave transition of the parent ${\rm EuFe_2As_2}$ leading to superconductivity which coexists with the ordered Eu$^{+2}$ moments. Recent neutron diffraction studies \cite{Jin2013,Nandi2014, Jin2015, Anand2015a} of the magnetic structure of Co-doped, Ir-doped and P-doped ${\rm EuFe_2As_2}$ reveal ferromagnetically (FM) ordered Eu moments with the moments oriented along the $c$~axis coexisting with the superconductivity.

Continuing our work on 122-type pnictides in the search for novel properties and ground states \cite{Anand2012a, Anand2012b, Pandey2013, Anand2013,Anand2014a, Anand2014b, Anand2014c, Anand2015b}, here we report a comprehensive investigation of the physical properties of ${\rm EuMn_2As_2}$ which crystallizes in the trigonal ${\rm CaAl_2Si_2}$-type structure (space group $P\bar{3}m1$) \cite{Ruhl1979}. This structure is quite different from the common body-centered tetragonal ${\rm ThCr_2Si_2}$-type structure of 122 compounds \cite{Zheng1988}. The formation of the ${\rm CaAl_2Si_2}$-type $AB_2X_2$ materials is restricted by a specific condition: the transition metals at the $B$-site must have a filled or half-filled $d$ electronic configuration ($d^0$, $d^5$ or $d^{10}$) and with few exceptions, the electron electrovalency count per formula unit is generally 16 [e.g.\ for ${\rm Ca^{+2}(Al^{+3})_2(Si^{-4})_2}$, $d^0$ ] or 12 [e.g.\ for ${\rm Ca^{+2}(Mn^{+2})_2(P^{-3})_2}$, $d^5$; ${\rm Ca^{+2}(Zn^{+2})_2(P^{-3})_2}$, $d^{10}$] \cite{Zheng1988,Klufers1977,Klufers1984}. In a ${\rm ThCr_2Si_2}$-type structure the $B$~site can be occupied by a variety of transition metals without any restrictions on the number of $d$ electrons. The ${\rm CaAl_2Si_2}$-structure compounds $A{\rm Zn_2Sb_2}$ ($A$ = Sr, Ca, Yb, Eu) exhibit small band gaps  with high carrier mobilities and low lattice thermal conductivities and hence have a high thermoelectric efficiency which identifies them as potentially useful thermoelectric materials \cite{Gascoin2005,Zhang2008,Toberer2010}.

The magnetic and transport properties of several ${\rm CaAl_2Si_2}$-type $AB_2X_2$ compounds with $A=$~Eu have been reported.  ${\rm EuZn_2Sb_2}$ is an intrinsic semiconductor, exhibits Eu AFM ordering with a N\'eel temperature $T_{\rm N}\approx13$~K and is a potential thermoelectric material \cite{Zhang2008, Toberer2010, Weber2006, May2012, Pomrehn2014}.  AFM ${\rm EuCd_2Sb_2}$ ($T_{\rm N} =7.4$~K) has a semimetallic or metallic  ground state \cite{Zhang2010,Goryunov2012}. ${\rm EuZn_2As_2}$ also exhibits AFM ordering ($T_{\rm N} =16.5$~K) and is semiconducting \cite{Goryunov2012}. ${\rm EuMn_2P_2}$ is an AFM insulator  with $T_{\rm N} =16.5$~K \cite{Payne2002}.  ${\rm EuMn_2Sb_2}$ exhibits both Eu and Mn AFM ordering; the Eu spins order at about 9~K and the Mn spins order at $\sim 128$~K  \cite{Schellenberg2010, FMSchll2010}.  Mn moment ordering has also been observed in semiconducting ${\rm CaMn_2Sb_2}$ \cite{Simonson2012} and ${\rm CaMn_2Bi_2}$ \cite{Gibson2015} with $T_{\rm N} = 85$~K and 154~K, respectively.

Very recently the crystallography and physical properties of trigonal ${\rm CaAl_2Si_2}$-type ${\rm CaMn_2As_2}$ and ${\rm SrMn_2As_2}$ single crystals have been investigated.  Both compounds are AFM semiconductors with $T_{\rm N} = 62$~K and 120~K, respectively \cite{Sangeetha2016, Das2016}.  A previous study of ${\rm SrMn_2As_2}$ single crystals found $T_{\rm N} = 125$~K \cite{Wang2011}.  The anisotropic magnetic susceptibility versus temperature $\chi(T)$ data for both compounds indicate that the easy plane for AFM ordering of the Mn$^{+2}$ spins $S = 5/2$ is the hexagonal $ab$~plane \cite{Sangeetha2016}.  Neutron diffraction measurements of ${\rm SrMn_2As_2}$ revealed that the Mn spins exhibit collinear AFM ordering in the $ab$~plane with propagation vector ${\bf k} = (0,0,0)$, i.e., the AFM and chemical unit cells are the same \cite{Das2016}. The $\chi(T)$ and neutron diffraction studies together indicate that the magnetism is quasi-two-dimensional with strong AFM correlations surviving to at least 900~K\@.

Herein we report the crystallography and physical properties of trigonal ${\rm CaAl_2Si_2}$-type ${\rm EuMn_2As_2}$ single crystals where both the Mn and Eu ions are magnetic.  One expects the compound to contain Mn$^{+2}$ ions with a $3d^5$ electronic configuration and $S=5/2$ and Eu$^{+2}$ ions with a $4f^7$ configuration and $S=7/2$.   Our $\chi(T)$, isothermal magnetization $M$ versus magnetic field $H$ and heat capacity $C_{\rm p}(T)$ measurements reveal an AFM ground state in ${\rm EuMn_2As_2}$. Two transitions at $T_{\rm N1} = 15.0$~K and $T_{\rm N2} = 5.0$~K are seen in $\chi(T)$ measured in low~$H$ associated with the Eu spins, where the latter is likely a spin reorientation transition. On the other hand, the $C_{\rm p}(T)$ and $\chi(T)$ measurements also show a transition at 142~K, which is attributed to AFM ordering of Mn moments since this $T_{\rm N}$ is similar to the above $T_{\rm N} = 120$~K for trigonal ${\rm SrMn_2As_2}$ and the Sr$^{+2}$ and Eu$^{+2}$ ions have nearly the same radius.  The electrical resistivity $\rho(T)$ data reveal an insulating ground state in ${\rm EuMn_2As_2}$ with an activation energy of 52~meV\@.

The physical properties of ${\rm EuMn_2As_2}$ are similar to those of ${\rm BaMn_2As_2}$ \cite{Singh2009a}; however, the latter compound forms in the body-centered tetragonal ${\rm ThCr_2Si_2}$-type structure. ${\rm BaMn_2As_2}$ is AFM below $T_{\rm N} = 625$~K where the Mn spins ($S = 5/2$) adopt a G-type (N\'eel-type) AFM structure (moments oriented along the $c$ axis) with an insulating ground state \cite{Singh2009a, An2009, Singh2009b, Johnston2011}.  A small 1.6\% substitution of K for Ba in ${\rm BaMn_2As_2}$ induces a metallic ground state \cite{Pandey2012}. In addition, higher K~substitution results in ferromagnetism in Ba$_{1-x}$K$_x$Mn$_2$As$_2$ $(x = 0.19,~0.26, 0.40)$ with half-metallic FM for $x=0.40$ \cite{Bao2012,Lamsal2013,Pandey2013b} and also for 60\% Rb-doped crystals \cite{Pandey2015}.

In view of the interesting results arising from hole doping of K and Rb for Ba in ${\rm BaMn_2As_2}$, we investigated and report here the effects of K~substitution for Eu on the properties of ${\rm EuMn_2As_2}$. We find metallic ground states in ${\rm Eu_{0.96}K_{0.04}Mn_2As_2}$ and ${\rm Eu_{0.93}K_{0.07}Mn_2As_2}$ single crystals.  The K~substitution also seems to affect the AFM structure. The experimental details are given in Sec.~\ref{ExpDetails} followed by the crystallography results in Sec.~\ref{Sec:crystallography}. Sections~\ref{Sec:EuMn2AS2}, \ref{Sec:EuKMn2As2_1} and~\ref{Sec:EuKMn2As2_2} are devoted to results and analyses on ${\rm EuMn_2As_2}$, ${\rm Eu_{0.96}K_{0.04}Mn_2As_2}$ and ${\rm Eu_{0.93}K_{0.07}Mn_2As_2}$, respectively, and a summary and discussion are given in Sec.~\ref{Conclusion}.

\section{\label{ExpDetails} Experimental Details}

Single crystals of ${\rm EuMn_2As_2}$ were grown using Sn flux. High-purity Eu (Ames Lab), and Mn (99.998\%), As (99.99999\%) and Sn (99.9999\%) from Alfa Aesar, were taken in a Eu:Mn:As:Sn = 1:2:2:25 molar ratio and placed in an alumina crucible which was then sealed in an evacuated quartz tube. The crystal growth was carried by heating the tube to 1100~$^\circ$C at a rate of 60~$^\circ$C/h, held for 35~h for homogenization, followed by slow cooling at  2.5~$^\circ$C/h to 800~$^\circ$C at which temperature the crystals were removed from the flux by centrifuging. Shiny platelike crystals of typical size $2.5 \times 2 \times 0.6$~mm$^3$ were obtained.

The single crystals of K-doped Eu$_{1-x}$K$_x$Mn$_2$As$_2$ were grown using self flux. Nominal mixtures Eu$_{1-x}$K$_x$ ($x=0.25$, 0.5, 0.75) [K (99.95\%), Alfa Aesar] and prereacted MnAs were taken in a 1:4 molar ratio which were placed in alumina crucibles and sealed in tantalum tubes which were then sealed in quartz tubes partially filled with argon ($\approx 1/4$ atm pressure). The doubly-sealed samples were heated to 1200~$^\circ$C at 60~$^\circ$C/h, held there for 25~h, heated to 1250~$^\circ$C in 1~h, held there for 6~h and then cooled slowly at 2 $^\circ$C/h to 1130~$^\circ$C where the flux was decanted with a centrifuge.  Shiny platelike crystals of typical size $2 \times 1.5 \times 0.5$~mm$^3$ were obtained for nominal $x=0.25$ and 0.5; however, no crystals were obtained for nominal $x=0.75$ using this temperature profile. The masses of the crystals used for the $\chi(T)$, $M(H)$ and $C_{\rm p}(T)$ measurements were typically $\sim7$~mg.

The single-phase nature and quality of the crystals were checked by high resolution images using a JEOL scanning electron microscope (SEM). An energy-dispersive x-ray (EDX) analyzer attachment to the SEM was used to determine the chemical compositions of the crystals which yielded the molar ratio Eu:Mn:As = 1.00(1)\,:\,2.00(2)\,:\,2.00(2) for the Sn-flux grown ${\rm EuMn_2As_2}$. For the self-flux grown K-doped ${\rm EuMn_2As_2}$ crystals the average chemical compositions obtained from the EDX analyses were ${\rm Eu_{0.96(1)}K_{0.04(1)}Mn_2As_2}$ for nominal $x=0.25$ and ${\rm Eu_{0.93(1)}K_{0.07(1)}Mn_2As_2}$ for nominal $x=0.5$ melts. The crystal structure was checked by powder x-ray diffraction (XRD) on crushed single crystals using Cu~K$_\alpha$ radiation and a Rigaku Geigerflex x-ray diffractometer.  The XRD data were analyzed by Rietveld refinement using the software {\tt FullProf} \cite{Rodriguez1993}.

${\rm SrZn_2As_2}$ is isotructural to \ema\ \cite{Mewis1980}.  Crystals were grown using Sn flux with a nominal molar ratio ${\rm SrZn_2As_2}$:Sn = 1:20. The crystal growth was carried out by heating at 60~$^\circ$C/h to 1150~$^\circ$C, holding there for 35~h, followed by slow-cooling to 790~$^\circ$C over 120~h at 3 $^\circ$C/h.  Rietveld refinement of powder XRD measurements on crushed crystals gave lattice parameters $a=4.2202(1)$~\AA\ and $c=7.2632(2)$~\AA\ and  atomic coordinates $z_{\rm As} =  0.7317(4)$ and $z_{\rm Zn} = 0.3704(3)$.  The crystal data are in good agreement with \cite{Mewis1980}.  Semiquantitative EDX analysis indicated the expected 1:2:2 stoichiomentry apart from a small Sr deficiency ($\sim3$\%) and a trace amount ($\sim 1.3$~mol\%) of Sn incorporated into the crystal structure.

The $M(T)$ measurements at fixed $H$ and $M(H)$ isotherm measurements for $H \leq 5.5$~T were performed using a Quantum Design, Inc., superconducting quantum interference device magnetic properties measurement system (MPMS). The high magnetic field $M(H)$ isotherm measurements for $H \leq 14$~T were performed using the vibrating sample magnetometer (VSM) option of a Quantum Design, Inc., physical properties measurement system (PPMS).  The $C_{\rm p}(T)$ was measured by a relaxation method using the heat capacity option of the PPMS\@. For the undoped and K-doped \ema\ crystals, the additional noise in the $C_{\rm p}$ data above 200~K is believed to be an experimental artifact.  The $\rho(T)$ data were  obtained by the standard four-probe ac technique using the the ac transport option of PPMS\@. The electrical leads to the ${\rm EuMn_2As_2}$ crystal were prepared by soldering 25~$\mu$m diameter Pt wire to the crystal whereas for Eu$_{1-x}$K$_x$Mn$_2$As$_2$ ($x=0.04$, 0.07) crystals silver epoxy was used to attach the Pt-wire leads.

\section{\label{Sec:crystallography} Crystallography}

\begin{figure}
\includegraphics[width=3.3in]{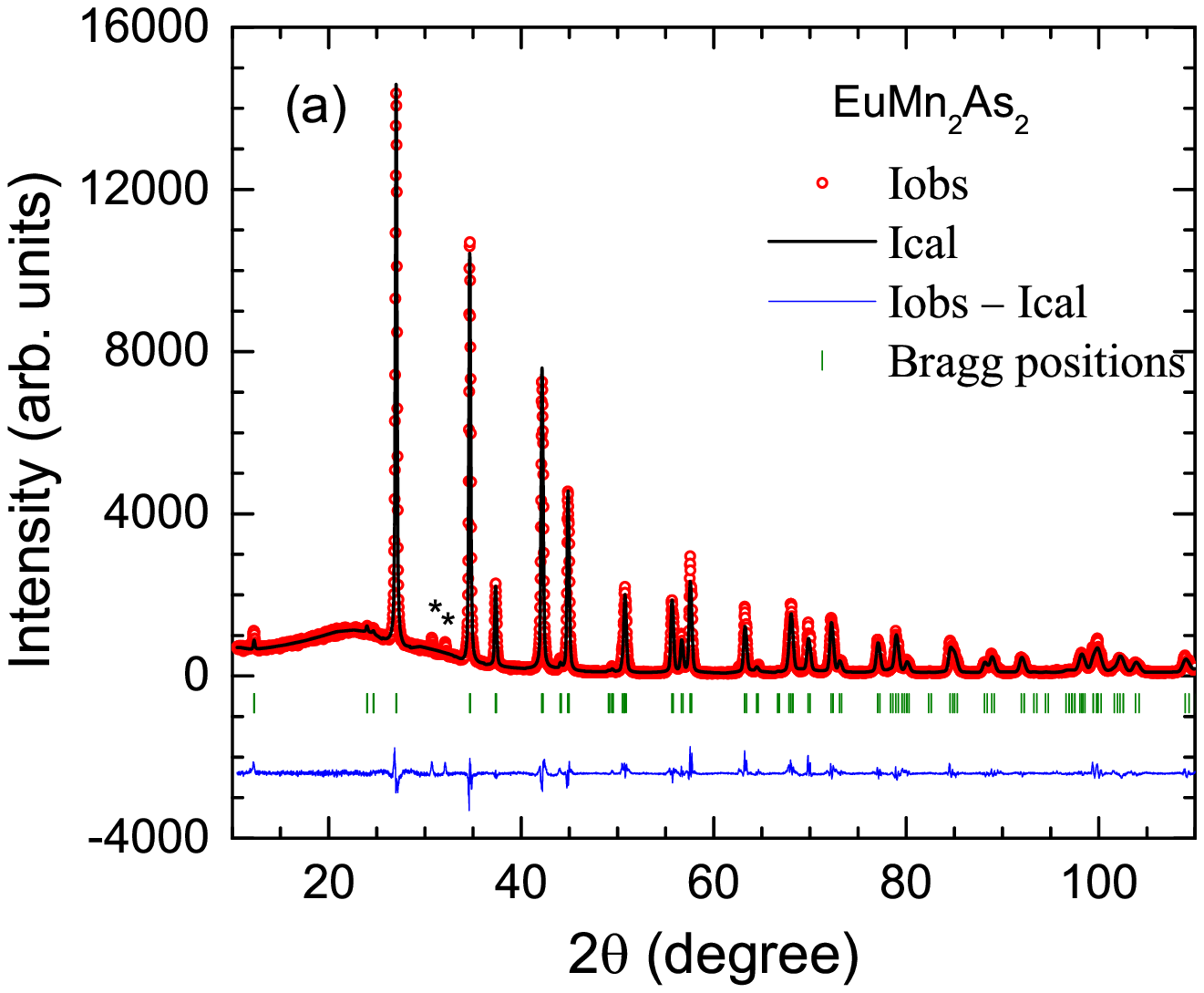}\vspace{0.1in}
\includegraphics[width=3.3in]{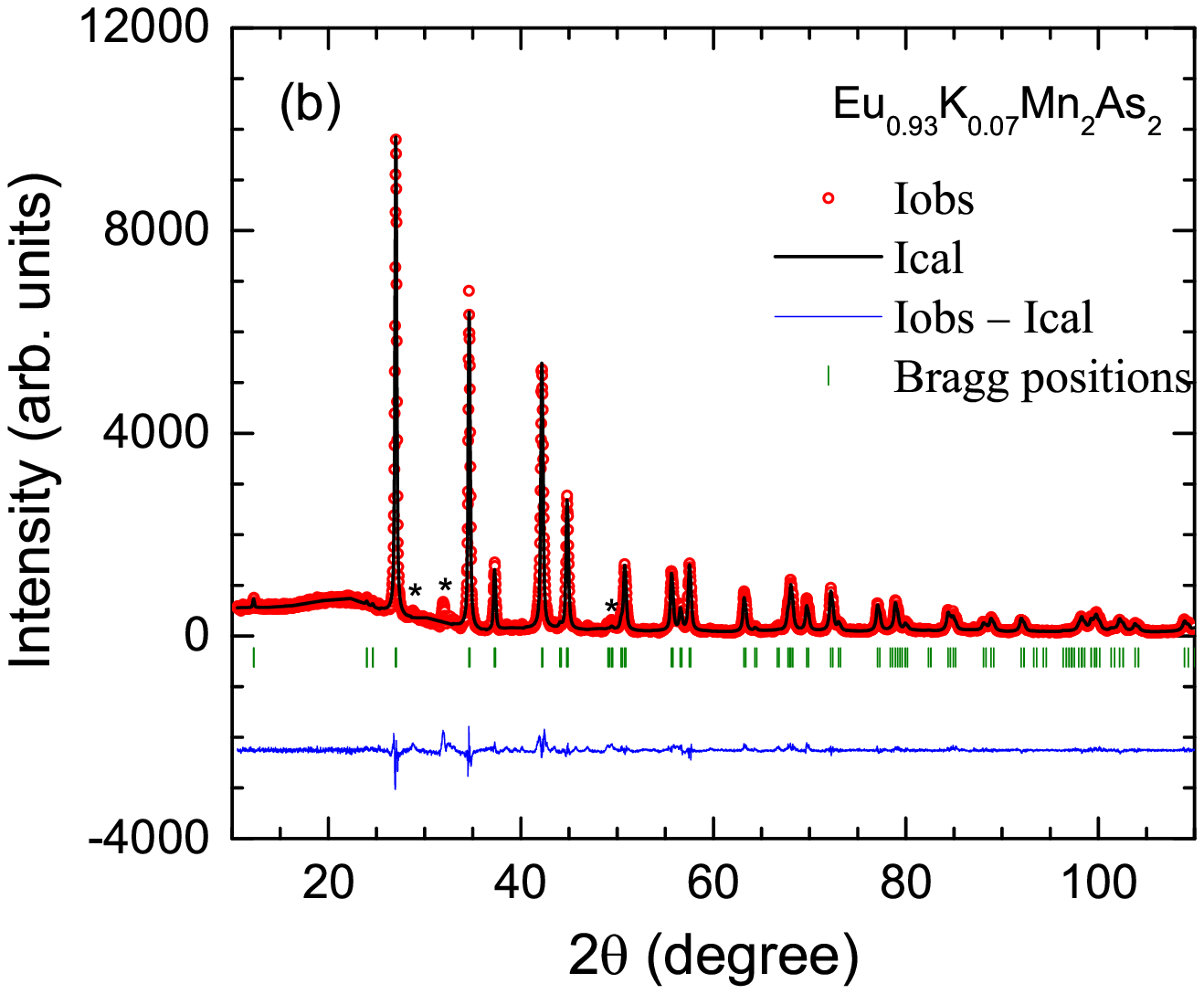}
\caption {(Colour online) Powder x-ray diffraction patterns of (a) ${\rm EuMn_2As_2}$ and (b) ${\rm Eu_{0.93}K_{0.07}Mn_2As_2}$ recorded at room temperature. The solid lines through the experimental points are the Rietveld refinement profiles calculated for the ${\rm CaAl_2Si_2}$-type trigonal structure (space group $P\bar{3}m1$). The short vertical bars mark the Bragg peak positions. The lowermost curves represent the differences between the experimental and calculated intensities. The unindexed peaks marked with stars correspond to peaks from the residual flux on the surface of the samples.}
\label{fig:EuMn2As2_XRD}
\end{figure}

\begin{table}
\caption{\label{tab:XRD1} Crystallographic and Rietveld refinement parameters obtained from powder XRD data for  crushed {Eu$_{1-x}$K$_x$Mn$_2$As$_2$} ($x=0$, 0.04, 0.07) single crystals crystallizing in the ${\rm CaAl_2Si_2}$-type trigonal structure (space group $P\bar{3}m1$, No.~164). The Wyckoff positions of Eu/K, Mn and As atoms in the hexagonal setting are $1a$ (0, 0, 0), $2d$ (1/3, 2/3, $z_{\rm Mn}$) and $2d$ (1/3, 2/3, $z_{\rm As}$), respectively.}
\begin{ruledtabular}
\begin{tabular}{lccc}
  & $x=0$ & $x=0.04$ & $x= 0.07$\\
\hline
\underline{Lattice parameters}\\
\hspace{0.8cm} $a$ (\AA)            		& 4.2846(1) & 4.2831(2) & 4.2839(2) \\	
\hspace{0.8cm} $c$ (\AA)          			& 7.2217(3) & 7.2295(3) & 7.2359(3) \\
\hspace{0.8cm} $c/a$          			& 1.6855(1) & 1.6879(2) & 1.6891(2) \\
\hspace{0.8cm} $V_{\rm cell}$  (\AA$^{3}$) 	& 114.81(1) & 114.86(2) & 115.00(2) \\
\hspace{0.8cm} $\rho_{\rm calc}$  (g/cm$^{3}$) 	& 5.95 & 5.89 & 5.83 \\
\underline{Atomic coordinates}\\
\hspace{0.8cm} $z_{\rm Mn}$                 & 0.3800(4) & 0.3779(4) & 0.3786(4) \\
\hspace{0.8cm} $z_{\rm As}$                 & 0.7343(3) & 0.7357(4) & 0.7344(3) \\
\underline{Refinement quality} \\
\hspace{0.8cm} $\chi^2$          & 4.66 & 5.08 & 4.16 \\	
\hspace{0.8cm} $R_{\rm p}$ (\%)  & 6.85 & 7.84 & 7.37 \\
\hspace{0.8cm} $R_{\rm wp}$ (\%) & 9.75 & 11.6 & 10.8 \\
\end{tabular}
\end{ruledtabular}
\end{table}

The room-temperature powder XRD data collected on crushed crystals of ${\rm EuMn_2As_2}$ and \ekzma\ are shown in Figs.~\ref{fig:EuMn2As2_XRD}(a) and~\ref{fig:EuMn2As2_XRD}(b), respectively. Rietveld refinements of the XRD patterns revealed the single-phase nature of the crystals and confirmed their ${\rm CaAl_2Si_2}$-type trigonal structure (space group $P\bar{3}m1$).  The Rietveld refinement profiles are shown by solid lines in Fig.~\ref{fig:EuMn2As2_XRD}. The weak unindexed peaks marked with stars in the patterns are extrinsic and arise from residual flux on the surface of the crystals. The crystallographic and refinement parameters are listed in Table~\ref{tab:XRD1}. The lattice parameters $a$ and $c$ are found to be in good agreement with the reported values \cite{Ruhl1979}.

\begin{figure} 
\includegraphics[width=3.3in]{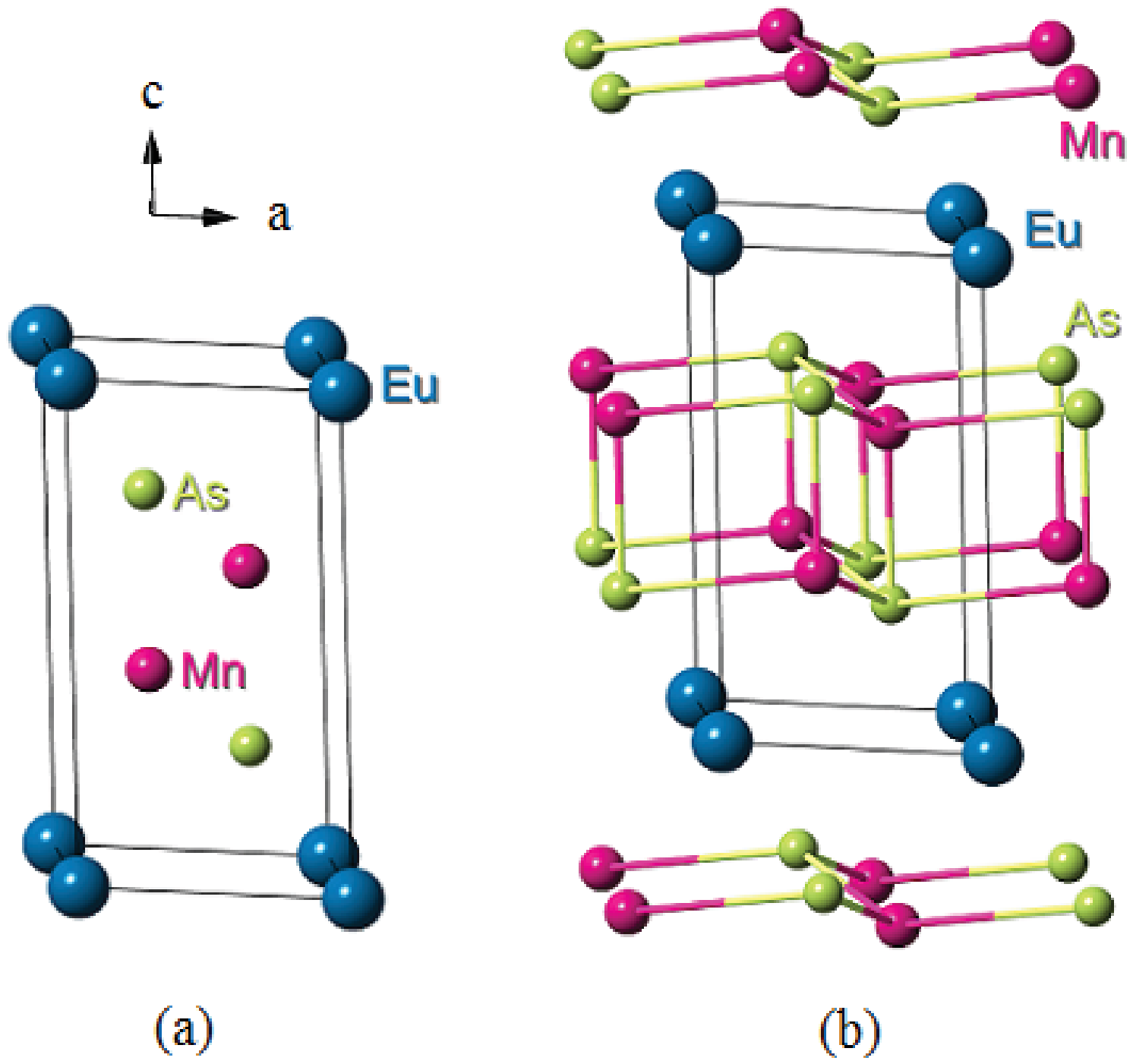}
\includegraphics[width=3.3in]{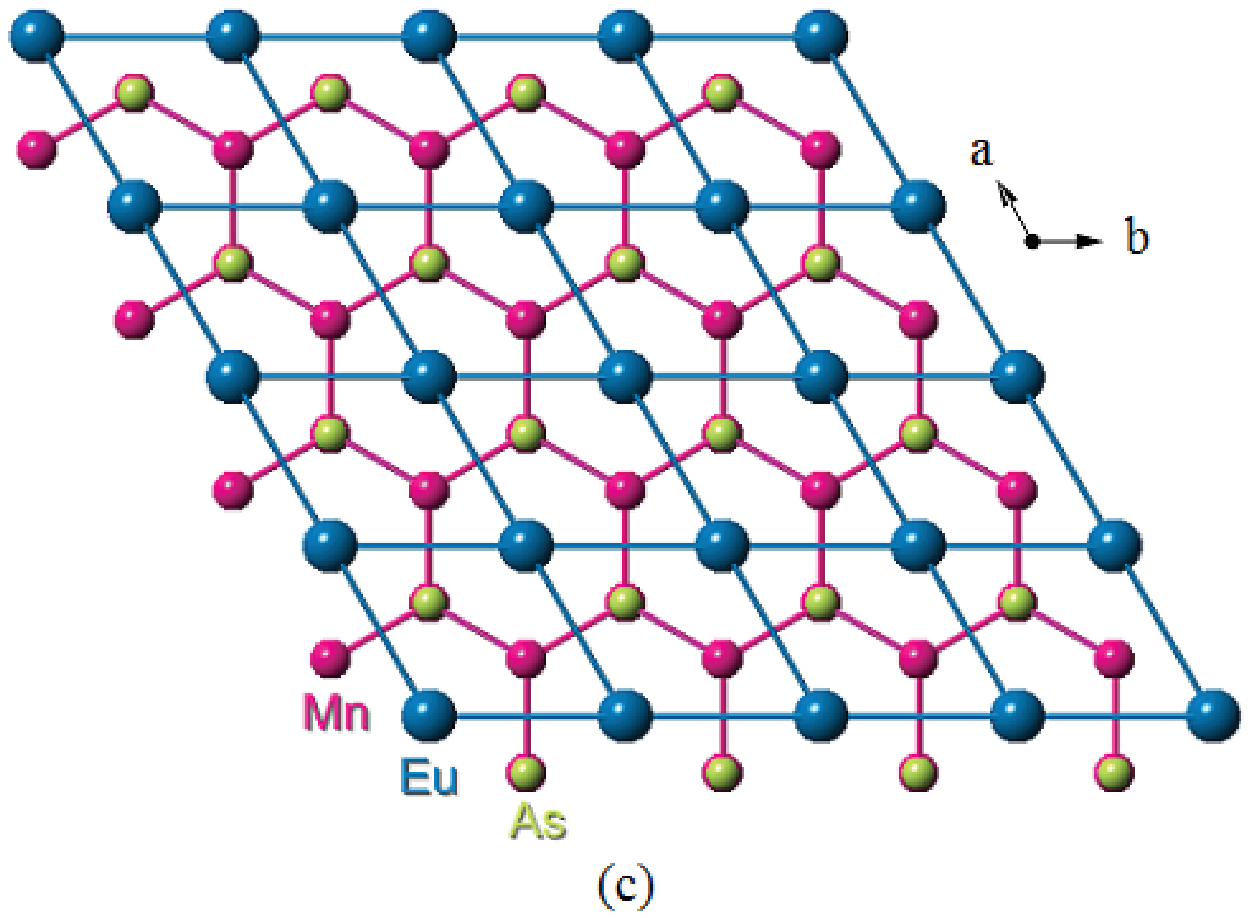}
\caption{(Color online) (a) ${\rm CaAl_2As_2}$-type trigonal crystal structure of ${\rm EuMn_2As_2}$ (space group $P\bar{3}m1$, No.~164). The spheres represent Eu, Mn and As atoms in decreasing order of size. (b)~Extended view of the ${\rm EuMn_2As_2}$ structure showing the Mn-As layers and $[{\rm Mn_2As_2}]^{-2}$ networks. (c)~A projection of the ${\rm EuMn_2As_2}$ structure onto the $ab$~plane showing the hexagonal arrangement of Eu, Mn and As atoms. The Eu atoms form triangular-lattice layers and the Mn atoms form a corrugated honeycomb lattice. }
\label{fig:EuMn2As2_Struct}
\end{figure}

Figure~\ref{fig:EuMn2As2_Struct}(a) shows the ${\rm CaAl_2Si_2}$-type trigonal structure of ${\rm EuMn_2As_2}$ in the hexagonal setting.  As can be seen from Fig.~\ref{fig:EuMn2As2_Struct}(b) the structure consists of alternating layers of Eu$^{+2}$ and $[{\rm Mn_2As_2}]^{-2}$ slabs stacked along the $c$ axis. The Eu atoms are arranged in simple-hexagonal (triangular) lattice layers that are separated by $[{\rm Mn_2As_2}]^{-2}$ slabs.  A $[{\rm Mn_2As_2}]^{-2}$ slab contains a corrugated Mn honeycomb lattice \cite{Zheng1986}. The Mn and As are both fourfold coordinated with different coordination environments. The Mn atoms are coordinated approximately tetrahedrally by four As atoms forming MnAs$_4$ tetrahedra (both up- and down-pointing), and each As atom is surrounded umbrella-like by four Mn atoms forming an AsMn$_4$ unit with a flipped-tetrahedron coordination \cite{Zheng1986,Burdett1990}. The structure can also be viewed as a slightly distorted hexagonal close packing of As atoms with Mn atoms occupying half the tetrahedral sites and Eu atoms occupying half the octahedral sites \cite{Burdett1990}. Thus the Eu atoms have a slightly distorted octahedral coordination with six nearest As neighbors forming EuAs$_6$ octahedra.

The ${\rm CaAl_2Si_2}$ structure, though different from the familiar ${\rm ThCr_2Si_2}$ structure, shares some common features with it.  In both $AB_2X_2$ structures, only one crystallographic site is occupied by each of the $A$, $B$ and $X$ atoms. Both structures can be generated by stacking of layers of $A$ and $[B_2X_2]^{-2}$ slabs along the $c$ direction. Both structures have $BX_4$ tetrahedra which are edge shared to form the $[B_2X_2]^{-2}$ slabs, however they have different arrangements in the two structures \cite{Zheng1988,Zheng1986}. In the ${\rm ThCr_2Si_2}$ structure $BX_4$ tetrahedra share four of their six edges whereas in the ${\rm CaAl_2Si_2}$ structure the $BX_4$ tetrahedra share only three edges. Furthermore, the $[B_2X_2]^{-2}$ networks are three-dimensional in the  ${\rm ThCr_2Si_2}$ structure but two-dimensional in the ${\rm CaAl_2Si_2}$ structure. In the ${\rm ThCr_2Si_2}$ structure the $B$ atoms form square-planar layers and are each coordinated by four other $B$ atoms at 90$^\circ$, whereas in the ${\rm CaAl_2Si_2}$ structure the $B$ atoms are coordinated by only three other $B$ atoms at 90$^\circ$ like the corner of a cube \cite{Brock1994}.

\section{\label{Sec:EuMn2AS2} Physical properties of E\lowercase{u}M\lowercase{n}$_2$A\lowercase{s}$_2$}

\subsection{\label{Sec:EuMn2AS2_MT_MH} Magnetization and Magnetic Susceptibility}

\subsubsection{Magnetism of the Eu Spins}

\begin{figure}
\includegraphics[width=3.3in]{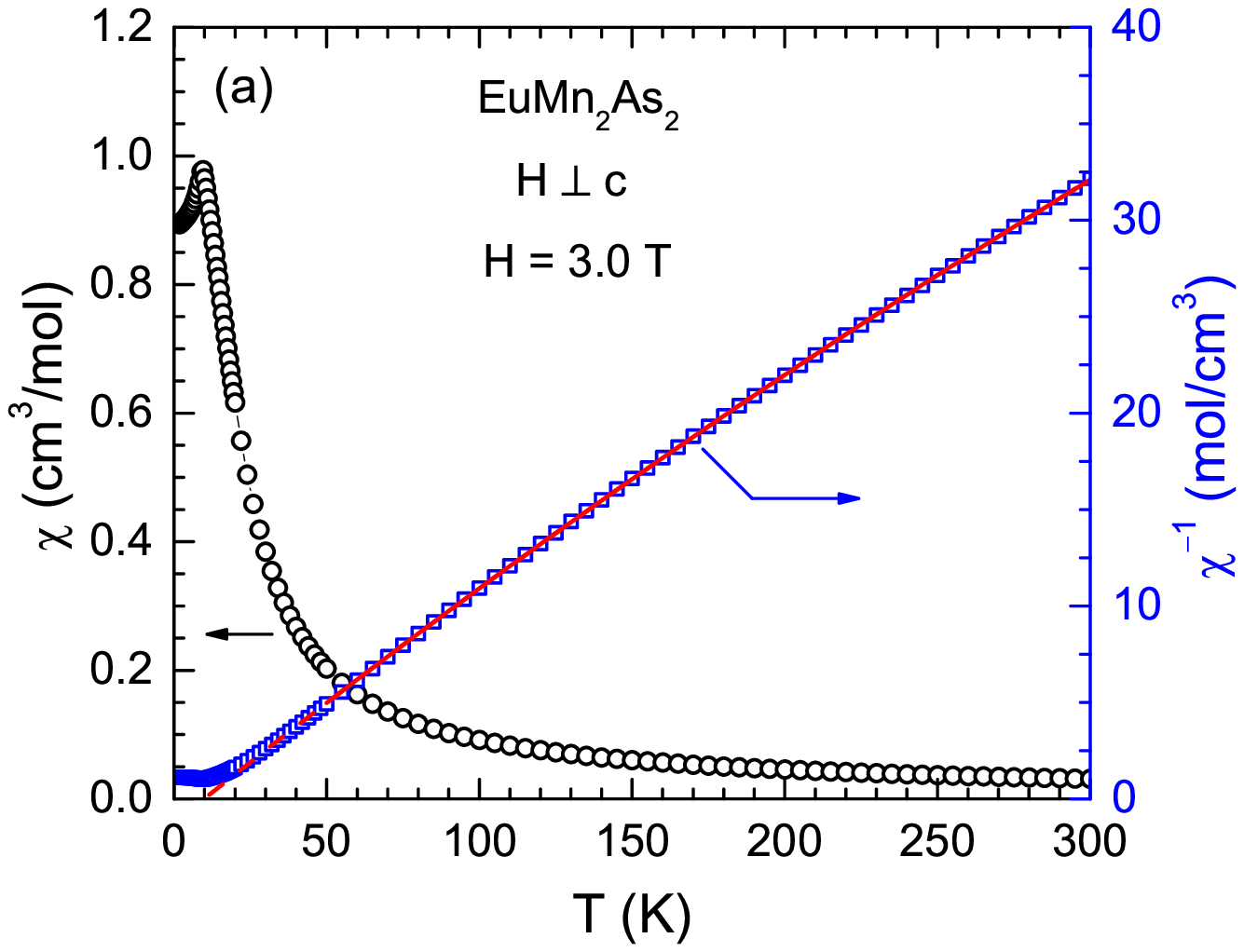}\vspace{0.1in}
\includegraphics[width=3.3in]{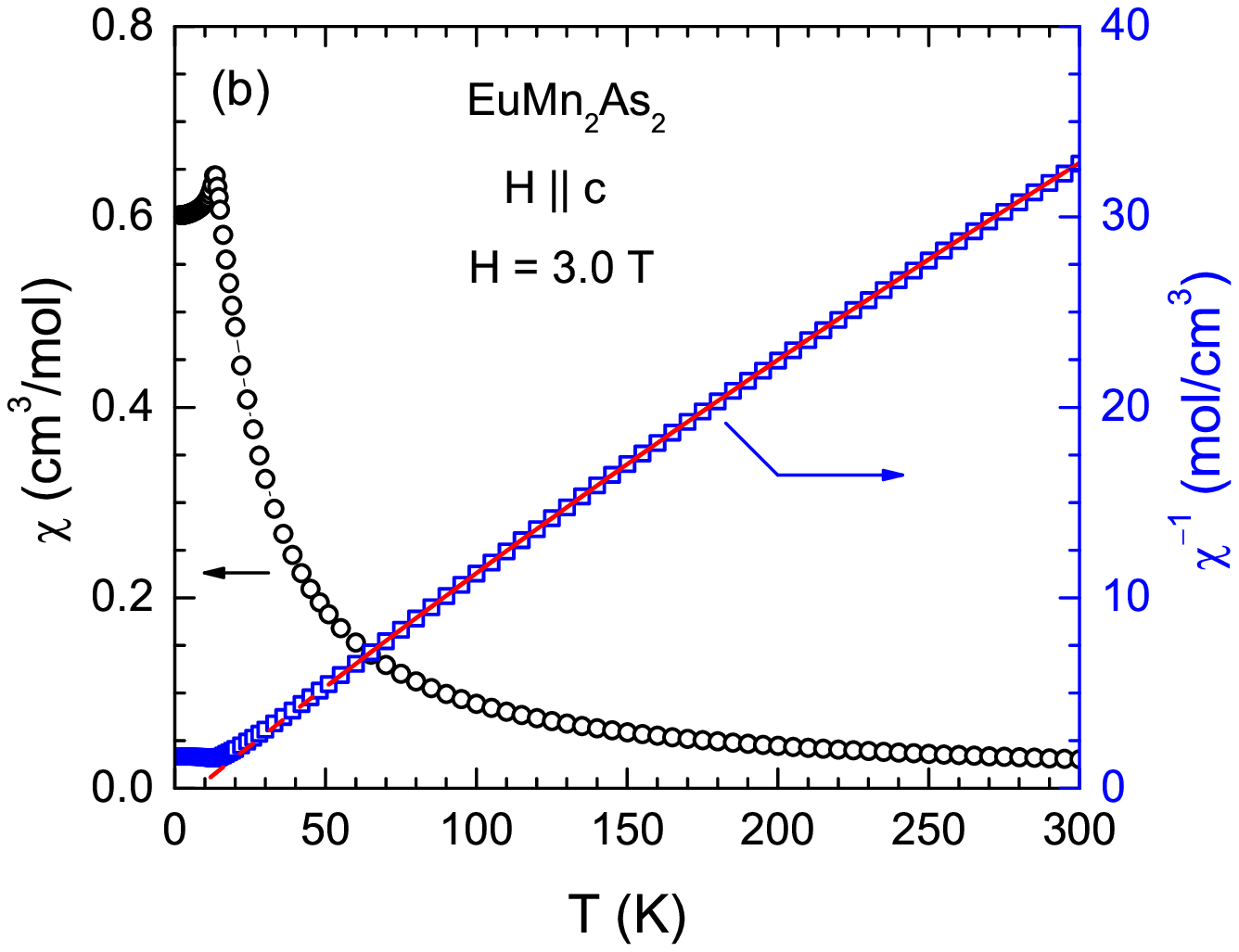}
\caption{(Color online) Zero-field-cooled magnetic susceptibility $\chi$ (left ordinate) and its inverse $\chi^{-1}$ (right ordinate) for an ${\rm EuMn_2As_2}$ crystal versus temperature $T$ for $1.8~{\rm K} \leq T \leq 300$~K in a magnetic field $H=3.0$~T applied (a) in the $ab$ plane ($\chi_{ab}, H \perp  c$) and (b) along the $c$ axis ($\chi_c, H \parallel c$). The solid red curves are fits of the $\chi^{-1}(T)$ data by the modified Curie-Weiss law~(\ref{eq:C-W}) over the $T$ range 50~K~$\leq T \leq$~300~K and the dashed curves are extrapolations.}
\label{fig:MTinv_EuMn2As2}
\end{figure}

The $\chi\equiv M/H$ and $\chi^{-1}$ measured in $H=3.0$~T for single-crystal \ema\ are shown for the $T$ range $1.8 ~{\rm K} \leq T \leq 300$~K with $H\perp c$ and $H\parallel c$ in Figs.~\ref{fig:MTinv_EuMn2As2}(a) and~\ref{fig:MTinv_EuMn2As2}(b), respectively. The $\chi(T)$ data for both spin directions reveal an AFM transition at $T_{\rm N1}=15.0$~K assigned to Eu$^{+2}$ spins $S=7/2$ as follows.

In the paramagnetic (PM) state of the Eu spins at $T \gtrsim 50$~K the $\chi(T)$ data follow the modified Curie-Weiss behavior
\begin{equation}
\chi(T) = \chi_0 + \frac{C}{T-\theta_{\rm p}}
\label{eq:C-W}
\end{equation}
where $\chi_0$ is a $T$-independent contribution to $\chi$, $C$ is the Curie constant and $\theta_{\rm p}$ is the Weiss temperature.  The fits of the $\chi^{-1}(T)$ data over the $T$ range 50~K~$\leq T \leq$~300~K by Eq.~(\ref{eq:C-W}) are shown as the red curves in Fig.~\ref{fig:MTinv_EuMn2As2}. The parameters $\chi_0,\ C$ and~$\theta_{\rm p}$ obtained from the fits are listed in Table~\ref{tab:CW}. The molar Curie constants are in agreement with the theoretical value of $7.88\, \mu_{\rm B}$/Eu for free Eu$^{+2}$ ions  with spin $S=7/2$ and spectroscopic splitting factor $g=2$. The Weiss temperatures $\theta_{\rm p}\approx 8$~K for both field directions suggest dominant FM interactions.  The value of $\chi_0$ for both field directions is comparable with the  average $\chi_{\rm ave}\sim 2.6\times10^{-3}~{\rm cm^3/mol}$ from 1.8 to 900~K for ${\rm SrMn_2As_2}$ which exhibits AFM ordering of the Mn$^{+2}$ spins $S=5/2$ at $T_{\rm N}=120$~K \cite{Sangeetha2016}.

\begin{figure}
\includegraphics[width=3.2in]{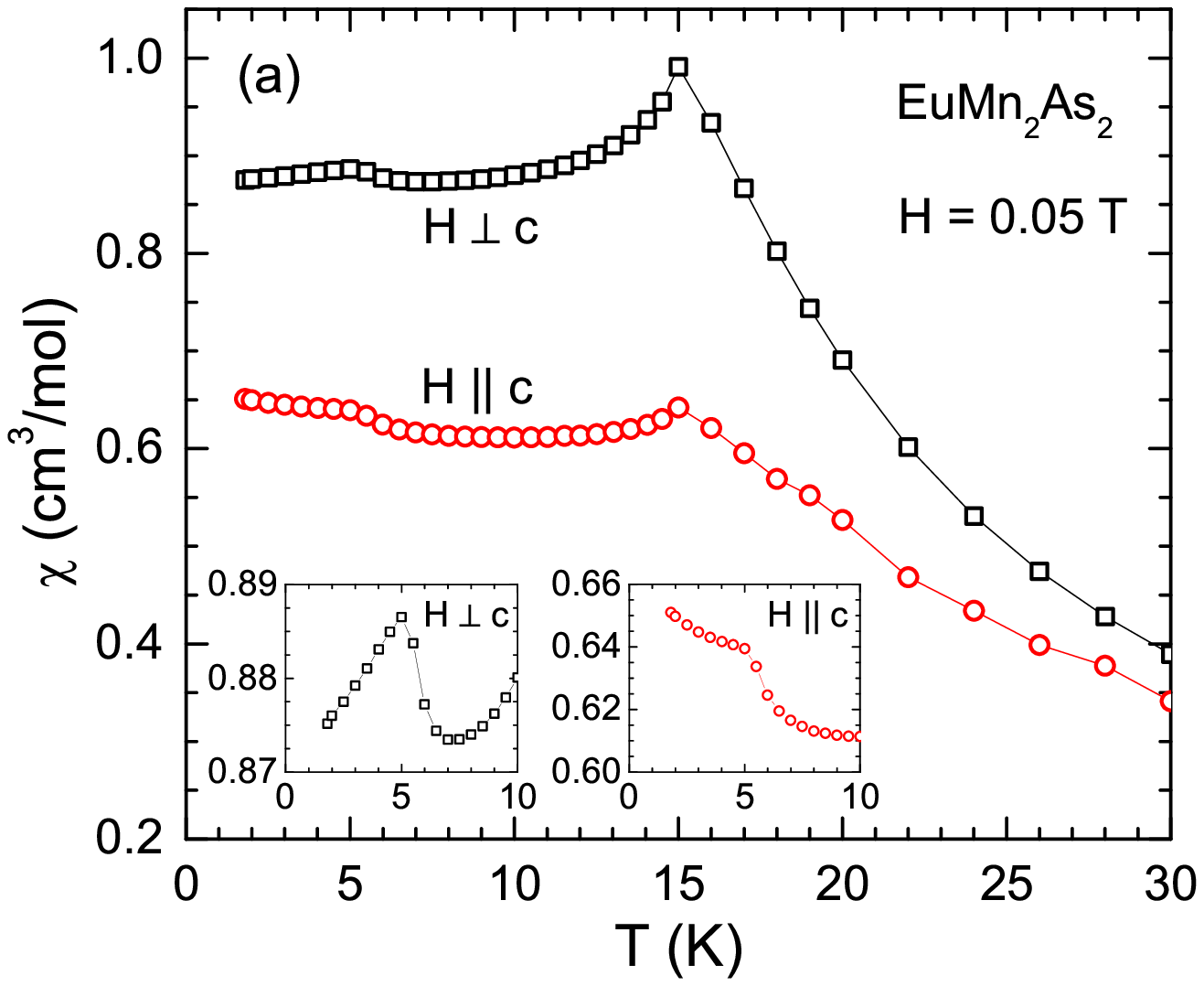}
\includegraphics[width=3.2in]{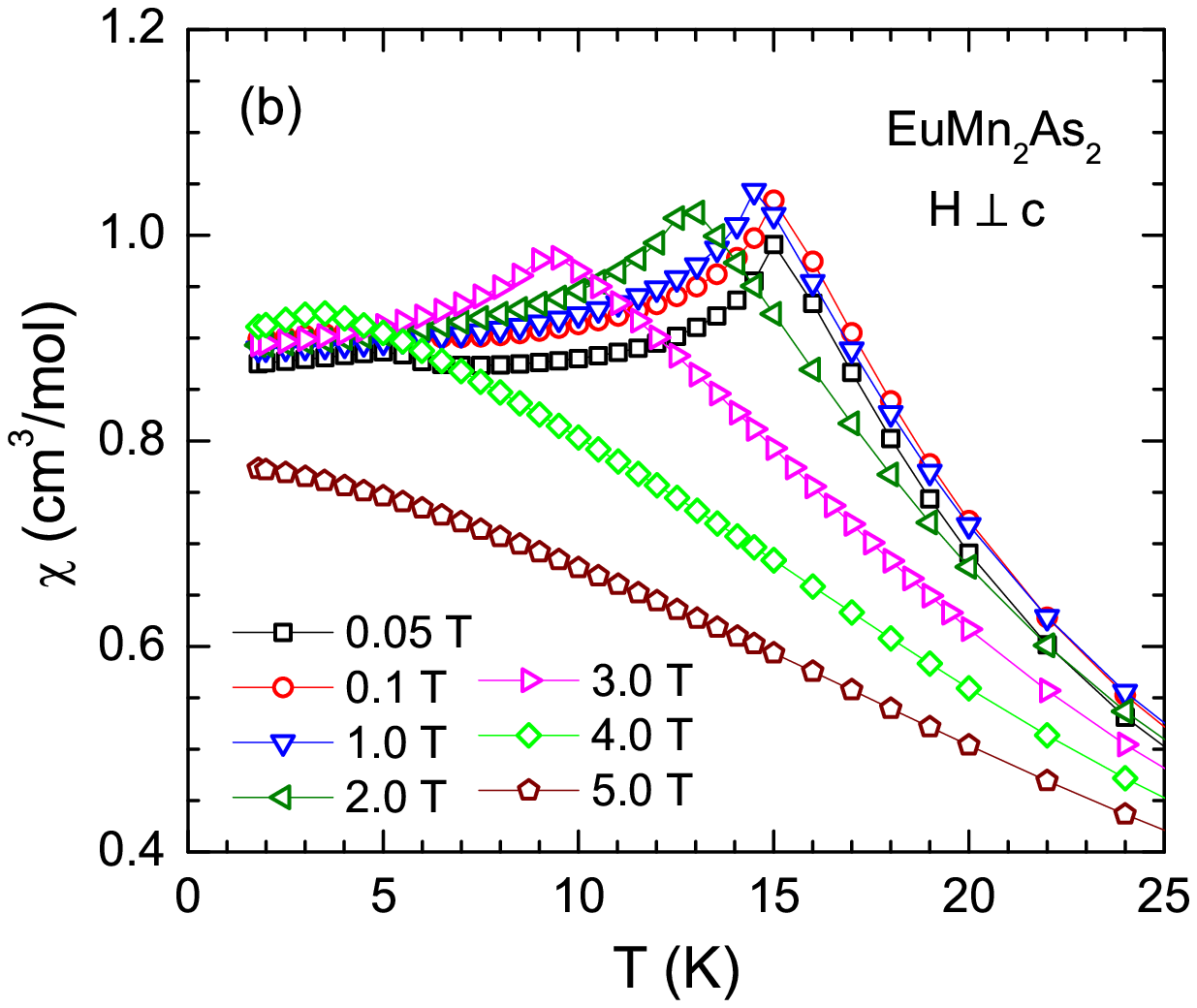}
\includegraphics[width=3.2in]{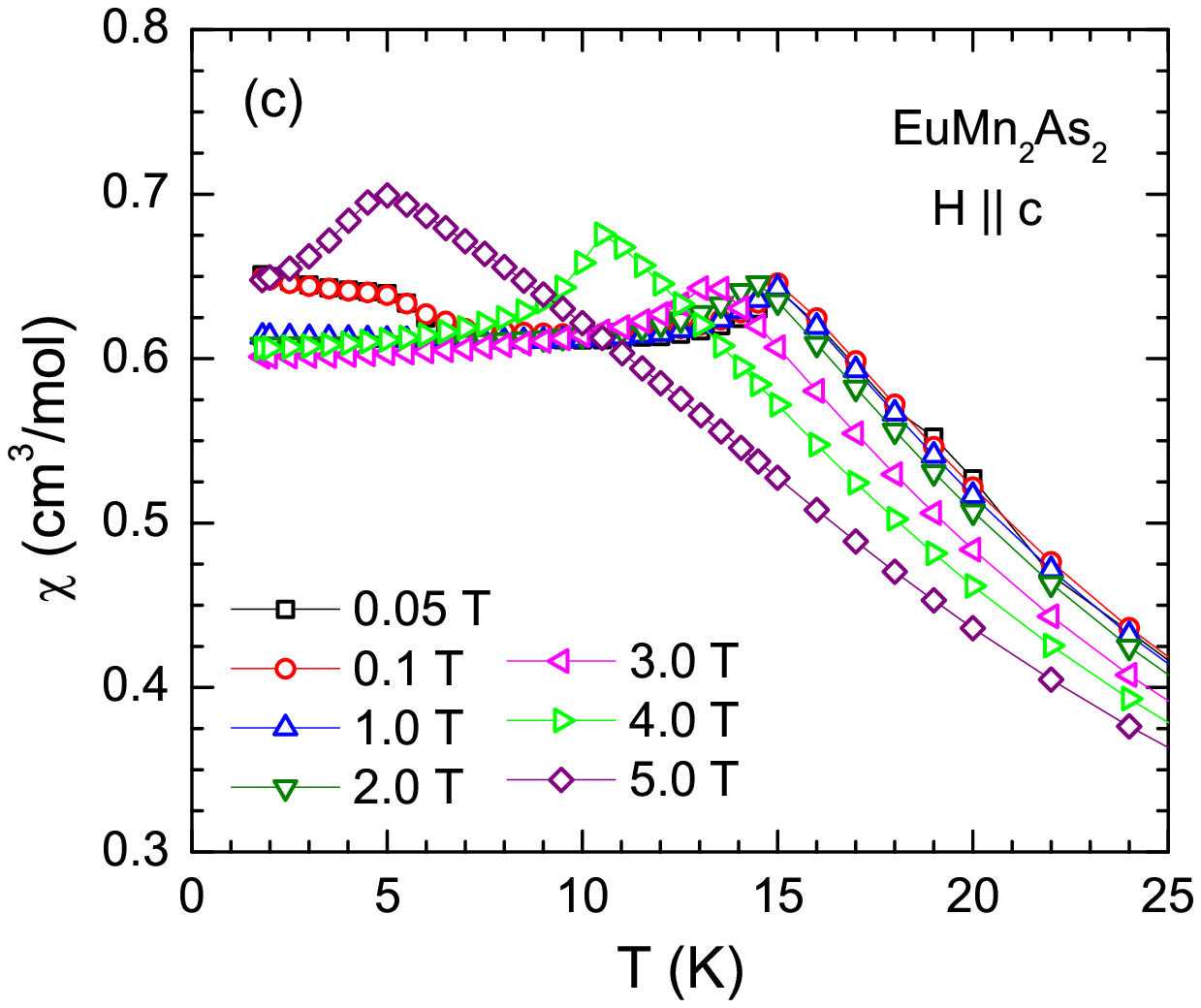}
\caption{(Color online) (a) Zero-field-cooled (ZFC) magnetic susceptibility $\chi$ of an ${\rm EuMn_2As_2}$ single crystal as a function of temperature $T$ for $1.8 ~{\rm K} \leq T < 30$~K measured in a magnetic field $H= 0.05$~T applied in the $ab$ plane ($\chi_{ab}, H \perp  c$) and along the $c$ axis ($\chi_c, H \parallel c$). Insets: Expanded views of $\chi(T)$ showing the anomalies near 5~K. (b) and~(c) ZFC $\chi_{ab}(T)$ and $\chi_c(T)$, respectively, for $1.8 ~{\rm K} \leq T \leq 25$~K measured at the indicated $H$ values.}
\label{fig:MT_EuMn2As2}
\end{figure}

The zero-field-cooled (ZFC) $\chi(T)$ data for an ${\rm EuMn_2As_2}$ crystal measured in different $H$ applied along the $c$~axis ($\chi_c,\ H \parallel c$) and in the $ab$~plane ($\chi_{ab},\  H \perp c$) are shown in Fig.~\ref{fig:MT_EuMn2As2} for $T\leq30$~K\@. At small $H = 0.05$~T, cusps are observed in Fig.~\ref{fig:MT_EuMn2As2}(a) for each of $\chi_{ab}(T)$ and $\chi_{c}(T)$ at $T_{\rm N1} = 15.0$~K and $T_{\rm N2} = 5.0$~K, where $T_{\rm N1}$ is the N\'eel temperature of the Eu spins and $T_{\rm N2}$ is likely an Eu spin-reorientation transition.  The insets of Fig.~\ref{fig:MT_EuMn2As2}(a) show expanded plots of the $\chi(T)$ data in the vicinity of $T_{\rm N2}$ to show this transition in more detail.  The ZFC and field-cooled (FC) $\chi(T)$ data do not exhibit any thermal hysteresis (not shown), consistent with second-order transitions at $T_{\rm N1}$ and~$T_{\rm N2}$.  The $\chi_c(T)$ data in Fig.~\ref{fig:MT_EuMn2As2}(a) are nearly independent of~$T$ for $T<T_{\rm N1}$, whereas the $\chi_{ab}(T<T_{\rm N1})$ data decrease by about 12\% on cooling below $T_{\rm N1}$.  {\it These results rule out collinear AFM ordering of the Eu spins along the $c$~axis, because in that case $\chi_c(T\to0) = 0$ {\rm \cite{Johnston2015}}, contrary to observation}.  For $T>T_{\rm N}$, the $\chi$ is also very anisotropic, with $\chi_c < \chi_{ab}$.

\begin{table*}
\caption{\label{tab:CW} Eu AFM ordering temperatures obtained from $\chi(T)$ data, Mn $T_{\rm N}$ from $C_{\rm p}(T)$ data, and the parameters obtained from fits of the $\chi(T)$ data for {Eu$_{1-x}$K$_x$Mn$_2$As$_2$} ($x=0$, 0.04, 0.07) by the modified Curie-Weiss law~(\ref{eq:C-W}).}
\begin{ruledtabular}
\begin{tabular}{lcccccccc}
Compound & Field     & Eu $T_{\rm N1}$  & Eu $T_{\rm N2}$  &  Mn $T_{\rm N}$ &Fit $T$-range & $\chi_0$     & $C$             &  $\theta_{\rm p}$   \\
         & direction & (K)           & (K)       & (K)  & (K)    & (cm$^3$/mol) & (cm$^3$\,K/mol) & (K)                 \\
\hline
${\rm EuMn_2As_2}$ &  $H \parallel c$ & 15.0 & 5.0 &  	 142 & $50 \leq T \leq 300$  & $3.7(3) \times 10^{-3}$  & 7.86(7)  & 7.5(8)   \\	
                   &  $H \perp c$     & 15.0 & 5.0 &   	& $50 \leq T \leq 300$  &  $3.7(4) \times 10^{-3}$  & 7.98(9)  & 9(2)      \\
${\rm Eu_{0.96}K_{0.04}Mn_2As_2}$ & $H \parallel c$  &  13.5 & 9.0 &   146	& $50 \leq T \leq 125$  & $7.2(9) \times 10^{-3}$  & 9.5(3)\footnotemark[1]  & 9(1)   \\	
                                  & $H \perp c$      &  13.5 & 9.0 &  & 	 $50 \leq T \leq 125$  & $7.4(9) \times 10^{-3}$  & 9.0(3)\footnotemark[1]  & 13(1)   \\	 

${\rm Eu_{0.93}K_{0.07}Mn_2As_2}$ & $H \parallel c$  & 14.5 & 12.5 &     150 & $75 \leq T \leq 225$  &  $1.05(3) \times 10^{-2}$ \footnotemark[2] & 7.9(1) & 18(2)  \\	
                                  & $H \perp c$ & &12.5   &  & $75 \leq T \leq 225$  &  $1.13(5) \times 10^{-2}$  \footnotemark[2]& 8.0(1) & 18(2) \\	
\end{tabular}
\end{ruledtabular}
\footnotetext[1]{These $C$ values are abnormally large, likely due to the limited $T$~range of the fit and to the two magnetic transitions at temperatures higher than the 50--125~K Curie-Weiss fitting temperature range that are observed in the $\chi^{-1}(T)$ data in Fig.~\ref{fig:MTinv_EuKMn2As2}.}
\footnotetext[2]{The $\chi_0$ is likely enhanced due to trace amounts of ferromagnetic MnAs impurity with Curie temperature $T_{\rm C}\approx 320$~K.}
\end{table*}

As seen from Figs.~\ref{fig:MT_EuMn2As2}(b) and \ref{fig:MT_EuMn2As2}(c), with increasing $H$ $T_{\rm N1}$ shifts to lower~$T$; e.g.\ at $H=3.0$~T, $T_{\rm N1}=15.0$~K decreases to 9.5~K ($\chi_{ab}$) and 13.0~K ($\chi_c$), consistent with an AFM transition at $T_{\rm N1}$. On the other hand, the $T_{\rm N2}$ which is observed in $\chi(T)$ at $H \leq 0.1$~T vanishes for $H \geq 0.5$~T, which is not obvious from Figs.~\ref{fig:MT_EuMn2As2}(b) and~\ref{fig:MT_EuMn2As2}(c) due to the overlapping data sets. 

\begin{figure}
\includegraphics[width=3.2in]{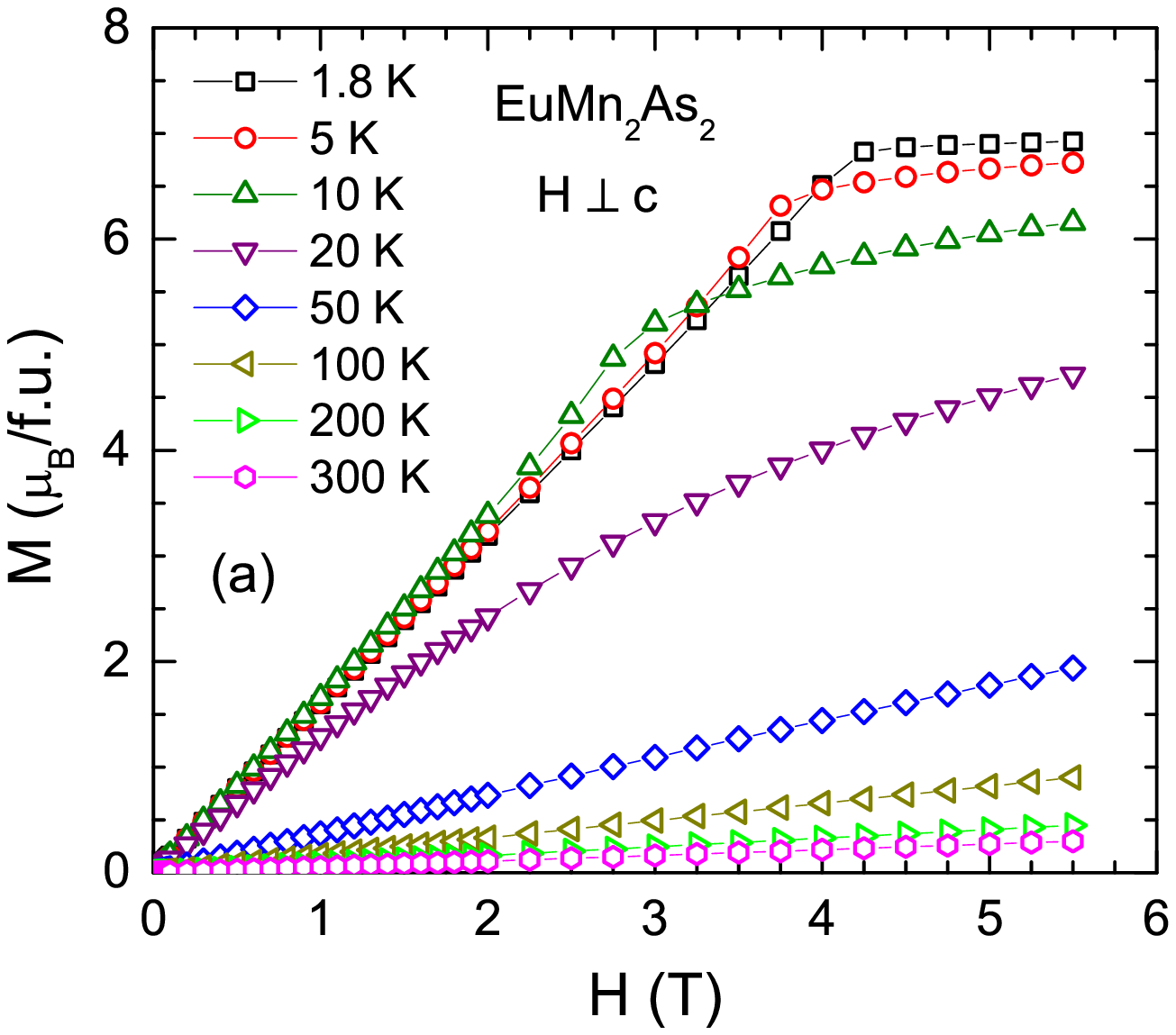}
\includegraphics[width=3.2in]{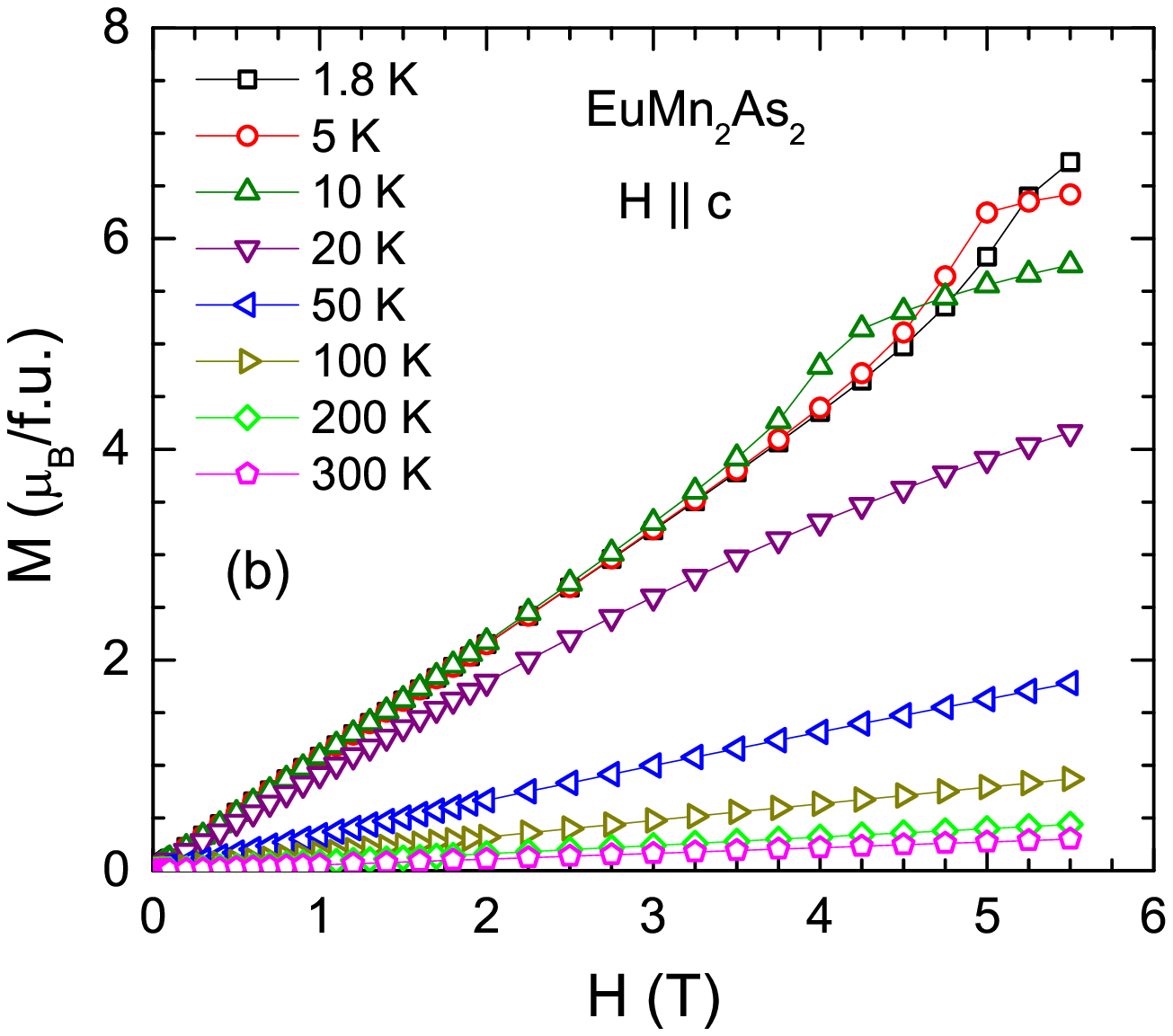}
\caption{(Color online) Isothermal magnetization $M$ of an ${\rm EuMn_2As_2}$  single crystal as a function of applied magnetic field $H$ measured at the indicated temperatures for $H$ applied (a) in the $ab$ plane ($M_{ab}, H \perp  c$) and (b) along the $c$ axis ($M_c, H \parallel c$).}
\label{fig:MH_EuMn2As2}
\end{figure}

\begin{figure}
\includegraphics[width=3.2in]{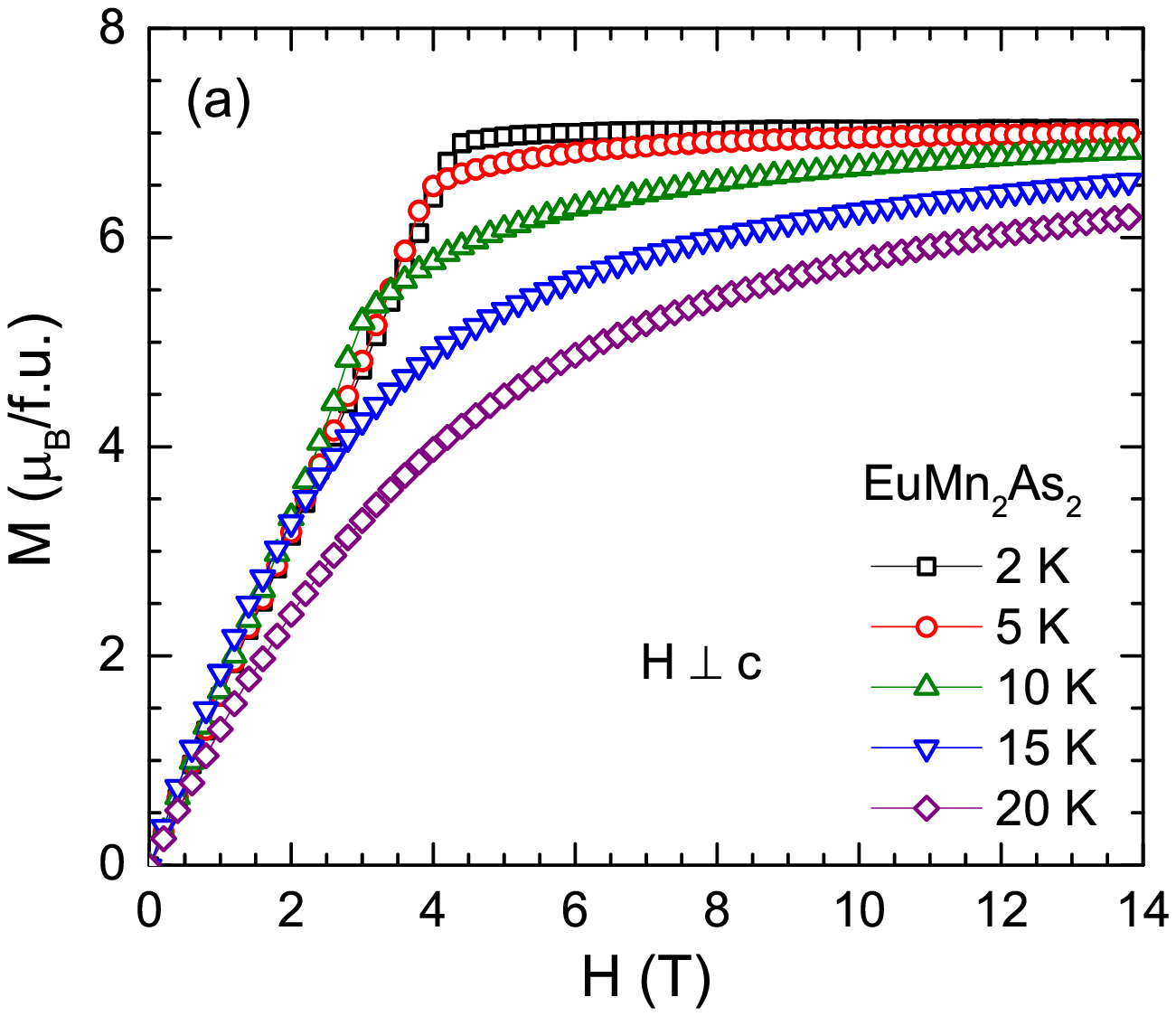}
\includegraphics[width=3.2in]{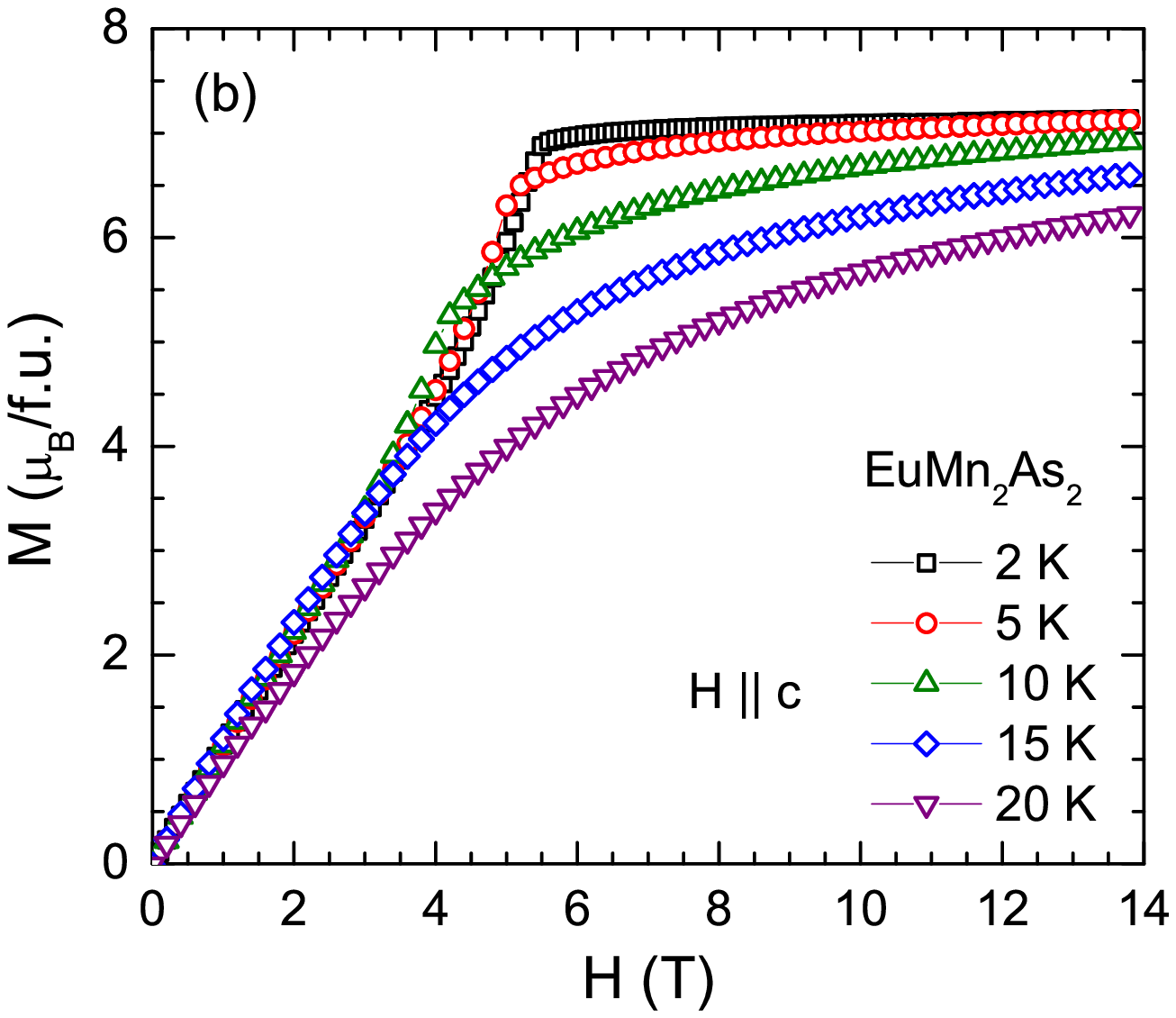}
\includegraphics[width=3.2in]{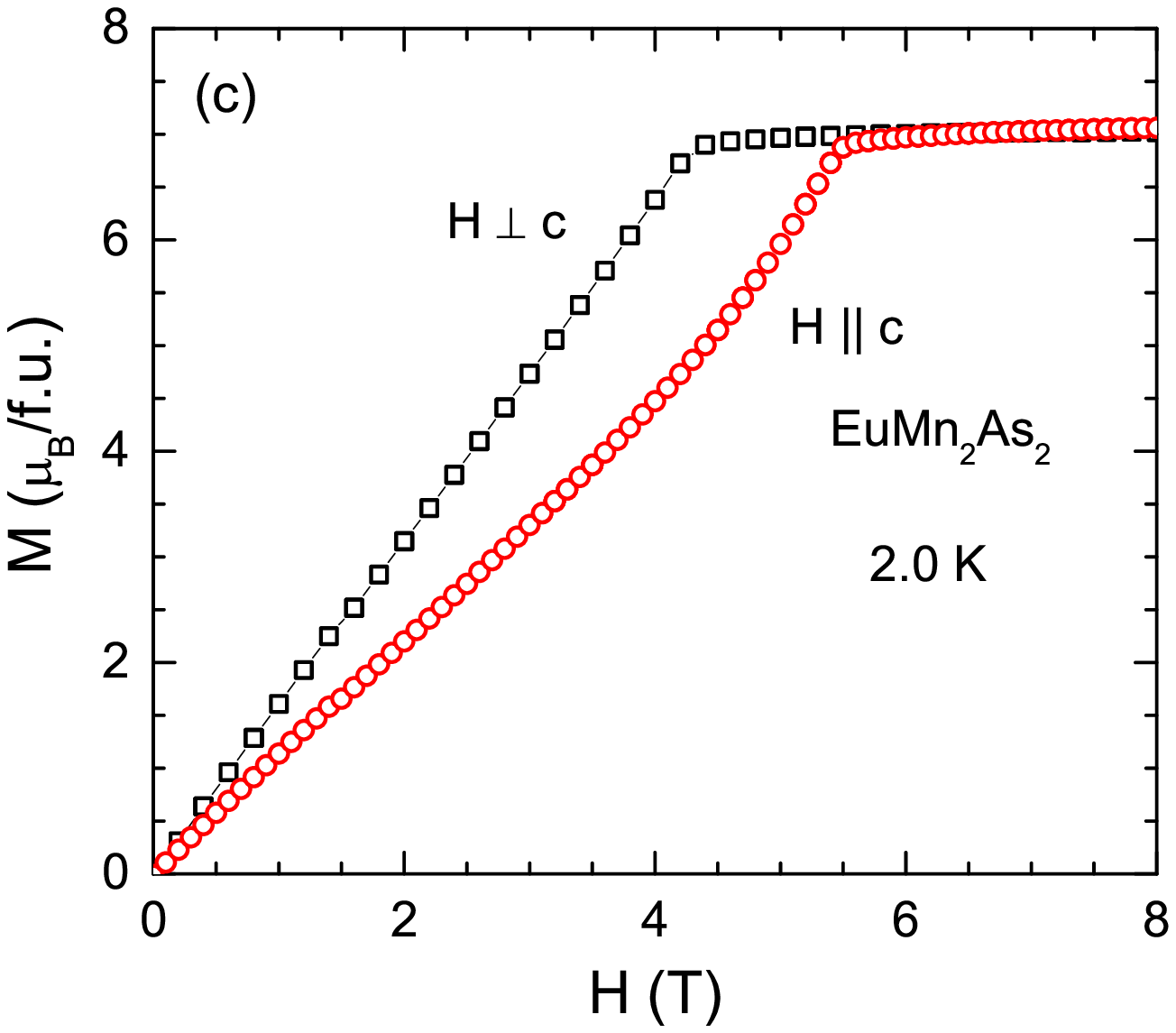}\vspace{-0.1in}
\caption{(Color online) High-field isothermal magnetization $M$ of an ${\rm EuMn_2As_2}$  single crystal as a function of applied magnetic field $H$ measured at the indicated temperatures for $H$ applied (a) in the $ab$ plane ($M_{ab}, H \perp  c$) and (b) along the $c$ axis ($M_c, H \parallel c$). (c) A comparison of $H \perp  c$ and $ H \parallel c$ $M(H)$ isotherms at 2~K\@.}
\label{fig:High_H_MH_EuMn2As2}
\end{figure}

\begin{figure}
\includegraphics[width=3.3in]{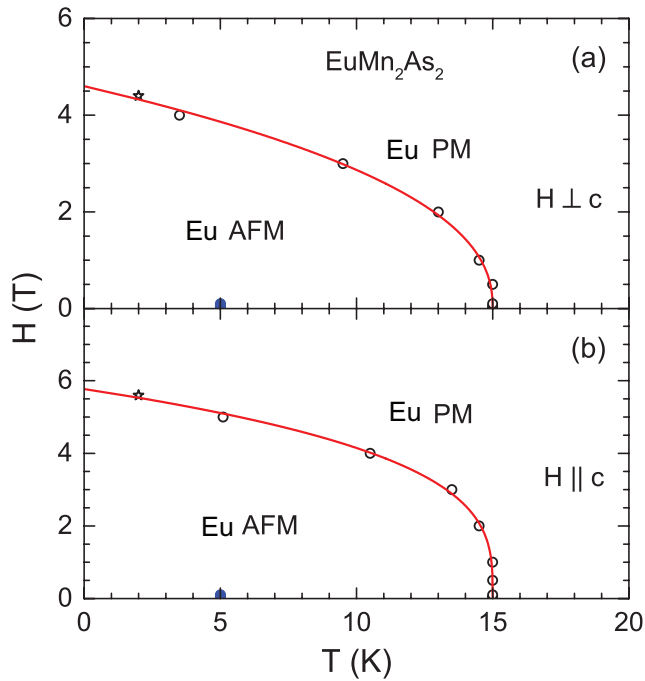}
\caption{(Color online) $H$--$T$ phase diagram for the Eu magnetic ordering in an ${\rm EuMn_2As_2}$ single crystal determined from the $H$ dependence of $T_{\rm N}$ from the magnetic susceptibility $\chi(T)$ data in Fig.~\ref{fig:MT_EuMn2As2} and $T$ dependence of isothermal magnetization $M(H)$ data in Fig.~\ref{fig:High_H_MH_EuMn2As2} for $H$ applied (a) in the $ab$~plane ($H \perp c$) and (b)~along the $c$ axis ($H \parallel c$). The solid curves are the fits to the data by $H=H_0(1-T/T_{\rm N})^{\alpha}$. The data points at $H=0.05$~T and $T=5$~K are for the spin-reorientation transition seen in Fig.~\ref{fig:MT_EuMn2As2}(a). The stars correspond to the point obtained from $M(H)$ measurements.}
\label{fig:H-T_EuMn2As2}
\end{figure}

The isothermal $M(H)$ data for $H \leq 5.5$~T of an ${\rm EuMn_2As_2}$ crystal are shown in Fig.~\ref{fig:MH_EuMn2As2} at eight different temperatures between 1.8~K and 300~K for $H$ applied along the $c$ axis ($M_c, H \parallel c$) and in the $ab$ plane ($M_{ab}, H \perp  c$). Consistent with the AFM ground state inferred from the $\chi(T)$ data above, the $M(H)$ isotherms at 1.8~K do not show any hysteresis between increasing and decreasing cycles of $H$ (not shown). It is seen from Figs.~\ref{fig:MH_EuMn2As2}(a) and \ref{fig:MH_EuMn2As2}(b) at $T = 1.8\ {\rm and}\ 5~{\rm K} <T_{\rm N1}=15.0$~K for $H \perp  c$ that the $M(H)$ isotherms are almost linear in $H$ until saturation is reached at the expected value $\mu_{\rm sat} = gS\mu_{\rm B}$/Eu $=7\,\mu_{\rm B}$/Eu, whereas for $H \parallel c$ the $M(H)$ isotherms are initially linear in $H$ for $H\lesssim4.0$~T but then show a weak S-type feature near 5~T before reaching saturation. {\it This anisotropy in the $M(H)$ isotherms suggests that the $c$ axis is the easy axis} \cite{Johnston2015}. However, this interpretation conflicts with the data in Fig.~\ref{fig:MT_EuMn2As2}(a) which demonstrates that collinear AFM ordering of the Eu spins does not occur along the $c$~axis .  {\it This inconsistency suggests that the AFM structure of the Eu spins in \ema\ is both noncollinear and noncoplanar.}

A more detailed examination of the anisotropic $M(H)$ behavior at $T\leq20$~K at high fields $H \leq 14$~T is  shown in Fig.~\ref{fig:High_H_MH_EuMn2As2}. The data in Fig.~\ref{fig:High_H_MH_EuMn2As2}(c) reveal that at 2.0~K the critical fields $H_{\rm c}$ for saturation of $M$ to $\mu_{\rm sat}$ for $H \perp c$ and $H \parallel c$ are 4.4~T and 5.6~T, respectively. Within the Weiss molecular field theory (MFT) \cite{Johnston2015}, this suggests that the $c$~axis is the easy axis for AFM ordering of the Eu spins.  This directly conflicts with the data in Fig.~\ref{fig:MT_EuMn2As2} that demonstrate that this is not correct.  The $H_{\rm c}$ decreases with increasing~$T$ as exemplified in Figs.~\ref{fig:High_H_MH_EuMn2As2}(a) and \ref{fig:High_H_MH_EuMn2As2}(b). A break in slope of $M(H)$ occurs at the critical field $H_{\rm c}(T)$ which is the field separating the AFM and PM states of the system.  At higher fields the $M(H)$ data exhibit negative curvature characteristic of the PM state at temperatures close to $T_{\rm N1}$.  This nonlinear behavior in the PM state  becomes proportional with increasing $T>T_{\rm N1}$ as seen in Fig.~\ref{fig:MH_EuMn2As2}.

The Eu-spin $H$--$T$ phase diagram delineated by $H_{\rm c}(T)$ in Fig.~\ref{fig:H-T_EuMn2As2} is obtained from the $\chi(T)$ data in Fig.~\ref{fig:MT_EuMn2As2} and the $M(H)$ data in Figs.~\ref{fig:MH_EuMn2As2} and~\ref{fig:High_H_MH_EuMn2As2}. The red curves in Fig.~\ref{fig:H-T_EuMn2As2} are fits of the $H_{\rm c}(T)$ data by the empirical function $H=H_0(1-T/T_{\rm N})^{\alpha}$ with $\alpha = 0.43$ and 0.30 for $H \perp  c$ and $H \parallel c$, respectively, with extrapolated $H_{\rm c}(T=0)$ values of~4.6 and 5.8~T, respectively.  In view of the high Mn $T_{\rm N}=142$~K and the large critical field $H_{\rm c}\sim 25$~T of the Mn spins at low~$T$ \cite{Sangeetha2016}, the Mn spins are likely still AFM~ordered over the $T$ and $H$ range of Fig.~\ref{fig:H-T_EuMn2As2}.  It was inferred above that AFM structure of the Eu spins is noncollinear and also noncoplanar.  One sees that the critical field for $T\to 0$  in Fig.~\ref{fig:H-T_EuMn2As2} is larger for $H\parallel c$ than for $H\perp c$.

\subsubsection{Magnetism of the Mn Spins}

The $\chi(T)$ data in Figs.~\ref{fig:MTinv_EuMn2As2} and~\ref{fig:MT_EuMn2As2} reveal AFM ordering of the Eu$^{+2}$ spins at $T_{\rm N1} = 15.0$~K, but no obvious anomaly is observed in $\chi(T)$ that can be associated with ordering of the nominal Mn$^{+2}$ spins $S = 5/2$.  The main difficulty is that the magnetism at all temperatures is expected to be dominated by that of the Eu spins.  However, the $C_{\rm p}(T)$ data presented below show a clear additional anomaly at 142~K\@. Furthermore, the data in Table~\ref{Tab:OrdTemps} below show that Mn-containing ${\rm CaAl_2Si_2}$-type $AB_2X_2$ compounds exhibit Mn AFM ordering temperatures between 63 and 154~K\@.  In particular, the compound ${\rm SrMn_2As_2}$ with nearly the same lattice parameters and unit cell volume as \ema,  and containing Sr$^{+2}$ with the same valence as Eu$^{+2}$, exhibits a Mn $T_{\rm N} = 120(2)$~K \cite{Sangeetha2016} which is close in temperature to the feature in $C_{\rm p}(T)$ at 142~K for \ema.

\begin{figure}
\includegraphics[width=3.4in]{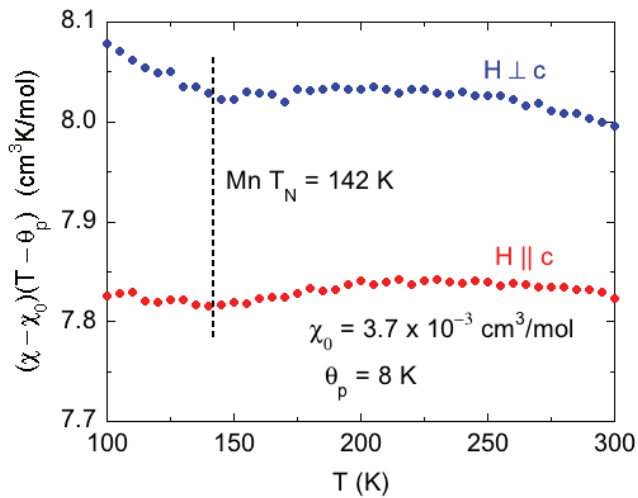}
\caption{(Color online) The quantity $(\chi-\chi_0)(T-\theta_{\rm p})$ versus temperature~$T$ in $H=3$~T for \ema\ with $H\perp c$ and $H\parallel c$.  The N\'eel temperature $T_{\rm N} = 142$~K of the Mn sublattice is apparent from the changes in slope of $(\chi-\chi_0)(T-\theta_{\rm p})$ upon traversing $T_{\rm N}$ with increasing~$T$\@.}
\label{Fig:EuMn2As2_MT_H3T(X-X0)(T-q)2}
\end{figure}

Therefore we looked for irregularities in $\chi(T)$ in the temperature range from 100 to 300~K\@.  To do this in an unbiased manner, we assumed that the ``background'' susceptibility for both $\chi_{ab}(T)$ and $\chi_c(T)$ is given by Eq.~(\ref{eq:C-W}) with the same parameters $\chi_0 = 3.7\times10^{-3}~{\rm cm^3/mol}$ and $\theta_{\rm p} = 8$~K obtained from the 50--300~K fits in Table~\ref{tab:CW}.  Therefore according to Eq.~(\ref{eq:C-W}) a plot of \mbox{$(\chi-\chi_0)(T-\theta_{\rm p})$} versus~$T$ should be the constant value~$C$\@.  Shown in Fig.~\ref{Fig:EuMn2As2_MT_H3T(X-X0)(T-q)2} are such plots for the two field directions in the 100--300~K temperature range.  One sees a clear change in slope in the $\chi_{ab}$ data on increasing $T$ through 142~K, and a more \mbox{subtle} slope change in $\chi_c(T)$\@.  Similar behaviors were observed for a ${\rm SrMn_2As_2}$ crystal on increasing $T$ through its $T_{\rm N} = 120$~K in $H=3$~T \cite{Sangeetha2016}.  We therefore infer that the Mn spins in \ema\ exhibit AFM ordering below $T_{\rm N} =  142$~K\@.  From the magnetization and $\chi$ data presented, the nature of the AFM structure of the ordered Mn moments below the Mn $T_{\rm N}$ in \ema\ is unclear.

\subsection{\label{Sec:EuMn2As2_HC} Heat Capacity}

\begin{figure}
\includegraphics[width=3.3in]{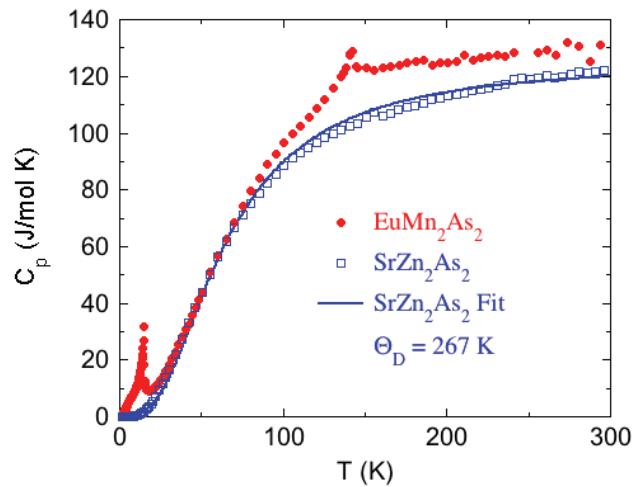}
\caption{(Color online) Heat capacity $C_{\rm p}$ of an ${\rm EuMn_2As_2}$ crystal and nonmagnetic reference compound ${\rm SrZn_2As_2}$ versus temperature $T$ in zero magnetic field. The blue curve is a fit of $C_{\rm p}(T)$ of ${\rm SrZn_2As_2}$ over the whole $T$ range by the Debye model of lattice heat capacity with $\Theta_{\rm D} = 267(2)$~K\@.}
\label{Fig:EuMn2As2_SrZn2As2_Cp}
\end{figure}

\begin{figure}
\includegraphics[width=3.4in]{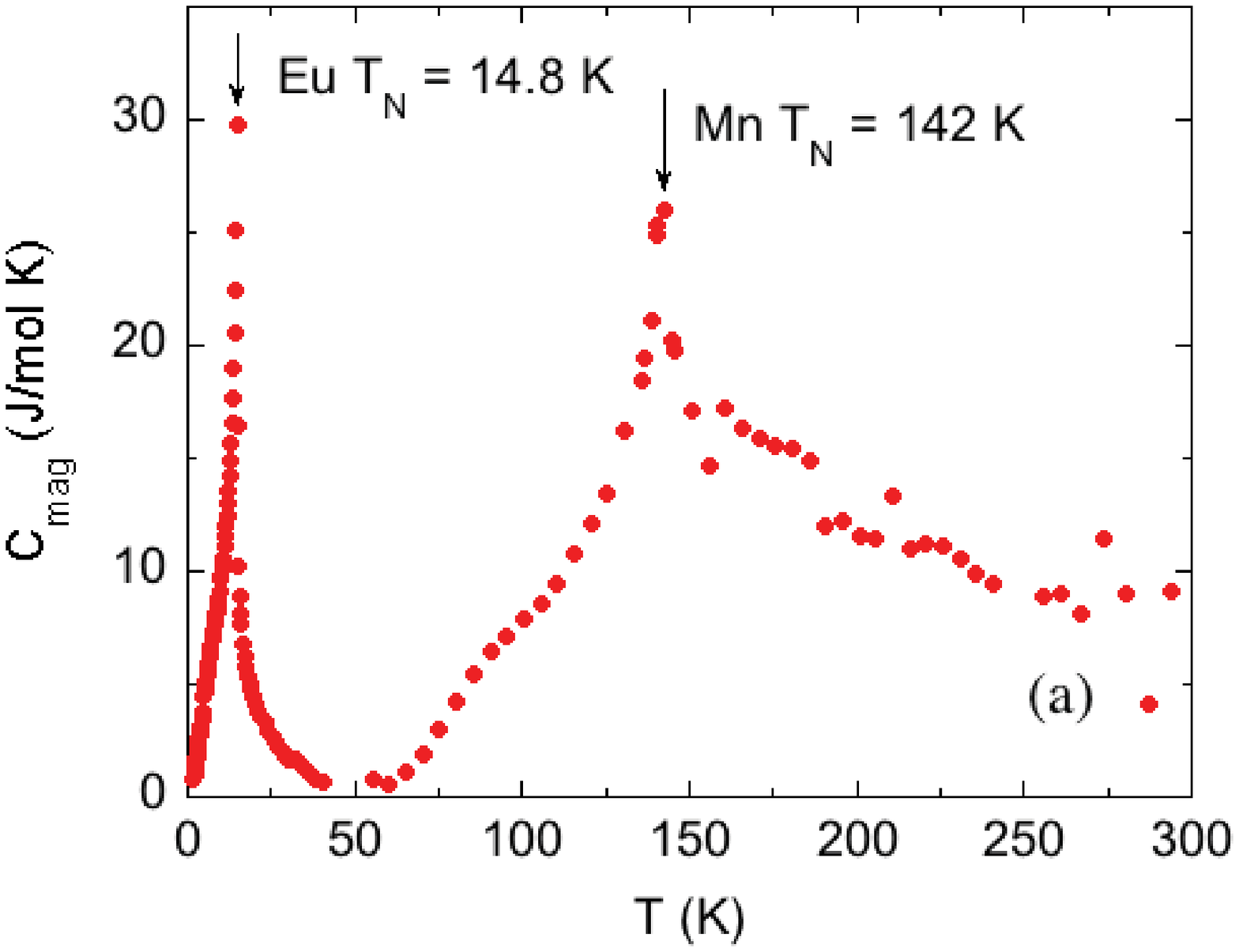}\vspace{0.0in}
\includegraphics[width=3.4in]{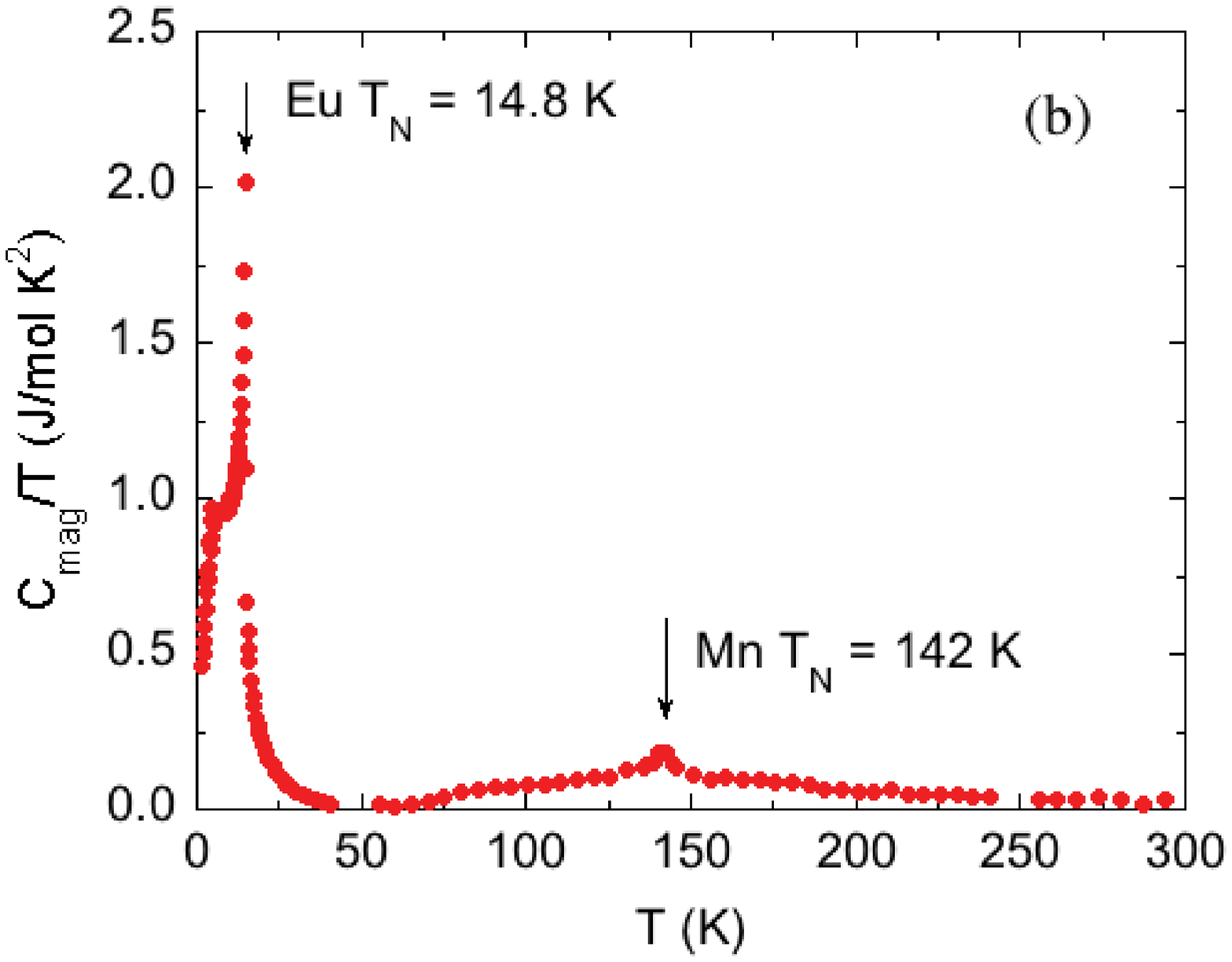}
\includegraphics[width=3.4in]{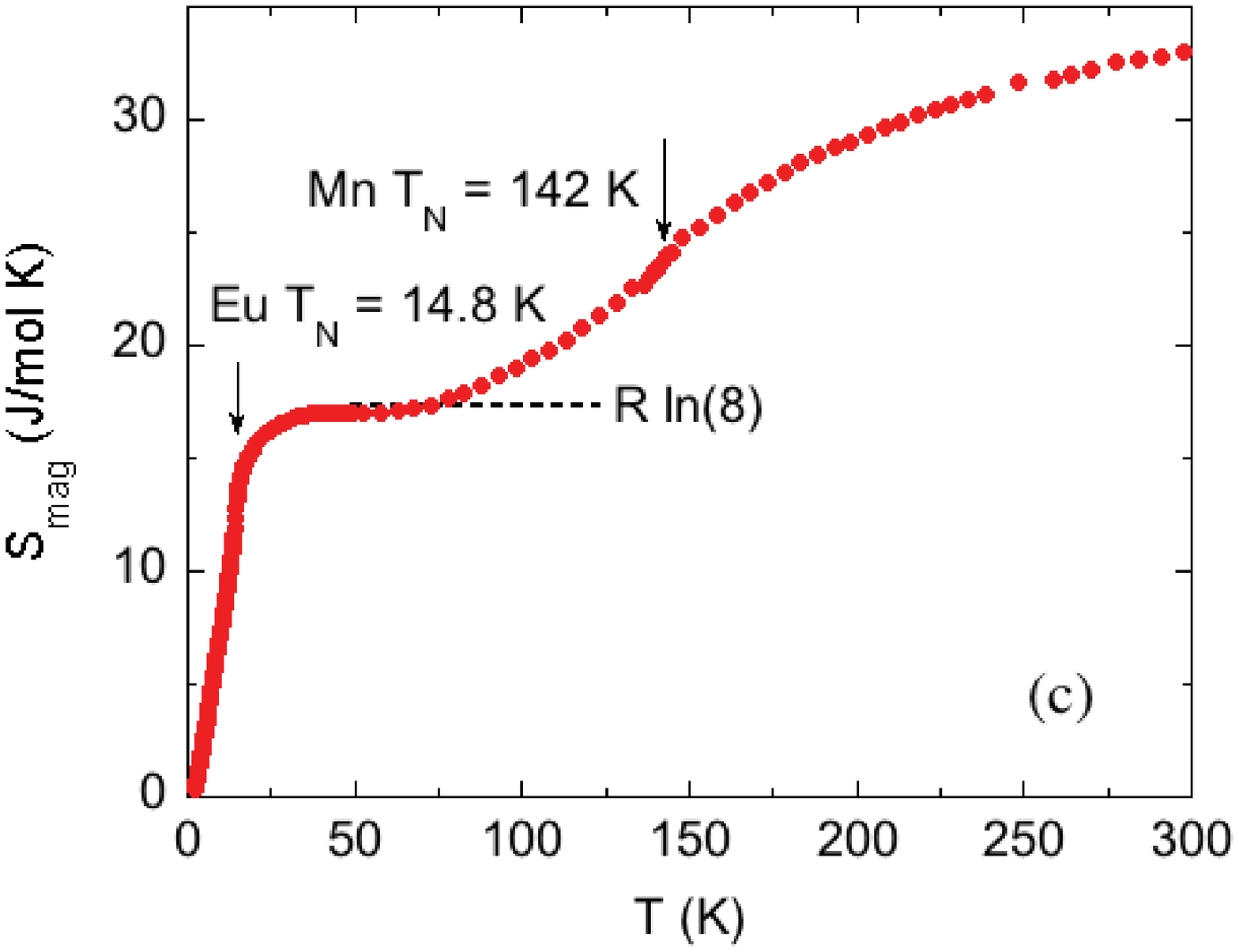}
\caption{(Color online) (a) Magnetic heat capacity $C_{\rm mag}$ of ${\rm EuMn_2As_2}$ versus temperature $T$ measured in $H=0$. (b)~$C_{\rm mag}/T$ versus $T$.  (c)~Magnetic entropy $S_{\rm mag}$ versus $T$\@. The AFM ordering temperatures $T_{\rm N}$ of the Eu and Mn spins are indicated in each panel.  The entropy of disordered Eu spins $R\ln(8)$ is indicated by the horizontal dashed line.  The continuing increase of $S_{\rm mag}(T)$ for $T>50$~K is due to the release of magnetic entropy of the long-range AFM-ordered Mn spins for $T<142$~K and of short-range-ordered  Mn spins for $T>142$~K.}
\label{Fig:EuMn2As2_Cmag}
\end{figure}

The $C_{\rm p}(T)$ data for an ${\rm EuMn_2As_2}$ crystal are shown in Fig.~\ref{Fig:EuMn2As2_SrZn2As2_Cp} together with those of a single crystal of the isostructural nonmagnetic reference compound ${\rm SrZn_2As_2}$ \cite{Mewis1980}. The $C_{\rm p}(T)$ of ${\rm SrZn_2As_2}$ was fitted by the Debye lattice heat capacity model \cite{Goetsch2012}, yielding a Debye temperature $\Theta_{\rm D} = 267(2)$~K\@.  The fit is shown as the blue curve in Fig.~\ref{Fig:EuMn2As2_SrZn2As2_Cp}.

Subtracting a spline fit of the $C_{\rm p}(T)$ of ${\rm SrZn_2As_2}$ from that of ${\rm EuMn_2As_2}$ yields the magnetic contribution $C_{\rm mag}(T)$ to $C_{\rm p}(T)$ of ${\rm EuMn_2As_2}$ as shown in Fig.~\ref{Fig:EuMn2As2_Cmag}(a).  The $C_{\rm mag}(T)$ shows a sharp $\lambda$-type anomaly at $T_{\rm N1} = 14.8$~K consistent with the long-range AFM ordering of the Eu spins observed at 15.0~K in $\chi(T)$ above.  In addition, the high-$T$ $C_{\rm mag}(T)$ data show a well-defined anomaly at 142~K, consistent with the anomaly found in the above $\chi(T)$ data in Fig.~\ref{Fig:EuMn2As2_MT_H3T(X-X0)(T-q)2} which we attribute to an AFM transition associated with the Mn spins.  We do not see any thermal hysteresis across the 142~K anomaly between the warming and cooling cycles of $C_{\rm p}(T)$ measurements (not shown), consistent with a second-order phase transition.  Similarly, a $C_{\rm p}$ anomaly at 120~K in addition to the Eu AFM transition at $T_{\rm N} \approx 10$~K was observed in ${\rm EuMn_2Sb_2}$ and attributed to AFM ordering of the Mn $S=5/2$ spins \cite{Schellenberg2010}.

The dependence of $C_{\rm mag}(T)/T$ on~$T$ is shown in Fig.~\ref{Fig:EuMn2As2_Cmag}(b).  After extrapolating the $C_{\rm mag}(T)$ data to $T=0$ using a $T^3$ function, the magnetic contribution $S_{\rm mag}(T)$ to the entropy was calculated using $S_{\rm mag}(T) = \int_0^T[C_{\rm mag}(T)/T]dT$, and the result is shown in Fig.~\ref{Fig:EuMn2As2_Cmag}(c).  For the Eu ordering, $S_{\rm mag}(T_{\rm N}) = 13.5$~J/mol\,K, which is 78\% of the high-$T$ limit $R\ln(2S+1) = R\ln(8)=17.3$~J/mol\,K\@.  This reduction of 22\% arises from dynamic short-range AFM ordering of the Eu spins existing at $T_{\rm N1}$.  The lost magnetic entropy is fully recovered by 40~K as shown in Fig.~\ref{Fig:EuMn2As2_Cmag}(c). The $S_{\rm mag}(T)$ thus indicates a valence state Eu$^{+2}$ with $S=7/2$ in ${\rm EuMn_2As_2}$ as also deduced above from the $\chi(T)$ data.

Taking the Mn spin to be $S=5/2$, the high-$T$ limit of the Mn spin entropy per mole of \ema\ is $2R\ln(6) = 29.8$~J/mol\,K\@. Thus the total high-$T$ spin entropy is expected to be 47.1~J/mol\,K\@.  The value at 300~K from Fig.~\ref{Fig:EuMn2As2_Cmag}(c) is only 32~J/mol\,K, so much of the magnetic entropy is still present as short-range AFM order of the Mn spins at that temperature.  This result is consistent with the inference in \cite{Sangeetha2016} that strong short-range AFM order of the Mn spins in ${\rm SrMn_2As_2}$ is still present even at 900~K\@.

\begin{figure}
\includegraphics[width=3.4in]{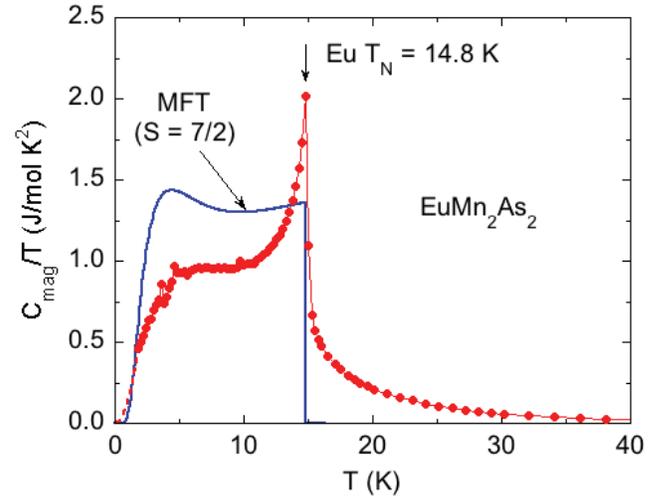}
\caption{(Color online) Magnetic heat capacity $C_{\rm mag}$ of an ${\rm EuMn_2As_2}$ crystal divided by temperature~$T$ versus $T$ below 40~K\@. The $T^3$ extrapolation to $T=0$ is the dashed red line.  The blue curve is the MFT prediction for Eu spin $S=7/2$ and $T_{\rm N} = 14.8$~K \cite{Johnston2011}.}
\label{Fig:EuMn2As2_CmagOnT_Eu}
\end{figure}

Shown in Fig.~\ref{Fig:EuMn2As2_CmagOnT_Eu} is an expanded plot of $C_{\rm mag}/T$ versus~$T$ for \ema.  The plot more clearly shows the sharp $\lambda$ peak at $T_{\rm N1} = 14.8$~K and a bulge at $T\sim5$~K\@.  Also shown is the MFT prediction for $S=7/2$ \cite{Johnston2011}.  The bulge is not associated with the Eu $T_{\rm N2} = 5.0$~K seen in $\chi(T)$ in Fig.~\ref{fig:MT_EuMn2As2}(a) likely arising from an Eu spin reorientation transition.  In particular, the bulge in the \ema\ data at $\sim 5$~K is also present in the MFT prediction and is a natural consequence of the combined effects of the increase in the exchange field and the decrease in the excited Zeeman level populations with decreasing~$T$\@.  As pointed out in \cite{Johnston2011}, within MFT the bulge is necessary so that the magnetic entropy at $T_{\rm N}$ increases with increasing spin~$S$ of the magnetic species.

\subsection{\label{Sec:EuMn2As2_Rho} Electrical Resistivity}

\begin{figure}
\includegraphics[width=3.3in]{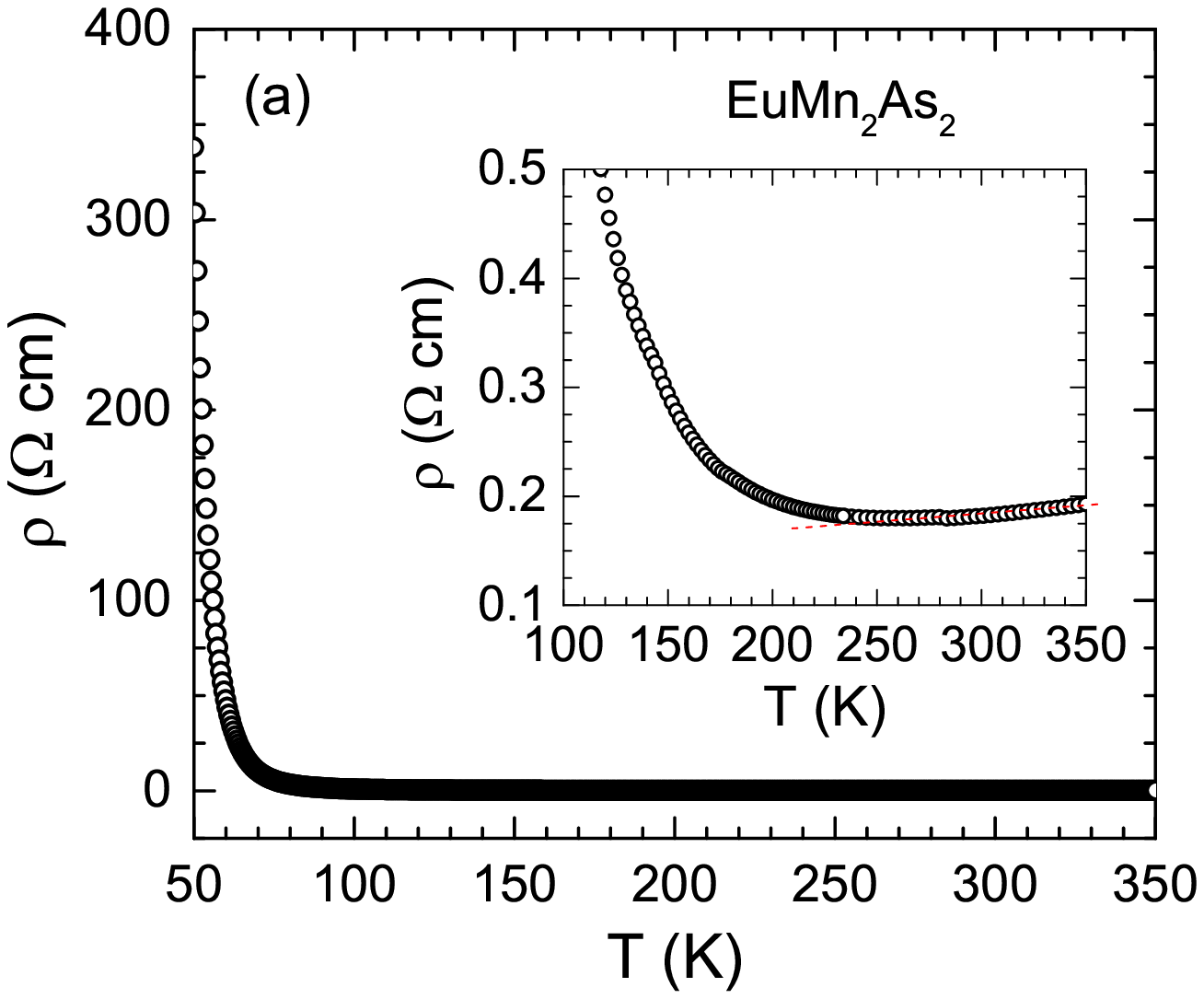}\vspace{0.1in}
\includegraphics[width=3.3in]{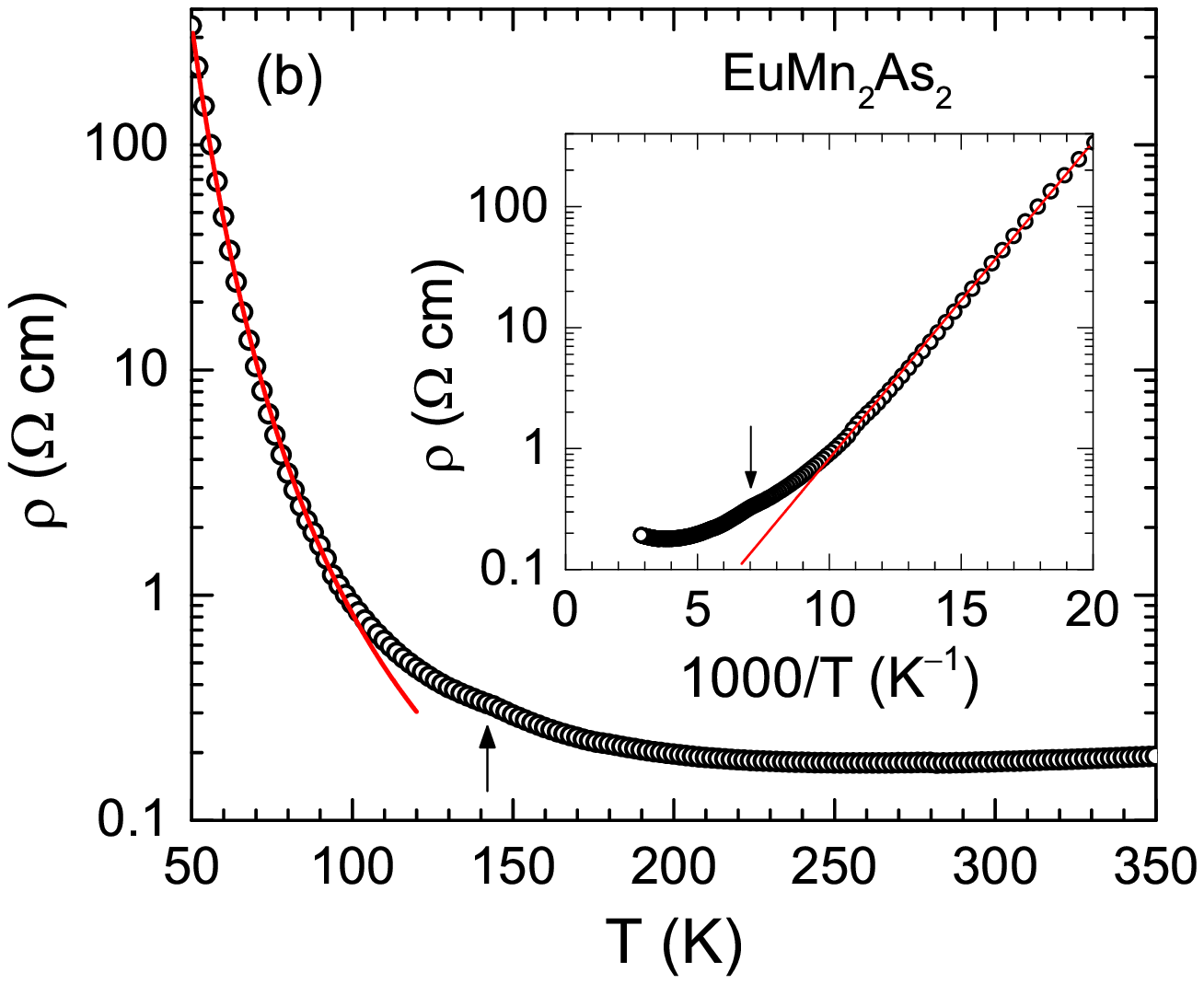}
\caption{(Color online) (a)~$ab$-plane electrical resistivity $\rho$ of an ${\rm EuMn_2As_2}$ single crystal as a function of temperature $T$ for $1.8~{\rm K} \leq T \leq 350$~K measured in zero magnetic field. Inset: Expanded view of  $\rho(T)$ data at $T > 100$~K\@. (b)~The $\rho(T)$ data in~(a) plotted on a semilog scale. The red curve is a  fit of $\rho(T)$ data with 50~K~$\leq T\leq 100$~K by Eq.~(\ref{Eq:rhoFit}). Inset:~$\log_{10}\rho$ versus $1000/T$. The straight red line is the same fit as in the main figure.  The vertical arrows in the main panel~(b) and its inset indicate a feature in the data at the Mn AFM ordering temperature $T_{\rm N} = 142$~K\@.}
\label{fig:rho_EuMn2As2}
\end{figure}

The $ab$-plane $\rho(T)$ of an ${\rm EuMn_2As_2}$ crystal is shown on linear scales in Fig.~\ref{fig:rho_EuMn2As2}(a) with an expanded plot above 100~K in the inset. The $\rho$ initially decreases slowly with decreasing~$T$ from 193~m$\Omega$\,cm at 350~K to 180~m$\Omega$\,cm at 250~K, slowly increases down to 100~K and then increases rapidly below this~$T$, yielding an insulating ground state in ${\rm EuMn_2As_2}$.

The $\rho(T)$ data plotted on a semilogarithmic scale in Fig.~\ref{fig:rho_EuMn2As2}(b) are described in the $T$ range 50~K~$\leq T\leq 100$~K by
\be
\rho(T) = \rho_0 \exp\left(\frac{\Delta}{k_{\rm B} T}\right),
\label{Eq:rhoFit}
\ee
where $k_{\rm B}$ is Boltzmann's constant, as shown by a fit of $\rho(T)$ data by Eq.~(\ref{Eq:rhoFit}) with activation energy $\Delta = 52.0(2)$~meV (red curve).  An alternative view of the fit on a semilog plot of $\rho$ versus $1000/T$ is shown as the red straight line in the inset of Fig.~\ref{fig:rho_EuMn2As2}(b). The activation energy for ${\rm EuMn_2As_2}$ is of the same order as for ${\rm BaMn_2As_2}$ ($\Delta = 54$~meV) \cite{Singh2009a}. In view of the activated behavior at low~$T$, we suggest that the metallic-like behavior of $\rho(T)$ at $T>250$~K arises from a competition between the carrier mobility and number density, such that with increasing~$T$ the mobility of the carriers decreases faster than the carrier concentration increases.

In addition, in Fig.~\ref{fig:rho_EuMn2As2}(b) and its inset one sees that the $\rho(T)$ data exhibit a bump near 142~K (vertical arrows) corresponding to the Mn $T_{\rm N}$ as observed above from the $C_{\rm p}(T)$ and $\chi(T)$ measurements.

\section{\label{Sec:EuKMn2As2_1} Physical properties of  E\lowercase{u}$_{0.96}$K$_{0.04}$M\lowercase{n}$_2$A\lowercase{s}$_2$}

\begin{figure}
\includegraphics[width=3in]{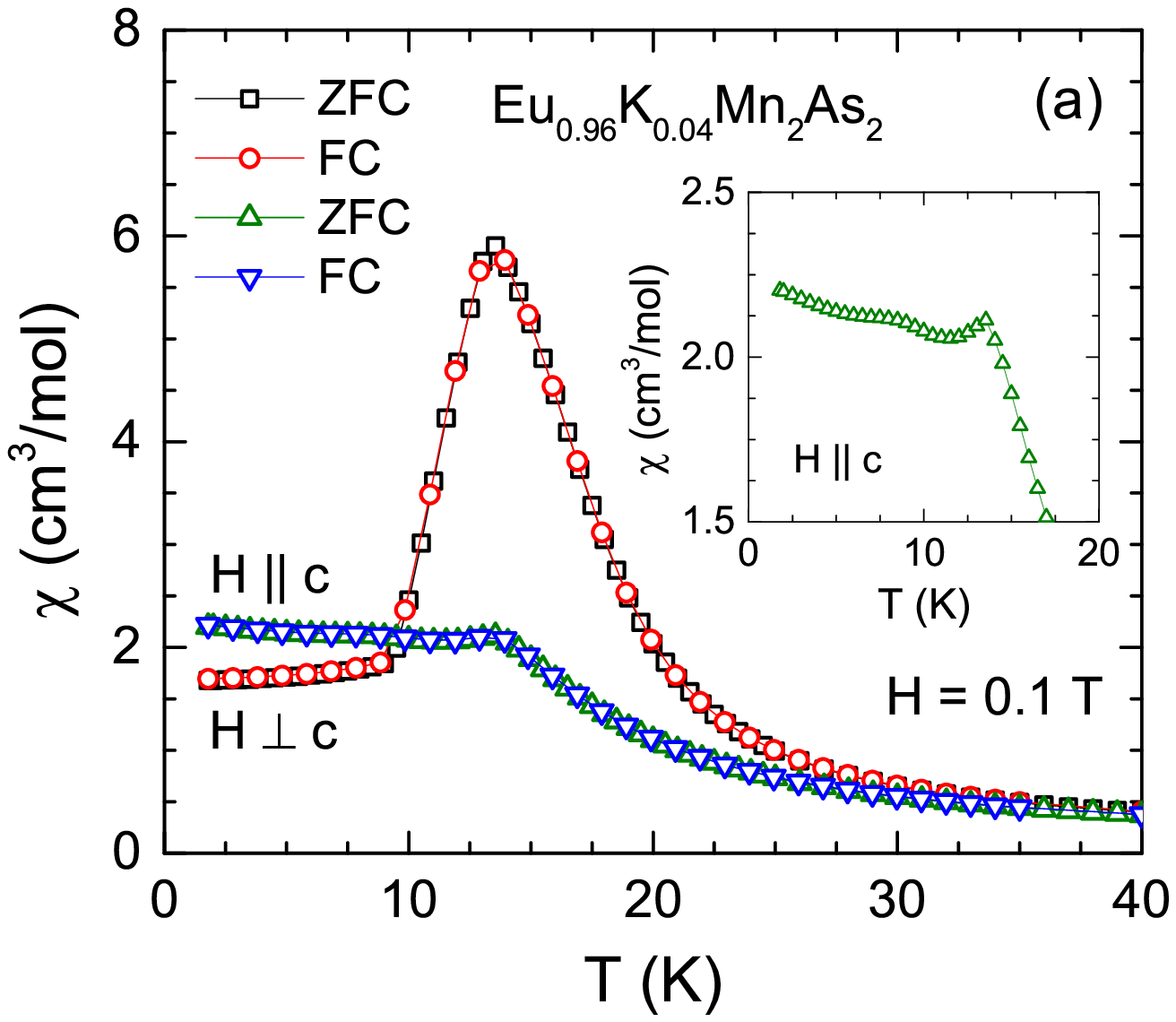}
\includegraphics[width=3in]{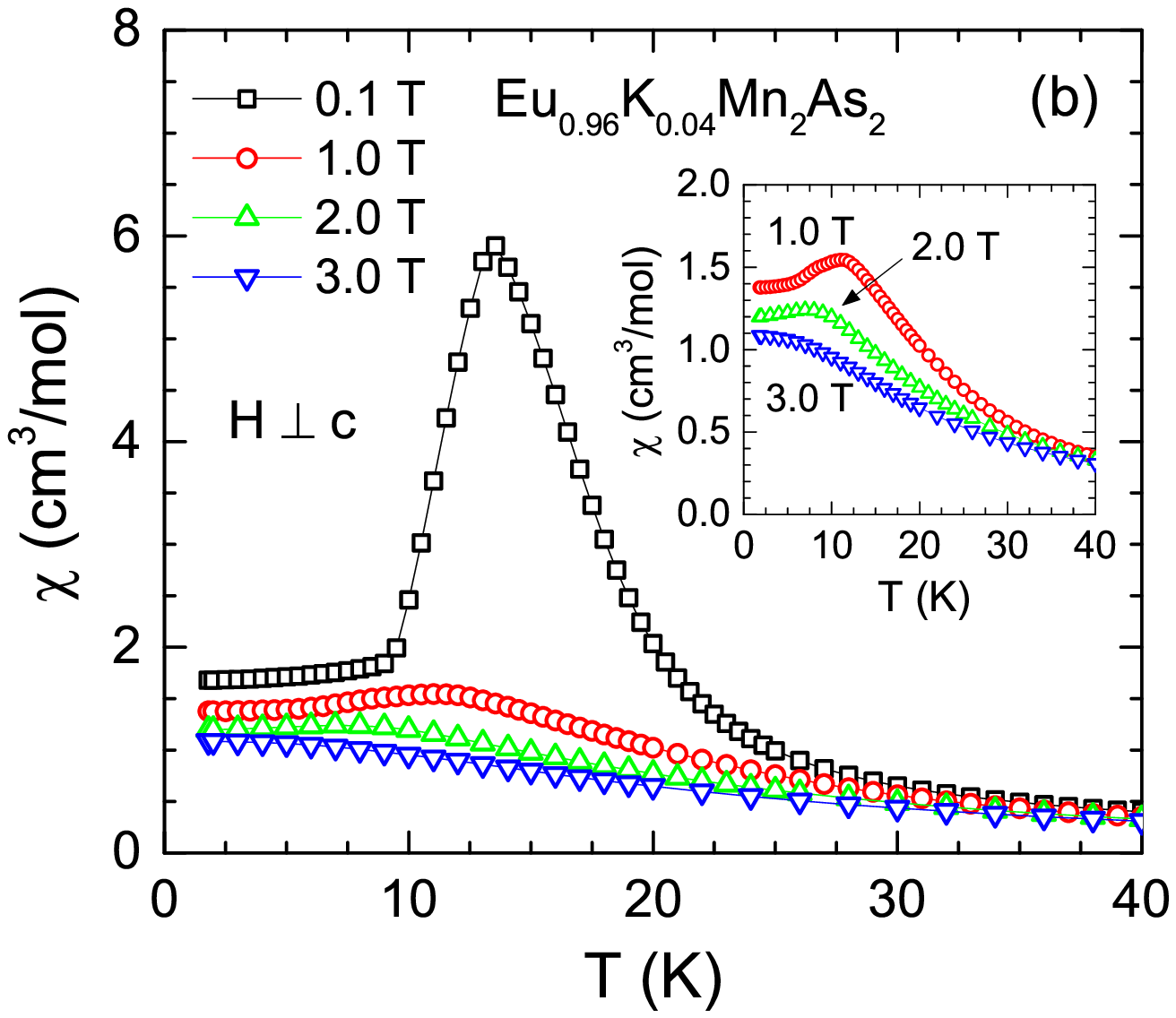}
\includegraphics[width=3in]{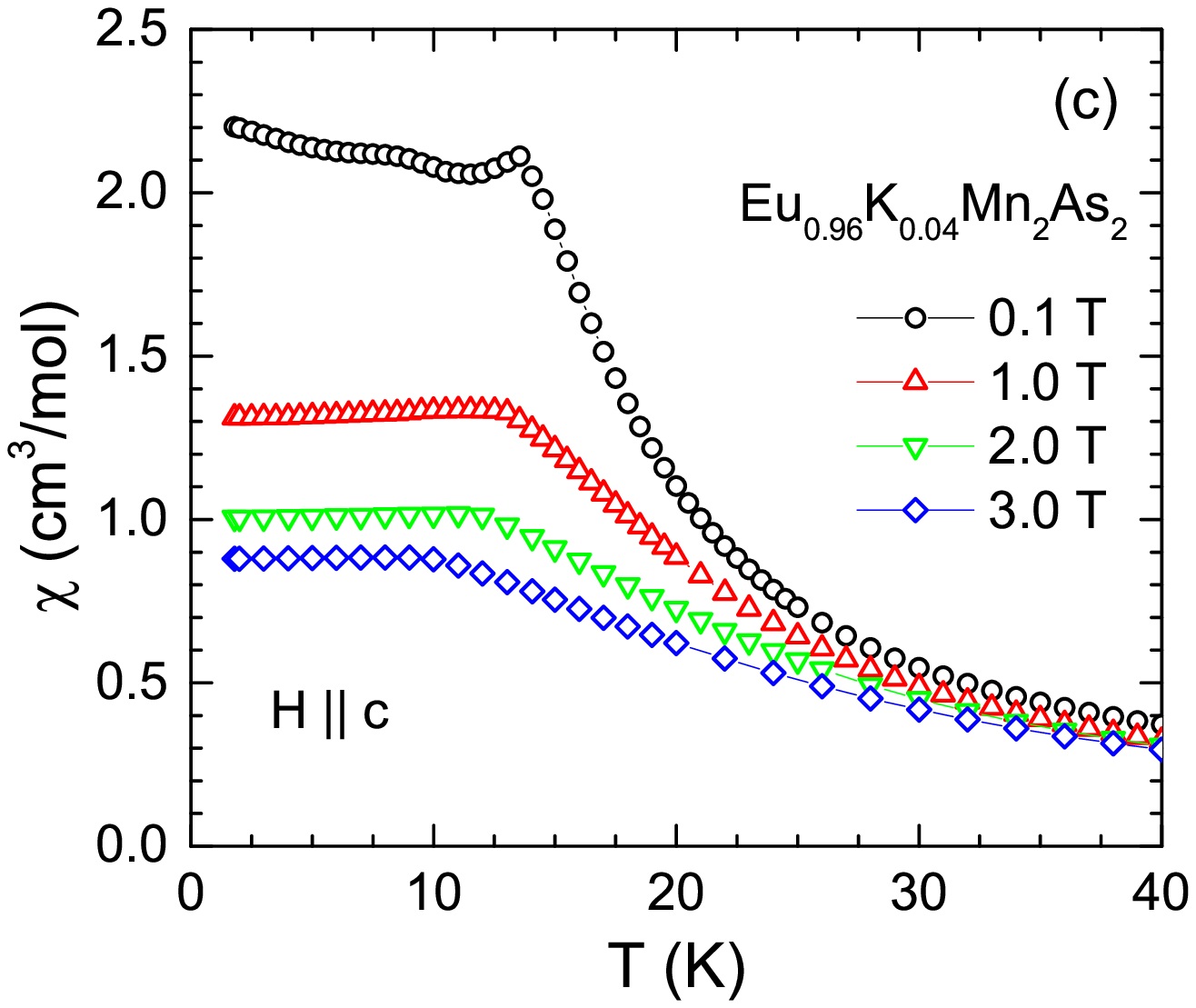}\vspace{-0.1in}
\caption{(Color online) (a) Zero-field-cooled (ZFC) and field-cooled (FC) magnetic susceptibility $\chi$ of an ${\rm Eu_{0.96}K_{0.04}Mn_2As_2}$ crystal versus temperature $T$ for $1.8~{\rm K} \leq T \leq 40$~K measured in a magnetic field $H= 0.1$~T applied in the $ab$ plane ($\chi_{ab}, H \perp  c$) and along the $c$ axis ($\chi_c, H \parallel c$). Inset: Expanded plot of $\chi(T)$ for $H \parallel c$ showing the anomaly near 9~K\@. (b) ZFC $\chi(T)$ for $1.8~{\rm K} \leq T \leq 40$~K measured at the indicated $ H \perp c$. Inset: Expanded plot of $\chi(T)$.  (c) ZFC $\chi(T)$ for $1.8~{\rm K} \leq T \leq 40$~K measured at the indicated $H \parallel c$ values.}
\label{fig:MT_EuKMn2As2}
\end{figure}

\begin{figure}
\includegraphics[width=3.3in]{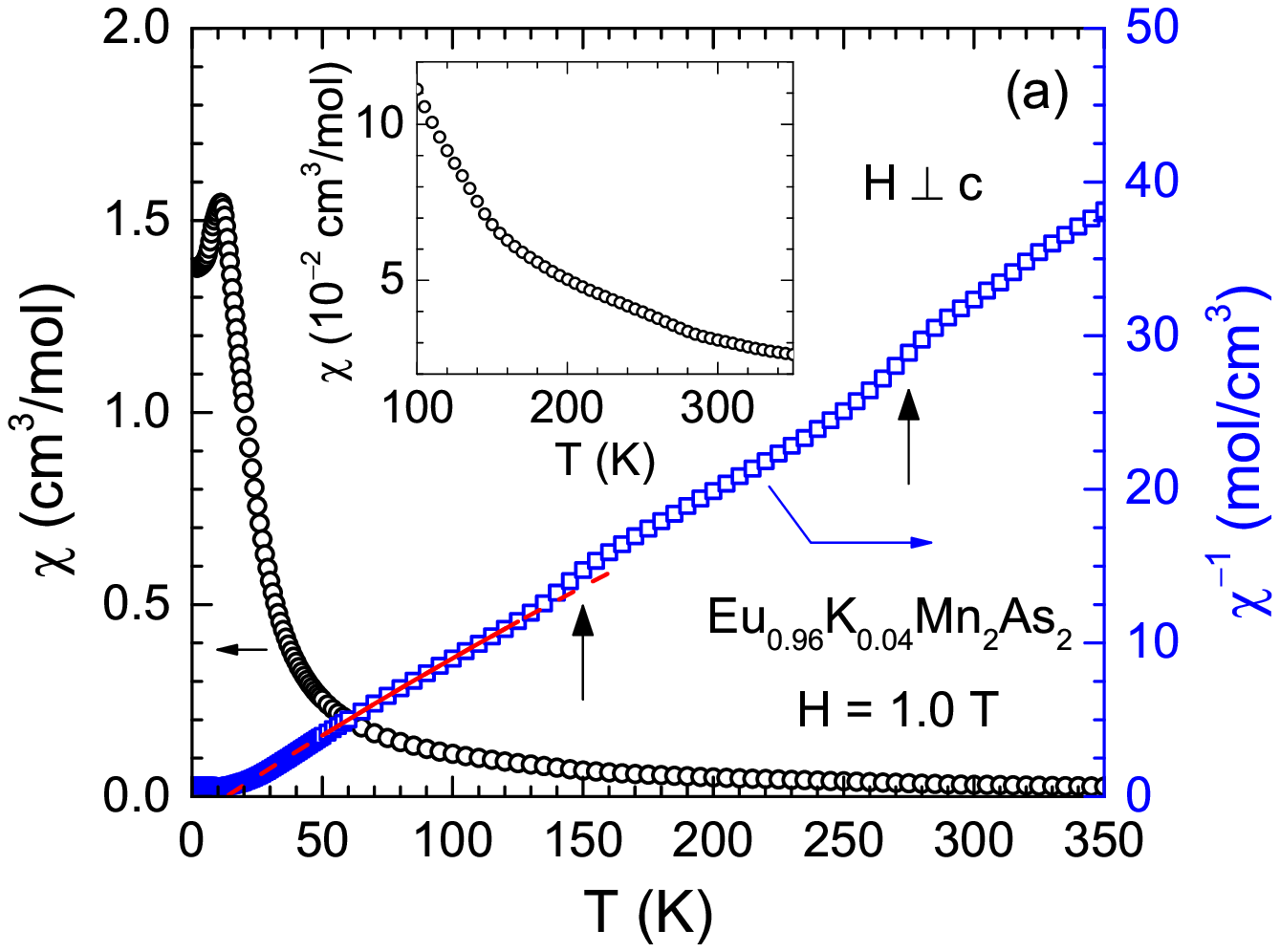}\vspace{0.1in}
\includegraphics[width=3.3in]{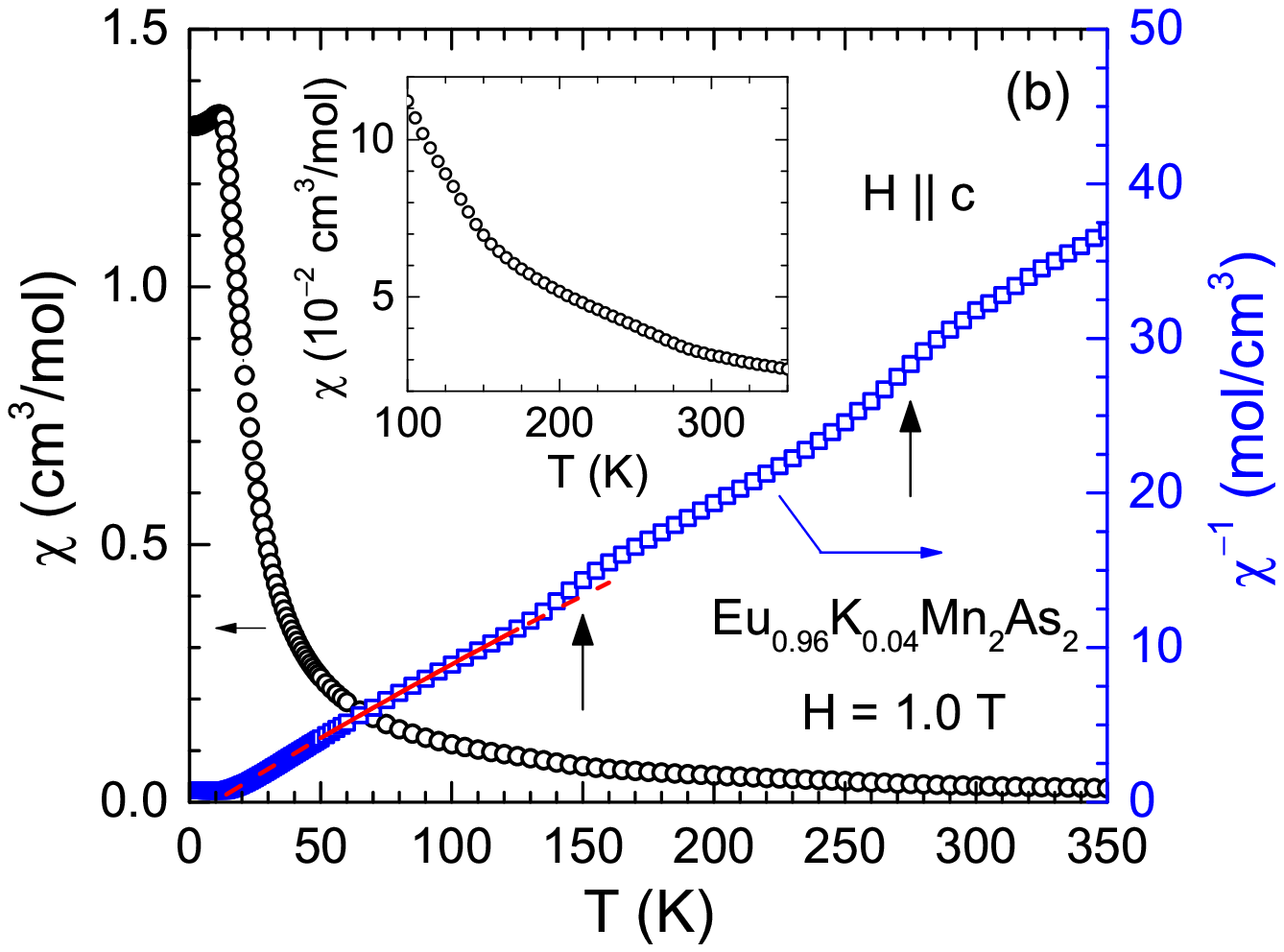}
\caption{(Color online) Zero-field-cooled magnetic susceptibility $\chi$ (left ordinate) and its inverse $\chi^{-1}$ (right ordinate) of an ${\rm Eu_{0.96}K_{0.04}Mn_2As_2}$ single crystal versus temperature $T$ for $1.8~{\rm K} \leq T \leq 350$~K measured in a magnetic field $H=1.0$~T applied (a) in the $ab$ plane ($\chi_{ab}, H \perp  c$) and (b) along the $c$ axis ($\chi_c, H \parallel c$). The red solid straight lines are the fits of $\chi^{-1}(T)$ data to the modified Curie-Weiss behavior in the $T$~range 50~K~$\leq T \leq$~125~K and the red dashed straight lines are extrapolations.  Arrows mark anomalies near 150~K and 275~K\@. Insets: Expanded plots of $\chi(T)$ for $100~{\rm K} \leq T \leq 350$~K\@.}
\label{fig:MTinv_EuKMn2As2}
\end{figure}

\subsection{\label{Sec:EuKMn2AS2_MT_MH} Magnetization and Magnetic Susceptibility}

The ZFC and FC $\chi(T)$ data for an ${\rm Eu_{0.96}K_{0.04}Mn_2As_2}$ crystal measured in $H=0.1$~T are shown in Fig.~\ref{fig:MT_EuKMn2As2}(a) for $T\leq 40$~K\@. These $\chi(T)$ data reveal well-defined anomalies due to AFM ordering at $T_{\rm N1} = 13.5$~K for both $H \parallel c$ and $H \perp c$ with no thermal hysteresis. In addition  Fig.~\ref{fig:MT_EuKMn2As2}(a) with $H \perp c$ and the inset with $H\parallel c$ exhibit some sort of spin-reorientation transition at $T_{\rm N2} = 9.0$~K\@. The AFM nature of the transition is confirmed by the shift of $T_{\rm N1}$ towards lower $T$ with increasing $H$ as shown in Figs.~\ref{fig:MT_EuKMn2As2}(b) and \ref{fig:MT_EuKMn2As2}(c).

The $\chi(T)$ in $H=0.1$~T is strongly anisotropic with $\chi_{ab} > \chi_c$ for $T>5.6$~K and $\chi_{ab} < \chi_c$ at lower~$T$ as seen in Fig.~\ref{fig:MT_EuKMn2As2}(a).  The $T$~dependence of the anisotropy is too strong to arise from the conventional shape-, magnetic-dipole and single-ion anisotropies \cite{Johnston2016}.  It may arise from $c$-axis FM correlations that grow with decreasing~$T$ faster than predicted by MFT and/or from interactions between the Eu and Mn spin sublattices.

Figure~\ref{fig:MTinv_EuKMn2As2} shows the $\chi(T)$ for $1.8~{\rm K} \leq T \leq 350$~K measured in $H=1.0$~T together with $\chi^{-1}(T)$\@. The $\chi^{-1}(T)$ shows two clear anomalies near 150~K and 275~K for both $H \parallel c$ and $H \perp c$. These anomalies can also be seen directly in the $\chi(T)$ data plotted on expanded scales in the insets of Figs.~\ref{fig:MTinv_EuKMn2As2}(a) and~\ref{fig:MTinv_EuKMn2As2}(b). We assign the 150~K anomaly to AFM ordering of the Mn spins, similar in temperature to $T_{\rm N} = 142$~K seen above in undoped ${\rm EuMn_2As_2}$. A more precise value $T_{\rm N} = 146$~K for the Mn AFM ordering is obtained below from $C_{\rm p}(T)$ data.  The origin of the 275~K anomalies in Fig.~\ref{fig:MTinv_EuKMn2As2} is unknown. The fits of the $\chi^{-1}(T)$ data by Eq.~(\ref{eq:C-W}) in the $T$ range 50~K~$\leq T \leq$~125~K are shown by the red lines in Fig.~\ref{fig:MTinv_EuKMn2As2} with fitting parameters listed in Table~\ref{tab:CW}.  One sees that $\chi_0$, $C$ and $\theta_{\rm p}$ all increase compared to the respective values for ${\rm EuMn_2As_2}$; the increases are likely spurious due to the multiple transitions at 150 and 275~K and the resulting small temperature range of the Curie-Weiss fit.

\begin{figure} 
\includegraphics[width=3.in]{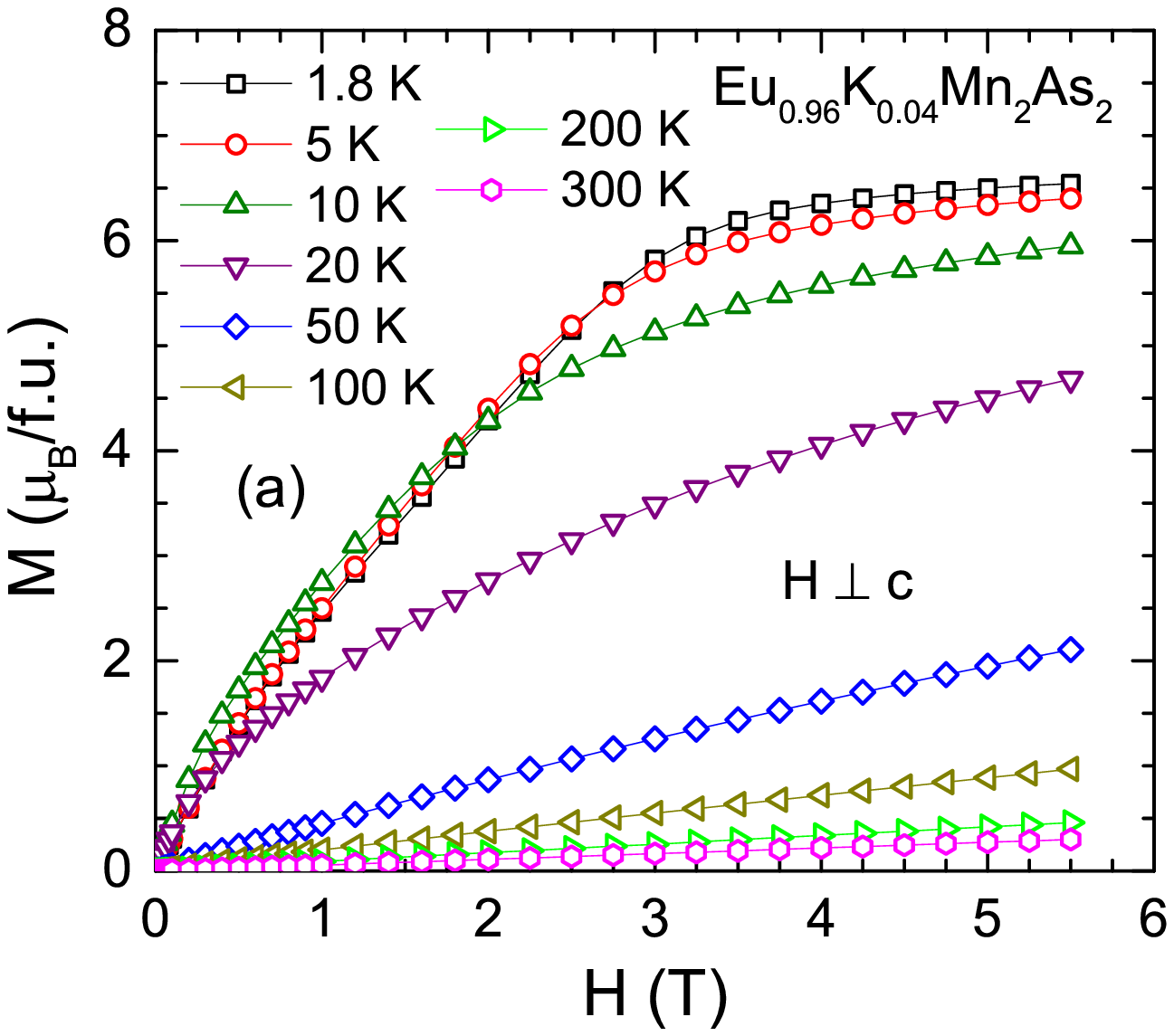}
\includegraphics[width=3.in]{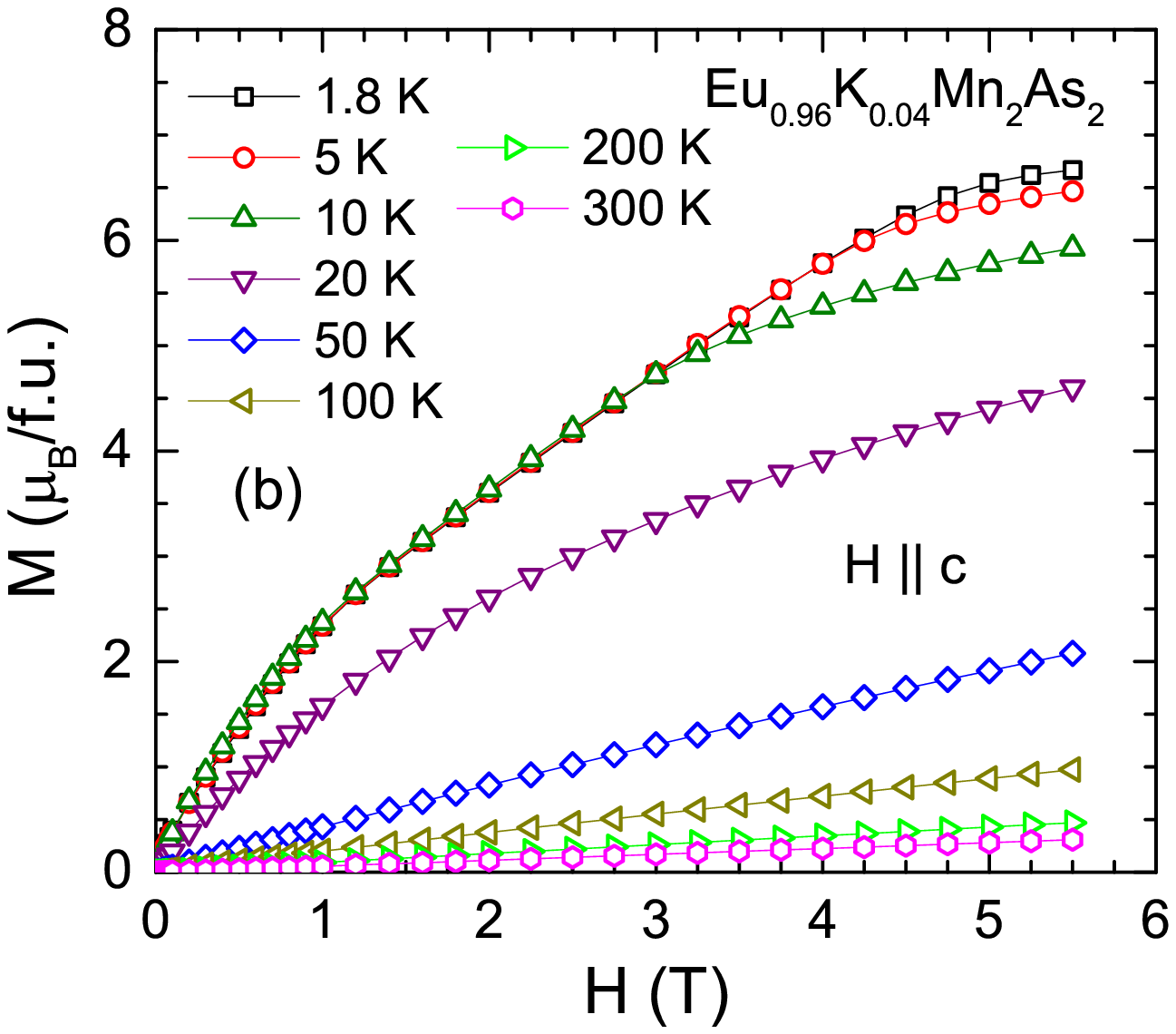}
\includegraphics[width=3.in]{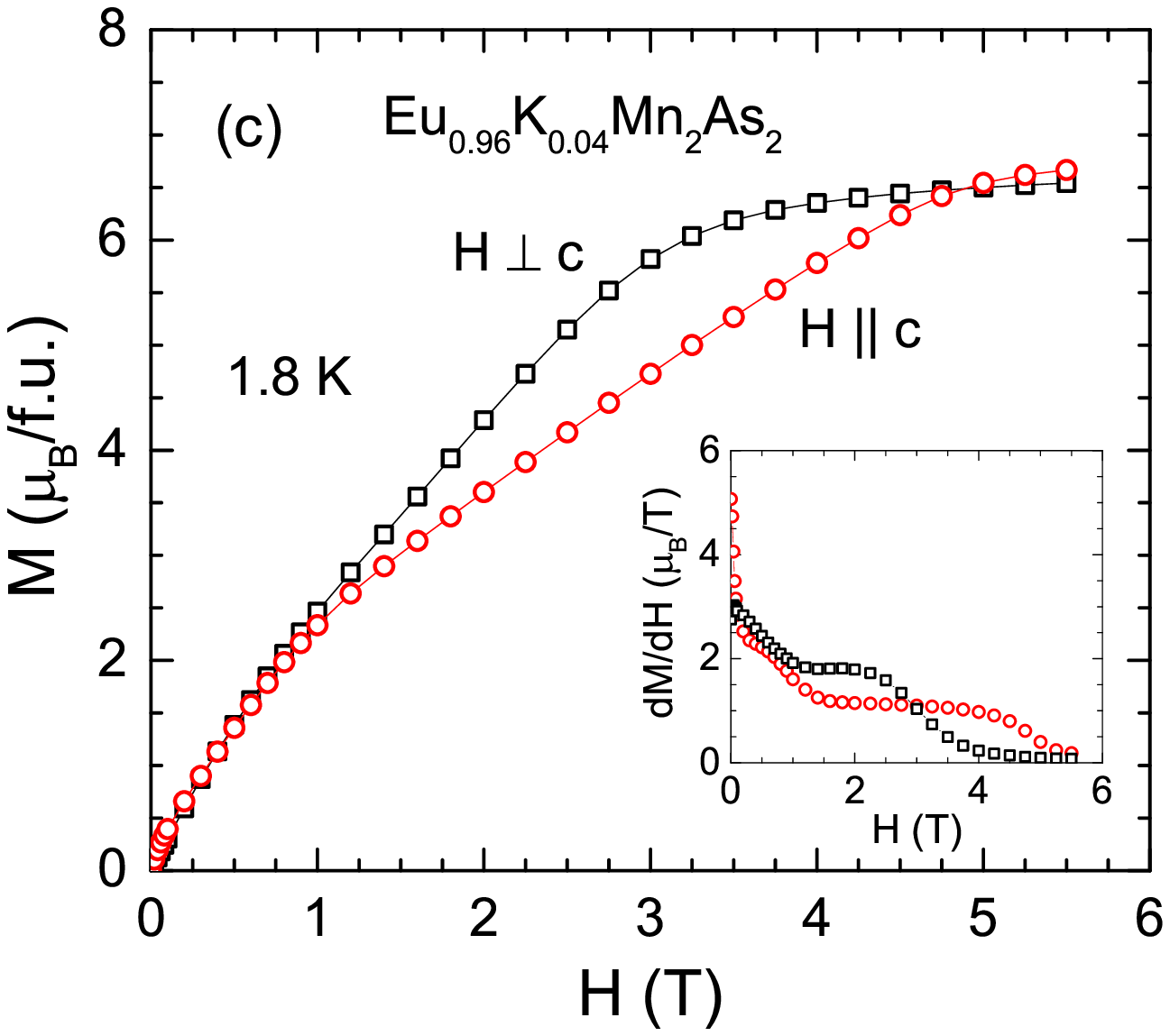}\vspace{-0.1in}
\caption{(Color online) Magnetization $M$ versus applied magnetic field $H$ isotherms for an ${\rm Eu_{0.96}K_{0.04}Mn_2As_2}$ crystal measured at the indicated temperatures for $H$ applied (a) in the $ab$ plane ($M_{ab}, H \perp  c$) and (b) along the $c$ axis ($M_c, H \parallel c$). (c)  Comparison of $H \perp  c$ and $ H \parallel c$ $M(H)$ isotherms at 1.8~K\@. Inset: $dM(H)/dH$ versus~$H$ at $T=1.8$~K\@.}
\label{fig:MH_EuKMn2As2}
\end{figure}

$M(H)$ isotherms for ${\rm Eu_{0.96}K_{0.04}Mn_2As_2}$ are shown in Fig.~\ref{fig:MH_EuKMn2As2}. Even at the lowest $T=1.8$~K, the $M(H)$ curves for both field directions show negative curvature over the whole $H$ range.  This suggests that the AFM structure of the Eu spins is both noncollinear and noncoplanar \cite{Johnston2015} as also deduced above for \ema.  Alternatively, the $M(H)$ behavior of the Eu spins might arise from interactions with the ordered Mn sublattice.  Furthermore, the $M(H)$ data at 1.8~K in Fig.~\ref{fig:MH_EuKMn2As2}(c) show metamagnetic transitions as more clearly illustated in plots of $dM/dH$ versus~$H$ in the inset.

As in the $\chi(T)$ data, the $M(H)$ isotherms show anisotropic behavior as shown for $T=1.8$~K in Fig.~\ref{fig:MH_EuKMn2As2}(c). The saturation magnetization for $H \perp  c$ and $H \parallel c$ at 1.8~K and 5.5~T are $\mu_{\rm sat} = 6.54\,\mu_{\rm B}$/f.u.\ and $\mu_{\rm sat} = 6.67\,\mu_{\rm B}$/f.u.\, respectively.  On a per Eu spin basis, these correspond to $\mu_{\rm sat} =6.81 \,\mu_{\rm B}$/Eu and $6.95 \,\mu_{\rm B}$/Eu, respectively, which are close to the expected $\mu_{\rm sat} =  7.0\,\mu_{\rm B}$ for Eu$^{+2}$. The critical fields for saturation of $M_{ab}$ and $M_c$ at 1.8~K are $H_{\rm c} \approx 3.5$~T and 5.0~T, respectively, indicating that the net anisotropy field seen by the Eu spins is strongest along the $c$~axis.

\subsection{\label{Sec:EuKMn2As2_HC}Heat Capacity}

\begin{figure} 
\includegraphics[width=3.2in]{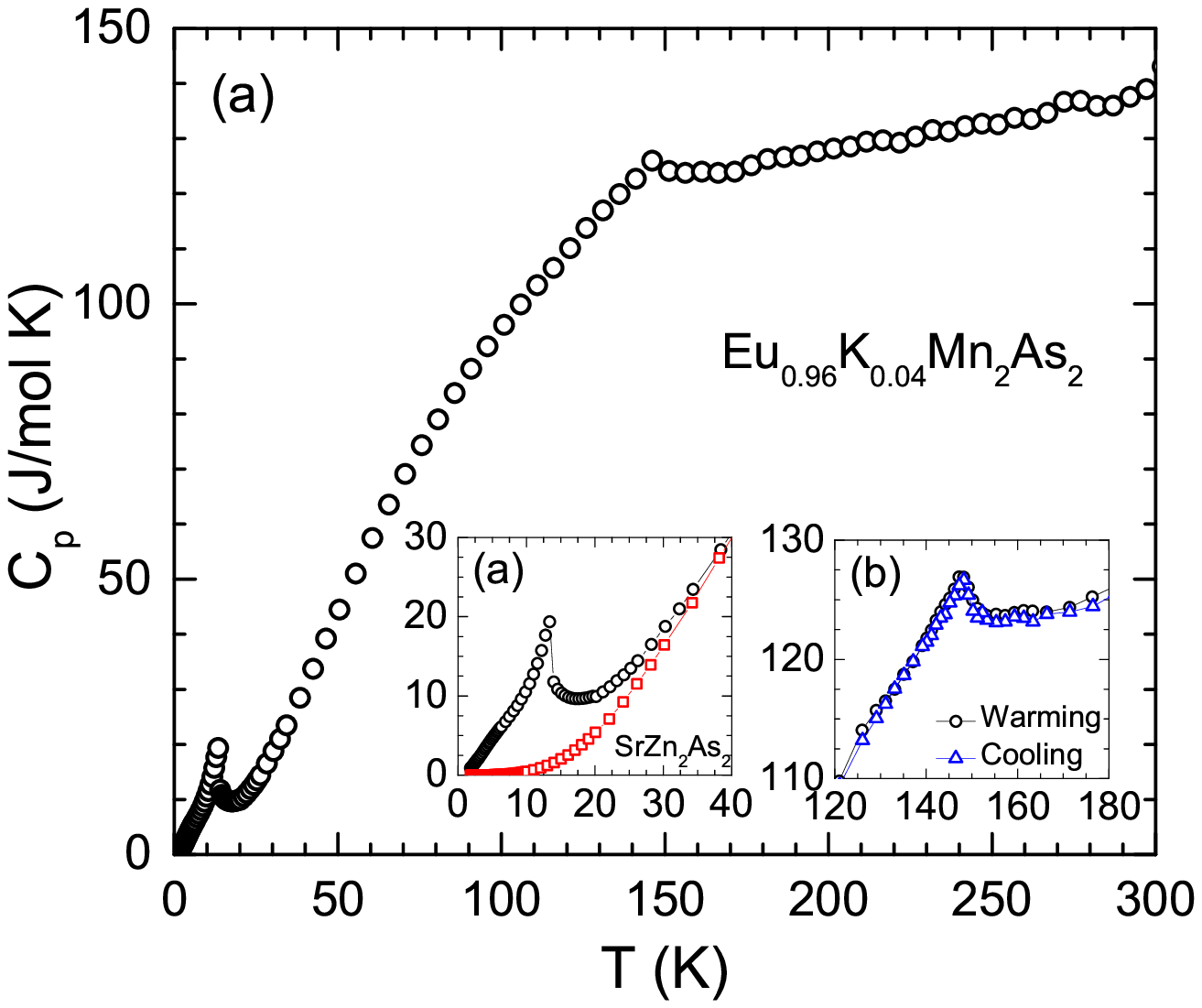}
\includegraphics[width=3.2in]{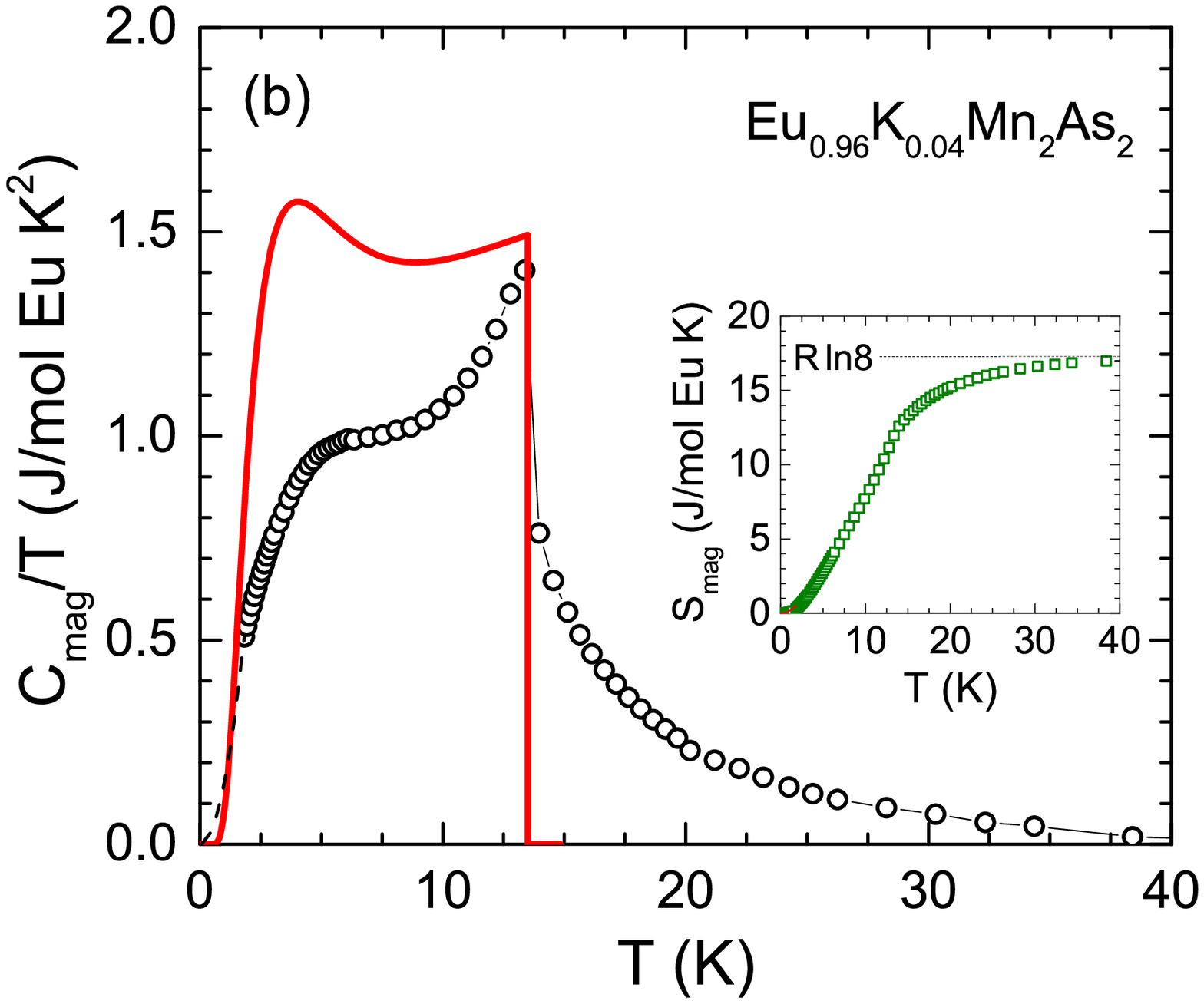}
\caption{(Color online) (a) Heat capacity $C_{\rm p}$ of an ${\rm Eu_{0.96}K_{0.04}Mn_2As_2}$ crystal versus temperature $T$ in $H=0$ for $1.8 ~{\rm K} \leq T \leq 300$~K\@. Insets: Expanded views of $C_{\rm p}(T)$ of (a) ${\rm Eu_{0.96}K_{0.04}Mn_2As_2}$ and ${\rm SrZn_2As_2}$) for $T < 40$~K and (b) ${\rm Eu_{0.96}K_{0.04}Mn_2As_2}$ for $120~{\rm K}\leq T\leq 180$~K\@. (b) Magnetic contribution $C_{\rm mag}$ to $C_{\rm p}$ plotted as $C_{\rm mag}(T)/T$ versus $T$ per mole of Eu spins. Inset: Magnetic contribution $S_{\rm mag}(T)$ to the entropy per mole of Eu spins.  The dashed curve is the extrapolation to $T=0$ and the solid red curve is the MFT prediction for $S=7/2$ and $T_{\rm N} = 13.5$~K\@.}
\label{fig:HC_EuKMn2As2}
\end{figure}

The $C_{\rm p}(T)$ data for ${\rm Eu_{0.96}K_{0.04}Mn_2As_2}$ are shown in Fig.~\ref{fig:HC_EuKMn2As2}. Like  ${\rm EuMn_2As_2}$, the low-$T$  $C_{\rm p}(T)$ data show a $\lambda$-type peak at the Eu  $T_{\rm N1} = 13.5$~K [inset (a) of Fig.~\ref{fig:HC_EuKMn2As2}(a)] without any anomaly near $T_{\rm N2}$. The high-$T$ $C_{\rm p}(T)$ data show a well-defined anomaly at 146~K without any thermal hysteresis [inset (b) of Fig.~\ref{fig:HC_EuKMn2As2}(a)] that is assigned to the Mn $T_{\rm N}$. The latter $T_{\rm N}$ is close to the Mn $T_{\rm N} \sim 150$~K seen in $\chi(T)$ in Fig.~\ref{fig:MTinv_EuKMn2As2} and is slightly higher than the Mn $T_{\rm N} = 142$~K determined above for \ema\ from $C_{\rm p}(T)$ data.

The magnetic contribution $C_{\rm mag}(T)$ per mole of Eu spins obtained after subtracting the lattice contribution is shown in Fig.~\ref{fig:HC_EuKMn2As2}(b) as $C_{\rm mag}(T)/T$ versus $T$\@. While an AFM transition at $T_{\rm N1}$ is clear from the $C_{\rm mag}(T)$ data, the presence of dynamic short-range AFM correlations up to 40~K  is also evident from the nonzero $C_{\rm mag}(T)$ at $T> T_{\rm N1}$. The MFT prediction for a second-order magnetic transition at $T_{\rm N1} = 13.5$~K for $S = 7/2$ \cite{Johnston2011} is shown by the red curve in Fig.~\ref{fig:HC_EuKMn2As2}(b). The $S_{\rm mag}(T)$ obtained from $C_{\rm mag}(T)/T$ is shown in the inset of Fig.~\ref{fig:HC_EuKMn2As2}(b).  At $T=40$~K, $S_{\rm mag}$ is close to $R\ln(8)$ expected for Eu$^{+2}$ with $S = 7/2$.  Due to the large magnetic contribution to $C_{\rm p}$ at low~$T$, it is not possible to extract a $\gamma T$ term associated with the conduction carriers identified from the $\rho(T)$ data that follow.

\subsection{\label{Sec:EuKMn2As2_Rho} Electrical Resistivity}

\begin{figure} 
\includegraphics[width=3.3in]{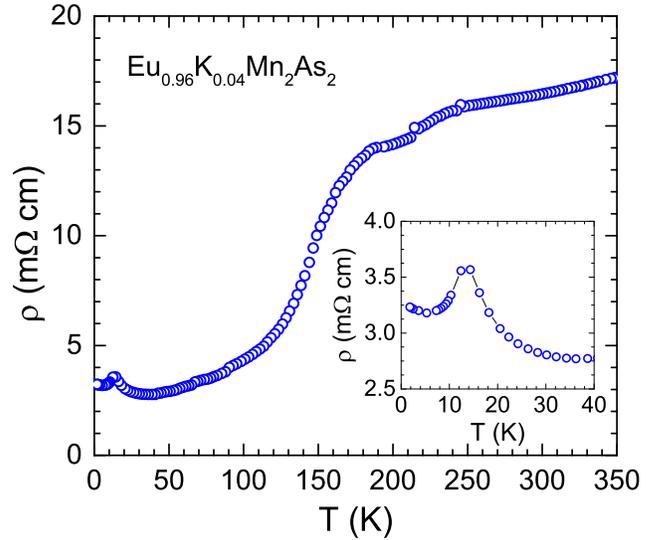}
\caption{(Color online) $ab$-plane electrical resistivity $\rho$ of an ${\rm Eu_{0.96}K_{0.04}Mn_2As_2}$ crystal versus temperature $T$ measured in $H=0$. Inset: Expanded view of $\rho(T)$ data at $T < 40$~K\@. The bumps in $\rho(T)$ at $\sim 210$~K are experimental artifacts.}
\label{fig:rho_EuKMn2As2}
\end{figure}

The $ab$-plane $\rho(T)$ of ${\rm Eu_{0.96}K_{0.04}Mn_2As_2}$ is shown in Fig.~\ref{fig:rho_EuKMn2As2}. The $\rho(T)$ exhibits metallic behavior and is qualitatively different from that of ${\rm EuMn_2As_2}$ which is semiconducting. The metallic state occurs as a result of hole~doping by partially replacing Eu$^{+2}$ by K$^{+1}$. This effect of K~substitution is similar to that in ${\rm BaMn_2As_2}$ where a 1.6\% K~substitution for Ba is sufficient to drive the system metallic \cite{Pandey2012}. The absolute value of $\rho$ in Fig.~\ref{fig:rho_EuKMn2As2} is high for a metal which  evidently arises from the low doped-hole concentration and/or low hole mobility.

The $\rho(T)$ in Fig.~\ref{fig:rho_EuKMn2As2} shows a sharp maximum in slope at $\approx 150$~K, close to the Mn $T_{\rm N} = 146$~K\@.  This strong  increase in $d\rho/dT$ on approaching $T_{\rm N}$ from above is very unusual.  The same behavior is found for ${\rm Eu_{0.93}K_{0.07}Mn_2As_2}$ in Fig.~\ref{fig:rho_EuKMn2As2_7p} below.  We suggest that this behavior arises from the decrease in spin-disorder scattering of the doped holes off the Eu local moments in the PM state above $T_{\rm N}$ and then a continuing decrease in $\rho$ occurs below $T_{\rm N}$ due to the long-range AFM order.  One does not normally observe a decrease in spin-disorder scattering in a metal containing local moments with short-range AFM order in the PM state, but it might be observable here because of the high magnitude of the resistivity which makes it more susceptible to such a perturbation.

Like $\chi(T)$ and $C_{\rm mag}(T)$, the $\rho(T)$ also shows a clear anomaly near the Eu $T_{\rm N1} = 13.5$~K as shown in the inset of Fig.~\ref{fig:rho_EuKMn2As2}. We suggest that the increase in $\rho$ with decreasing~$T$ just above the Eu $T_{\rm N1}$ arises from the formation of superzone energy pseudogaps in the Brillouin zone due to short-range incommensurate AFM order of the Eu spins \cite{Elliott1963,Elliott1964,Ellerby1998}.   Indeed, AFM superzone gap formation below the AFM $T_{\rm N}$ of the Eu$^{+2}$ spins-7/2 was recently inferred in the related Eu compound ${\rm EuPd_2As_2}$ with the tetragonal ${\rm ThCr_2Si_2}$-type structure \cite{Anand2014c}.  The decrease in $\rho$ below $T_{\rm N1}$ might then arise from a decrease in spin-disorder scattering.

\section{\label{Sec:EuKMn2As2_2} Physical properties of  E\lowercase{u}$_{0.93}$K$_{0.07}$M\lowercase{n}$_2$A\lowercase{s}$_2$}

\subsection{\label{Sec:EuKMn2AS2_7p_MT_MH} Magnetization and Magnetic Susceptibility}

\begin{figure}
\includegraphics[width=3in]{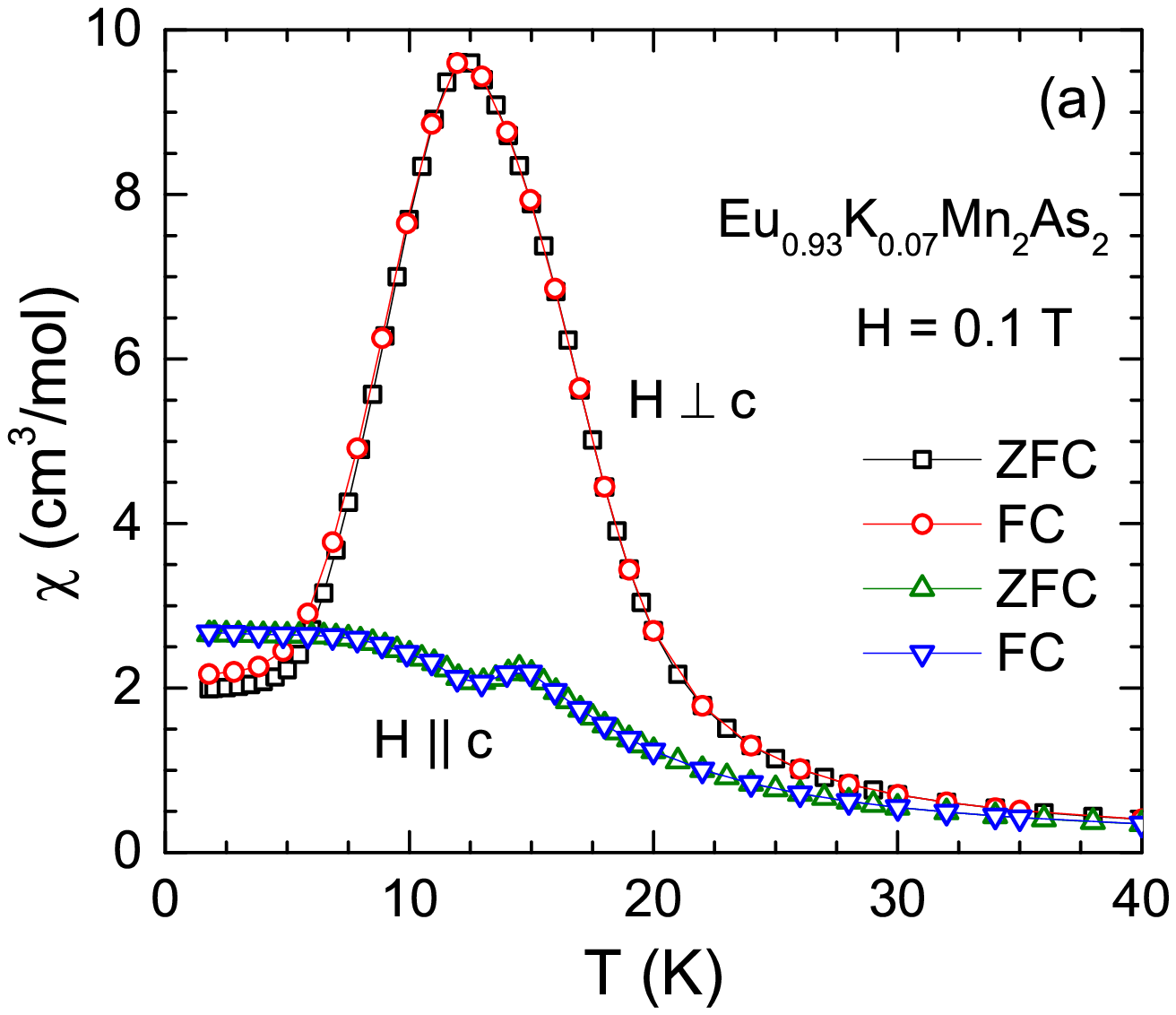}
\includegraphics[width=3in]{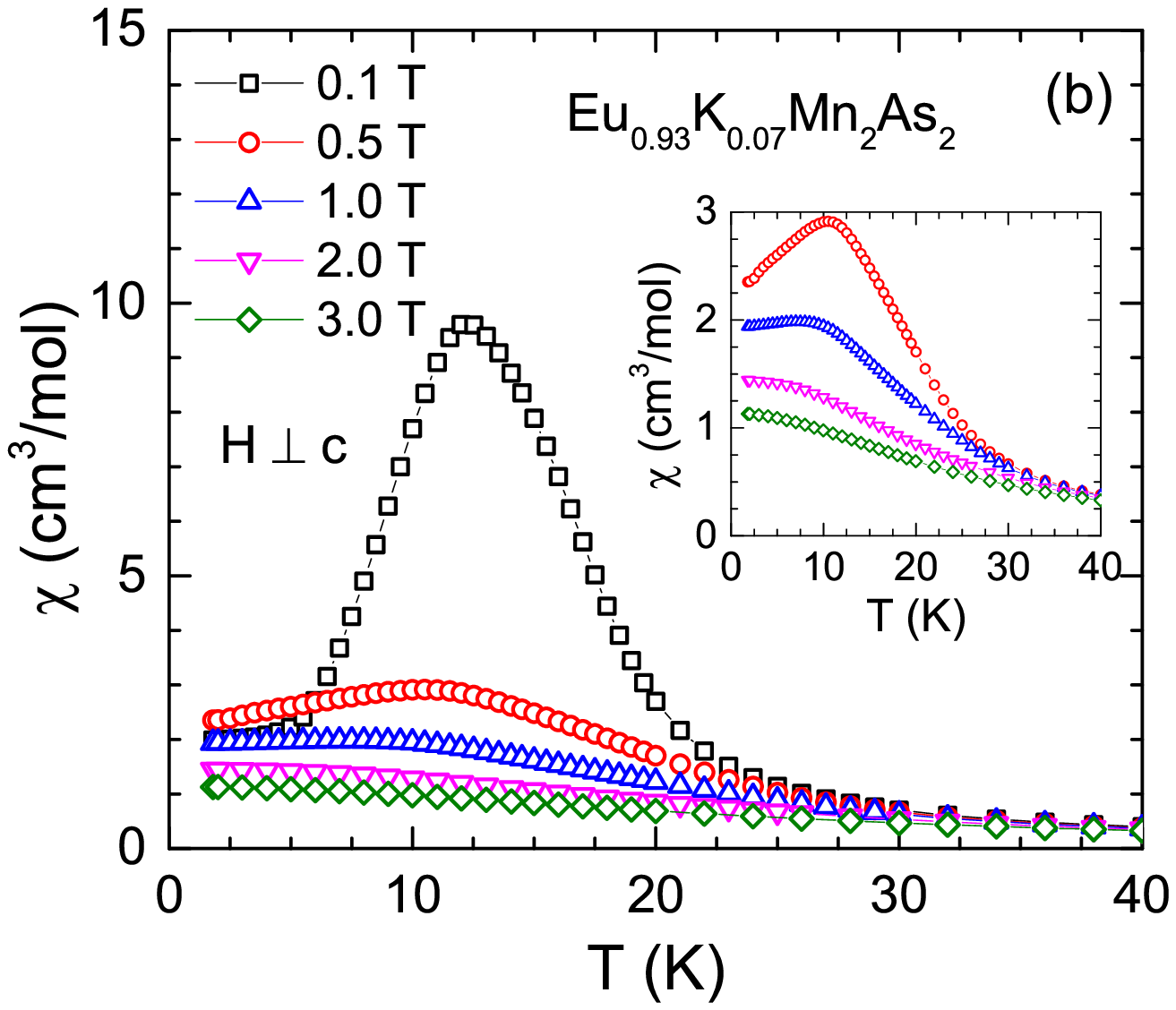}
\includegraphics[width=3in]{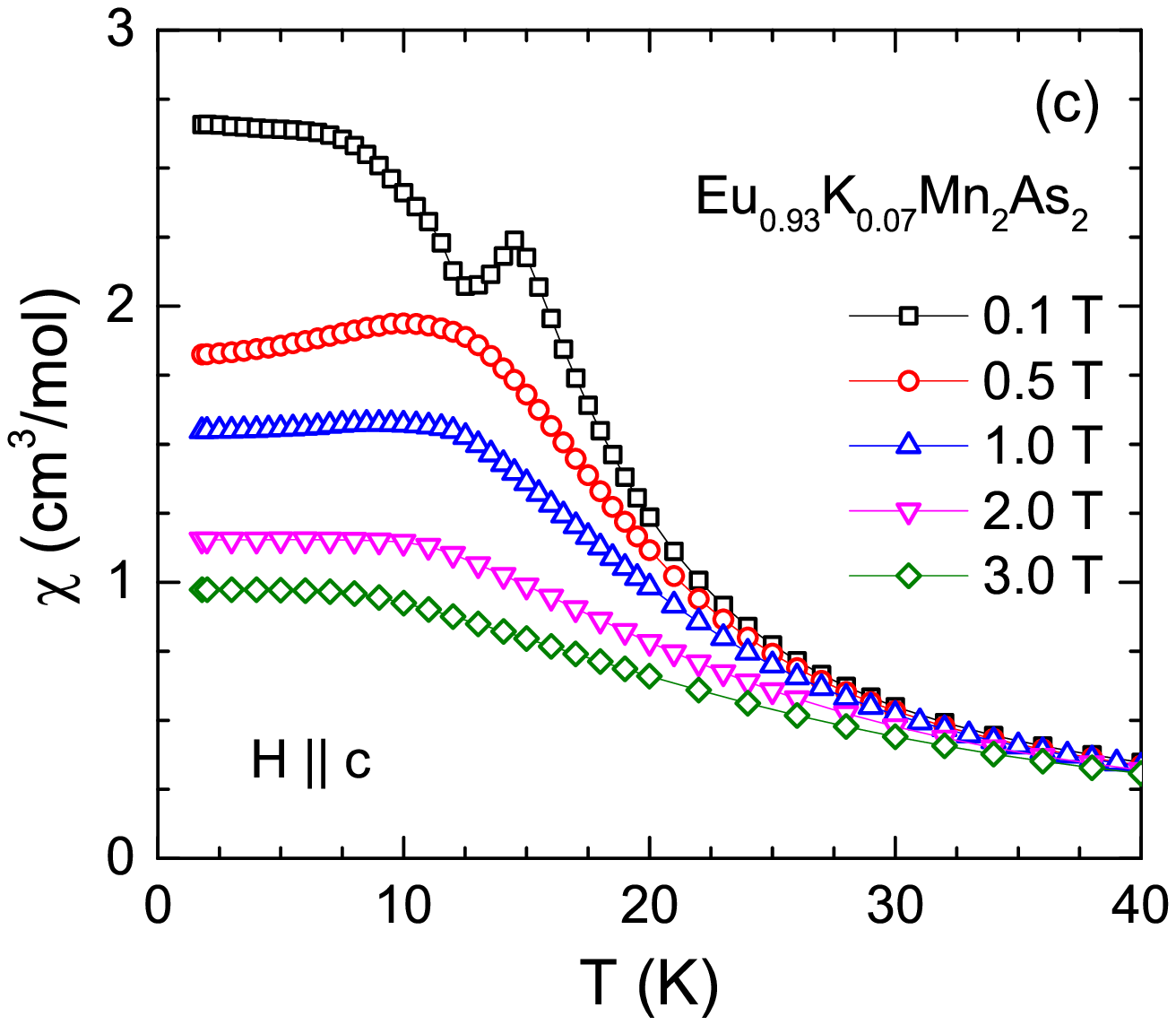}
\caption{(Color online) (a) Zero-field-cooled (ZFC) and field-cooled (FC) magnetic susceptibility $\chi$ of an ${\rm Eu_{0.93}K_{0.07}Mn_2As_2}$ crystal versus temperature $T$ for $1.8~{\rm K} \leq T \leq 40$~K measured in a magnetic field $H= 0.1$~T applied in the $ab$ plane ($\chi_{ab}, H \perp  c$) and along the $c$ axis ($\chi_c, H \parallel c$). (b) ZFC $\chi(T)$ for $1.8~{\rm K} \leq T \leq 40$~K measured at the indicated $ H \perp c$. Inset: Expanded plot of $\chi(T)$.  (c) ZFC $\chi(T)$ for $1.8~{\rm K} \leq T \leq 40$~K measured at the indicated $H \parallel c$.}
\label{fig:MT_EuKMn2As2_7p}
\end{figure}

\begin{figure} 
\includegraphics[width=\columnwidth]{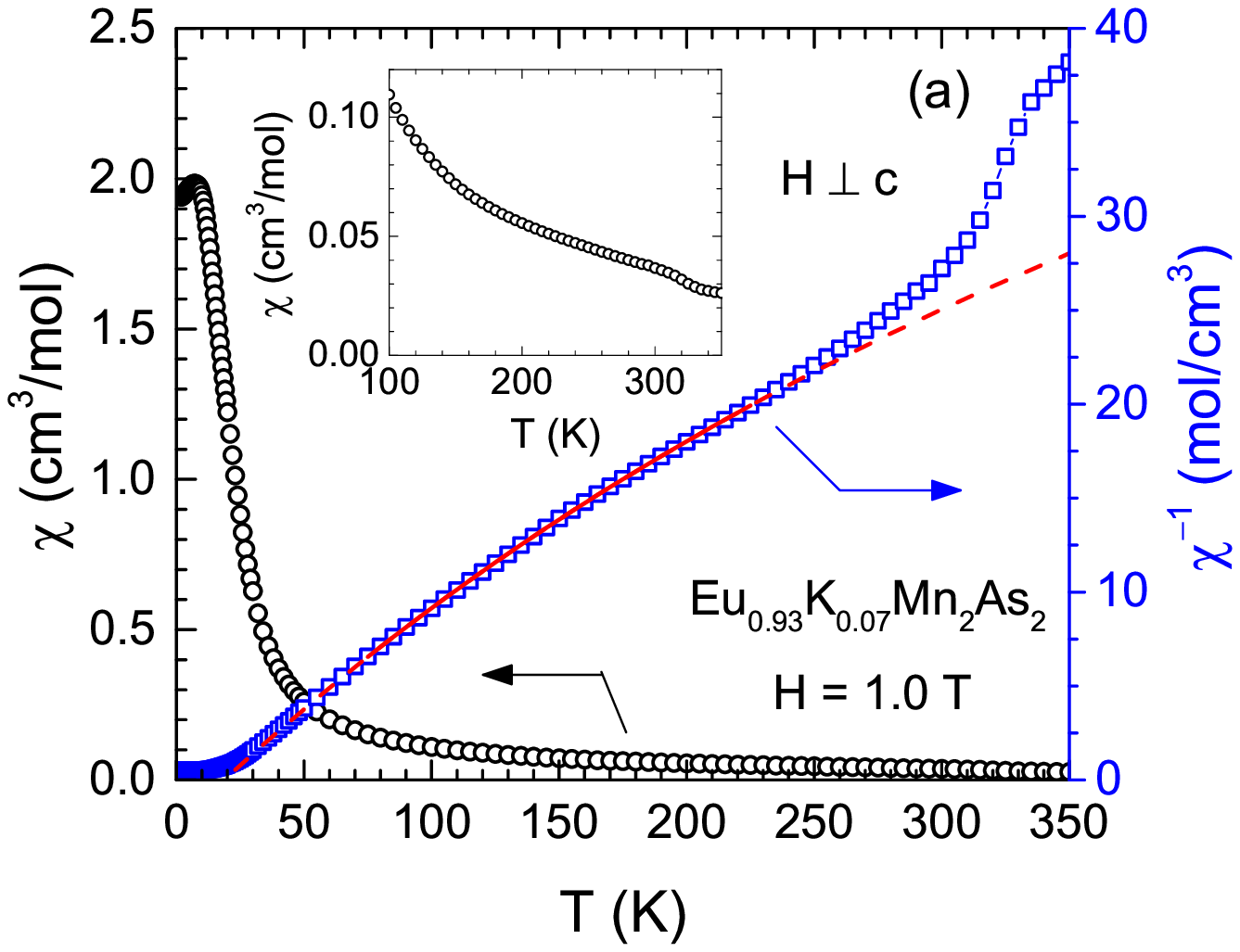}\vspace{0.1in}
\includegraphics[width=\columnwidth]{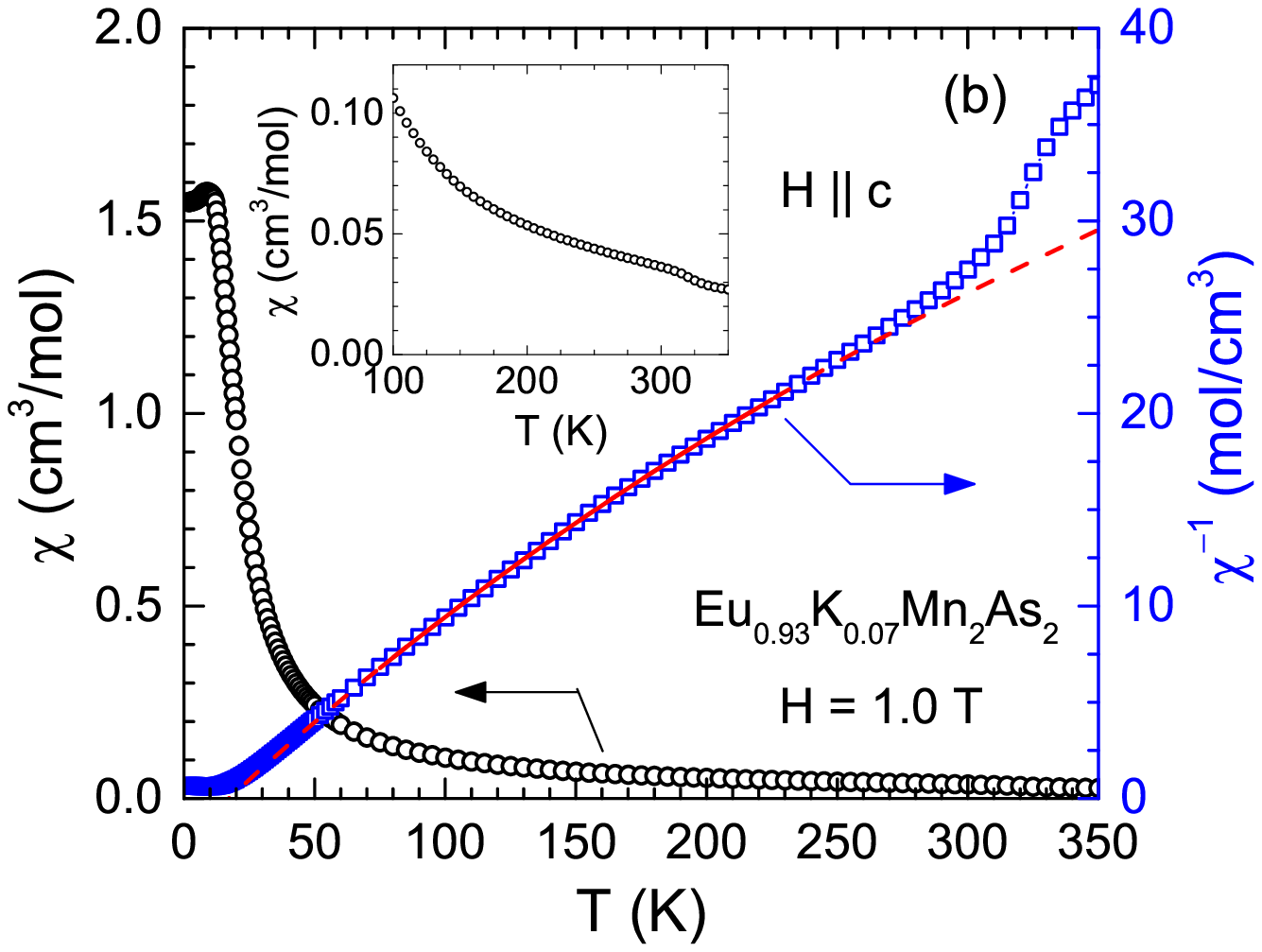}
\caption{(Color online) Zero-field-cooled magnetic susceptibility $\chi$ (left ordinate) and its inverse $\chi^{-1}$ (right ordinate) of an ${\rm Eu_{0.93}K_{0.07}Mn_2As_2}$ crystal versus temperature $T$ for $1.8~{\rm K} \leq T \leq 350$~K measured in a magnetic field $H=1.0$~T applied (a) in the $ab$ plane ($\chi_{ab}, H \perp  c$) and (b) along the $c$ axis ($\chi_c, H \parallel c$). The solid curves are fits of the $\chi^{-1}(T)$ data by the modified Curie-Weiss behavior in the $T$ range 75~K~$\leq T \leq$~225~K and the dashed curves are the extrapolations. Insets: Expanded plots of $\chi(T)$.}
\label{fig:MTinv_EuKMn2As2_7p}
\end{figure}

The ZFC and FC $\chi(T\leq 40~{\rm K})$ data for an ${\rm Eu_{0.93}K_{0.07}Mn_2As_2}$ single crystal in $H=0.1$~T for $H \parallel c$ and $H \perp c$ are shown in Fig.~\ref{fig:MT_EuKMn2As2_7p}(a). For $H \parallel c$ the data show a cusp at $T_{\rm N1} =14.5$~K arising from AFM ordering of the Eu$^{+2}$ spins followed by a sharp minimum at $T_{\rm N2} = 12.5$~K and then an increase, eventually tending to a nearly constant value below 6~K [see also Fig.~\ref{fig:MT_EuKMn2As2_7p}(c)]. For $H \perp c$  a single peak is seen at $T_{\rm N2} = 12.5$~K\@.  This mismatch between the temperatures of the highest transition $T_{\rm N1} = 14.5$~K  for $H\parallel c$ and $T_{\rm N2} = 12.5$~K for $H\perp c$ obtained in the same field is unusual and is likely a result of the data in Fig.~\ref{fig:MT_EuKMn2As2_7p}(a) not being in the low-field limit.  No hysteresis is observed between the ZFC and FC $\chi$ for either field direction. From Figs.~\ref{fig:MT_EuKMn2As2_7p}(b) and~\ref{fig:MT_EuKMn2As2_7p}(c), the AFM transitions at $T_{\rm N1}$ and~$T_{\rm N2}$ become ill-defined for $H\geq 0.5$~T and hence we infer that long-range AFM transitions may not occur for $H\geq 0.5$~T\@.  $C_{\rm p}(T)$ measurements versus field are needed to  clarify this issue.

The $\chi(T)$ in Fig.~\ref{fig:MT_EuKMn2As2_7p}(a) is strongly anisotropic with $\chi_{ab} > \chi_c$ for $T > 5.9$~K, whereas $\chi_{ab} < \chi_c$ for $T < 5.9$~K\@. As with ${\rm Eu_{0.97}K_{0.03}Mn_2As_2}$ above, this $T$-dependent anisotropy is too strong to be explained by the usual shape, magnetic-dipole and single-ion anisotropies for Heisenberg spins.  We infer that FM correlations beyond MFT and/or interactions with the ordered Mn sublattice contribute to the strong anisotropy.

The $\chi(T)$ and $\chi^{-1}(T)$ data in the $T$ range $1.8~{\rm K} \leq T \leq 350$~K measured in $H=1$~T are shown in Figs.~\ref{fig:MTinv_EuKMn2As2_7p}(a) and~\ref{fig:MTinv_EuKMn2As2_7p}(b) for $H \perp c$ and $H \parallel c$, respectively. The high-$T$ $\chi(T)$ (insets) and especially $\chi^{-1}(T)$ data in Fig.~\ref{fig:MTinv_EuKMn2As2_7p} for both field orientations show a strong deviation from the extrapolated modified Curie-Weiss behavior (see below) above $\sim250$~K, undoubtedly due to FM ordering of a trace amount of MnAs impurity below its Curie temperature of 320~K as seen previously in ${\rm BaMn_2As_2}$ crystals \cite{Singh2009a}. From the data in Fig.~\ref{fig:MTinv_EuKMn2As2_7p} we estimate the saturation magnetization for the MnAs impurities to be $\approx{\rm 9~G~cm^3/mol} = 0.0016~\mu_{\rm B}$/f.u., which corresponds to 0.046~mol\% of MnAs impurities using the saturation moment of $\approx 3.5~\mu_{\rm B}$/f.u.\ \cite{Haneda1977, Saparov2012} for MnAs.  This very low impurity level is not detectable by XRD\@.

The fits of the $\chi^{-1}(T)$ data in Fig.~\ref{fig:MTinv_EuKMn2As2_7p} by the modified Curie-Weiss law in Eq.~(\ref{eq:C-W}) over the temperature range 75~K~$\leq T \leq$~225~K are shown by the solid red curves in Fig.~\ref{fig:MTinv_EuKMn2As2_7p} and the extrapolations are shown as dashed red curves. The parameters obtained from the fits are listed in Table~\ref{tab:CW}. The values of $C$ are close to the value expected for Eu$^{+2}$. Contrary to $\chi^{-1}(T)$ of ${\rm Eu_{0.96}K_{0.04}Mn_2As_2}$ in Fig.~\ref{fig:MTinv_EuKMn2As2}, there is no obvious anomaly in the $\chi^{-1}(T)$ at $\sim 150$~K in Fig.~\ref{fig:MTinv_EuKMn2As2_7p} that can be associated with Mn AFM ordering. However, this transition does appear at 150~K in the $C_{\rm p}(T)$ and $\rho(T)$ data below.

\begin{figure} 
\includegraphics[width=3in]{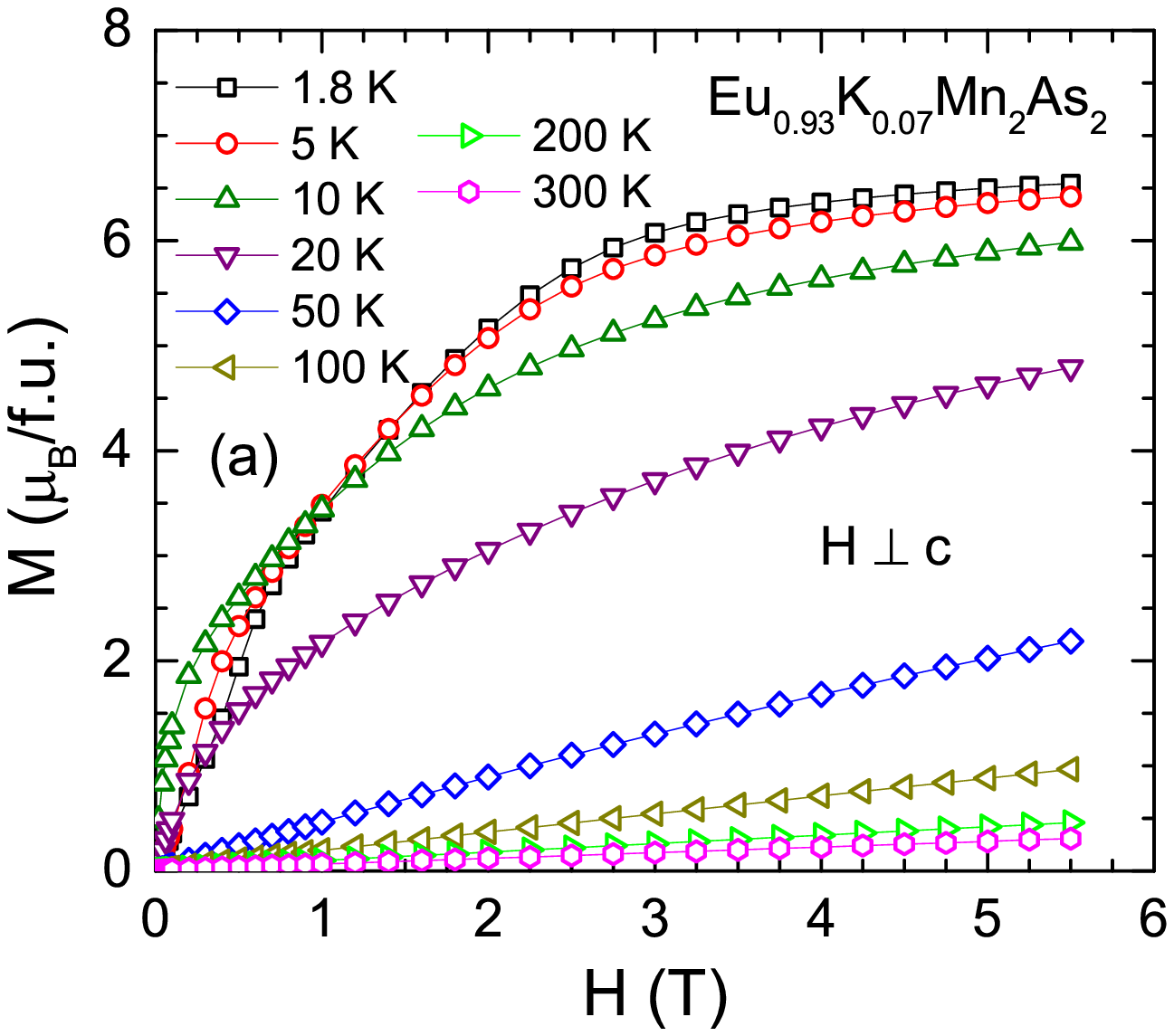}\vspace{0.1in}
\includegraphics[width=3in]{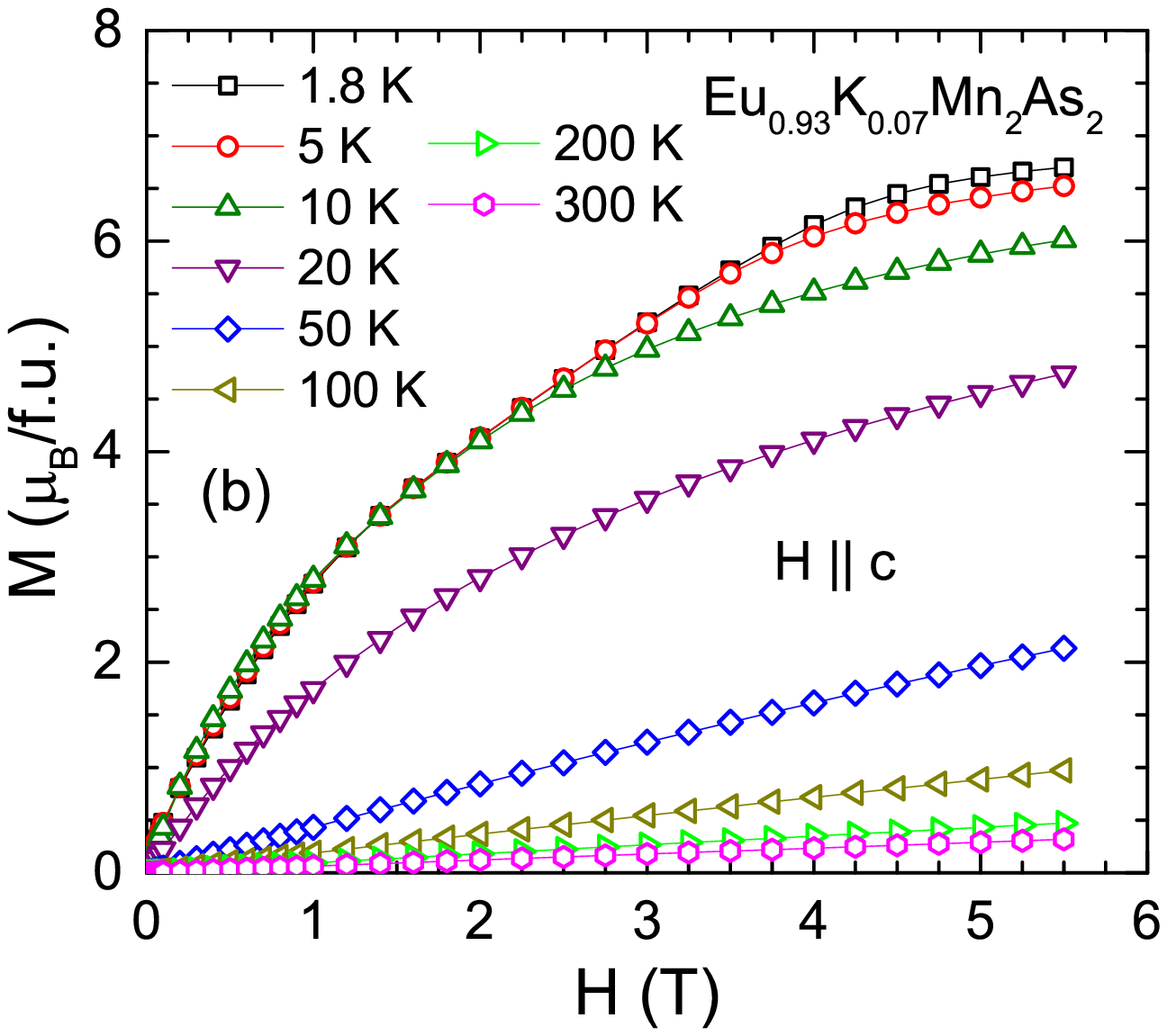}\vspace{0.1in}
\includegraphics[width=3in]{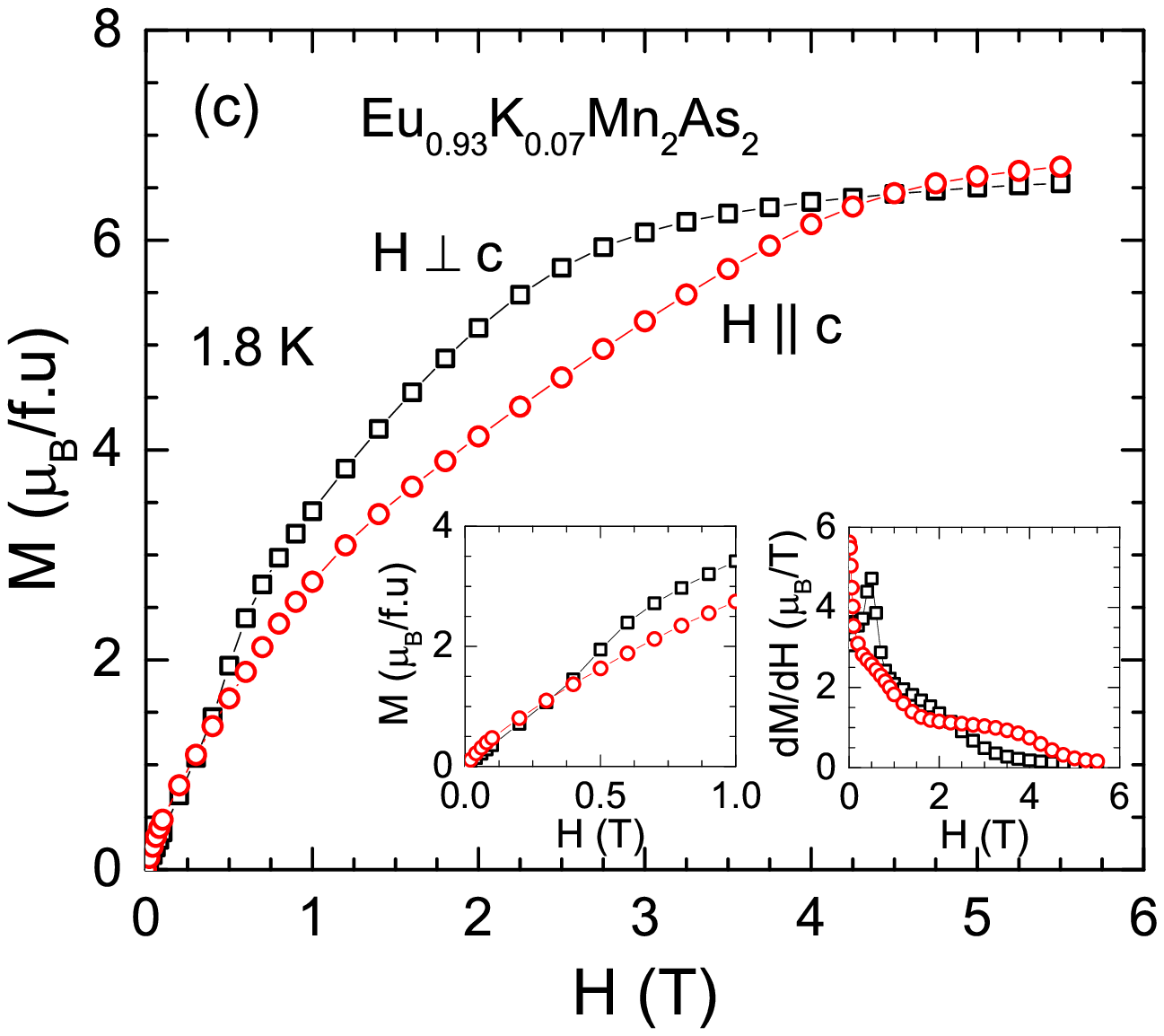}
\caption{(Color online) Magnetization $M$ versus applied magnetic field~$H$ isotherms for an ${\rm Eu_{0.93}K_{0.07}Mn_2As_2}$ crystal measured at the indicated temperatures for $H$ applied (a) in the $ab$ plane ($M_{ab}, H \perp  c$) and (b) along the $c$ axis ($M_c, H \parallel c$). (c) Comparison of $H \perp  c$ and $ H \parallel c$ $M(H)$ isotherms at 1.8~K\@. Left inset: $M(H<1$~T)\@. Right inset:  $dM/dH$ versus $H$\@.}
\label{fig:MH_EuKMn2As2_7p}
\end{figure}

Isothermal $M(H)$ data for ${\rm Eu_{0.93}K_{0.07}Mn_2As_2}$ are shown in Figs.~\ref{fig:MH_EuKMn2As2_7p}(a) and~\ref{fig:MH_EuKMn2As2_7p}(b) for $H\perp c$ and $H\parallel c$, respectively. The isotherms are somewhat similar to those of ${\rm Eu_{0.96}K_{0.04}Mn_2As_2}$. For $H \perp c$ and $T<T_{\rm N1}$, the $M(H)$ isotherms exhibit strong negative curvature until the magnetization reaches saturation.  For $H\perp c$ the isotherm at 10~K has a large initial slope, reminiscent of FM, which results in crossing over the 1.8 and 5~K isotherms at a higher field of $\sim1$~T\@.  This behavior is unusual.  At 1.8~K the $M_{ab}(H)$ data saturate above $H= 3.0$~T to $M_{\rm s}^{ab} = 6.54\,\mu_{\rm B}$/f.u.\ which yields $7.03 \,\mu_{\rm B}$/Eu in agreement with expectation. On the other hand, at 1.8~K $M_c = 6.70\,\mu_{\rm B}$/f.u.\ at 5.5~T, which yields a moment of $7.20 \,\mu_{\rm B}$/Eu which is slightly larger than the value expected for Eu$^{+2}$.

The isotherms at 1.8~K are shown separately in \ref{fig:MH_EuKMn2As2_7p}(c).  Here one can identify structures in the $M(H)$ isotherms that are more clearly seen at low field in the left inset and in $dM/dH$ versus~$H$ in the right inset, suggesting metamagnetic transitions.  For $H \parallel c$, one see a very fast decrease in $dM/dH$ with $H$, suggesting a possible small FM component to the ordering. Then a bump occurs in $dM_c/dH$ at $\sim 3.5$~T associated with the saturation behavior of $M_c$\@.  For $H\perp c$, the data show a sharp peak at about 0.5~T that is not reflected in the $H\parallel c$ data, suggesting a different type of metamagnetic transition than at $\sim 3.5$~T in $M_c(H)$. Thus the $H$~dependence of the AFM structure is complex and anisotropic.  The ordered Mn moments are likely involved in the field dependences of the magnetization at low~$T$\@.
 
\subsection{\label{Sec:EuKMn2As2_7p_HC} Heat Capacity}

\begin{figure} 
\includegraphics[width=3.3in]{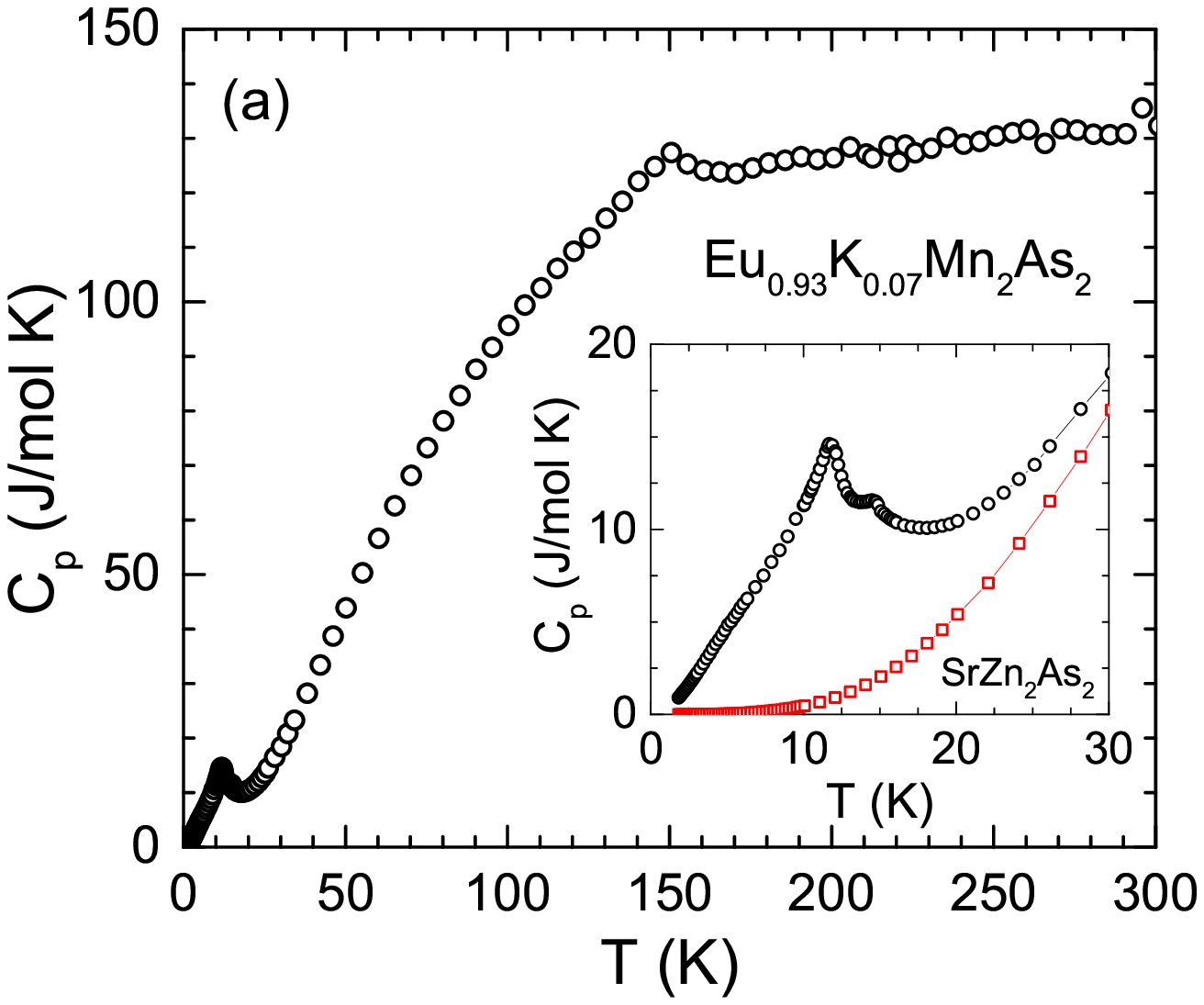}
\includegraphics[width=3.3in]{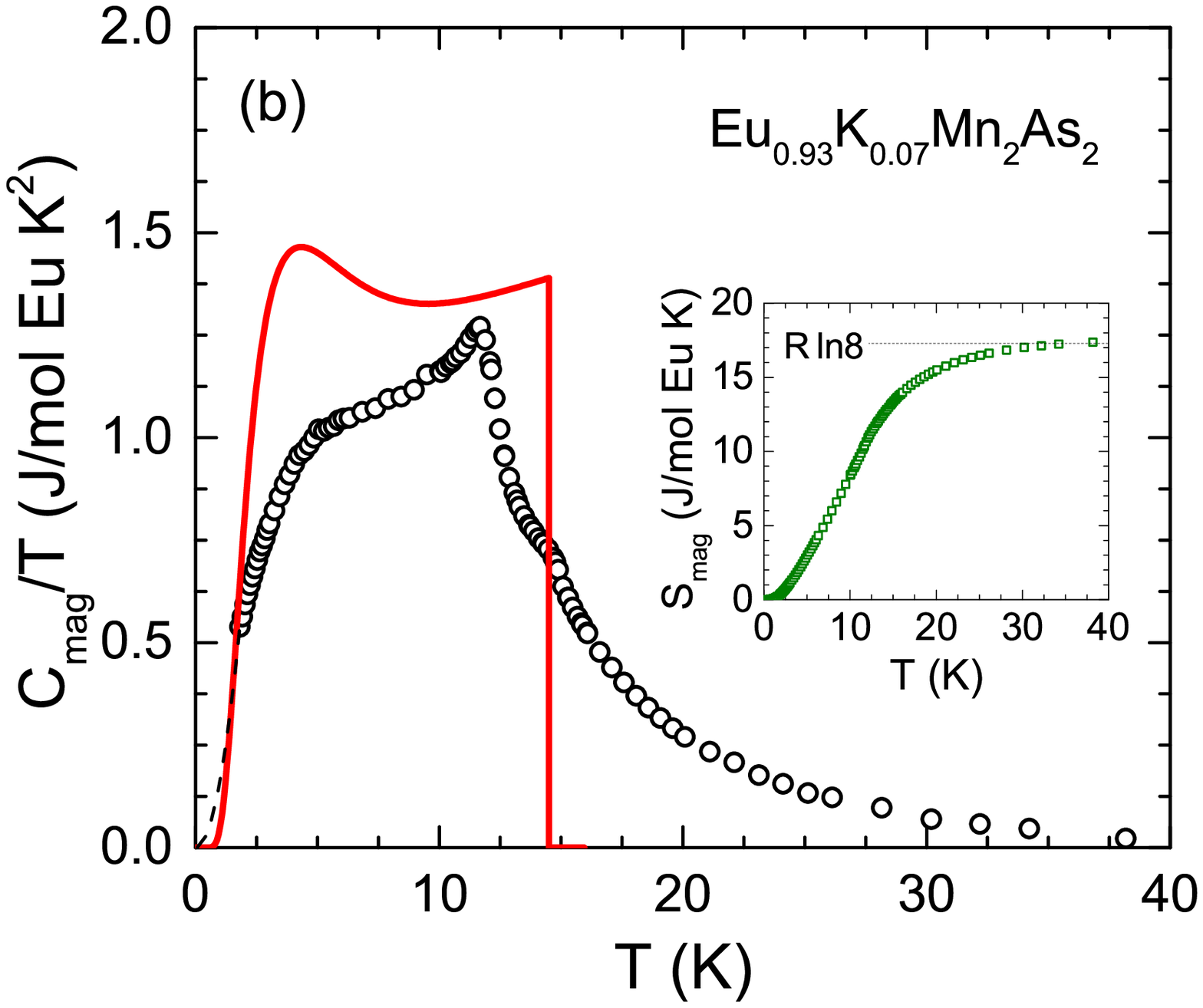}
\caption{(Color online) (a) Heat capacity $C_{\rm p}$ of an ${\rm Eu_{0.93}K_{0.07}Mn_2As_2}$ crystal versus temperature $T$ for $1.8~{\rm K} \leq T \leq 300$~K measured in zero magnetic field. Inset: Expanded plot of $C_{\rm p}(T)$ along with the $C_{\rm p}(T)$ of the nonmagnetic reference compound ${\rm SrZn_2As_2}$. (b) Magnetic contribution $C_{\rm mag}$ to $C_{\rm p}$ plotted as $C_{\rm mag}(T)/T$ versus $T$ per mole of Eu. Inset: Magnetic contribution $S_{\rm mag}(T)$ to the entropy per mole of Eu. The dashed curve is the extrapolation to $T=0$ and the solid red curve is the MFT prediction for $S=7/2$ and $T_{\rm N} = 14.5$~K\@. }
\label{fig:HC_EuKMn2As2_7p}
\end{figure}

\begin{figure} 
\includegraphics[width=3.3in]{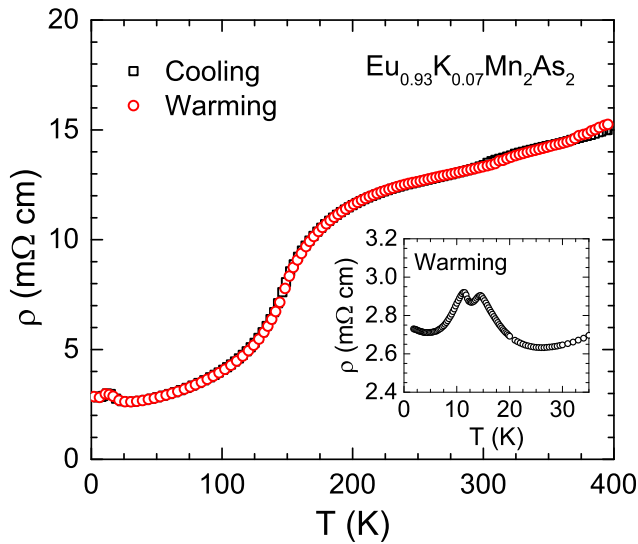}
\caption{(Color online) $ab$-plane electrical resistivity $\rho$ of an ${\rm Eu_{0.93}K_{0.07}Mn_2As_2}$ crystal versus temperature $T$ for $1.8~{\rm K} \leq T \leq 395$~K measured in zero magnetic field. Inset: Expanded plot of  $\rho(T)$ data for $T < 35$~K\@.}
\label{fig:rho_EuKMn2As2_7p}
\end{figure}

The $C_{\rm p}(T)$ data for an ${\rm Eu_{0.93}K_{0.07}Mn_2As_2}$ crystal are shown in Fig.~\ref{fig:HC_EuKMn2As2_7p}(a). For this material two well-defined anomalies in $C_{\rm p}(T)$ occur at $T_{\rm N1} = 14.5$~K and $T_{\rm N2} = 11.9$~K that are related to Eu-moment ordering, as shown in the inset of Fig.~\ref{fig:HC_EuKMn2As2_7p}(a), which are consistent with the transitions $T_{\rm N1}$ and $T_{\rm N2}$ seen in $\chi(T)$ above. The observation of two anomalies in the $C_{\rm p}(T)$ of ${\rm Eu_{0.93}K_{0.07}Mn_2As_2}$ is an interesting effect of increasing the K concentration. The $C_{\rm p}(T)$ of ${\rm Eu_{0.96}K_{0.04}Mn_2As_2}$ and ${\rm EuMn_2As_2}$ showed only one anomaly at 13.5~K and 15.0~K, respectively, although they both show two anomalies in the $\chi(T)$ data. This difference suggests a change in magnetic structure brought about by increasing the K~concentration to 7\% doping.   

The $C_{\rm mag}(T)$ per mole of Eu spins shown as $C_{\rm mag}(T)/T$ versus $T$ in Fig.~\ref{fig:HC_EuKMn2As2_7p}(b) were derived as previously and exhibit short-range AFM correlations above $T_{\rm N1}$. The MFT prediction for $T_{\rm N1} = 14.5$~K and $S = 7/2$ \cite{Johnston2011} is shown in Fig.~\ref{fig:HC_EuKMn2As2_7p}(b). The $S_{\rm mag}(T)$ per mole of Eu spins derived from $C_{\rm mag}(T)$ is shown in the inset of Fig.~\ref{fig:HC_EuKMn2As2_7p}(b), where the high-$T$ limit is close to $R\ln8$ as expected for $S=7/2$.

At high~$T$ the $C_{\rm p}(T)$ data in Fig.~\ref{fig:HC_EuKMn2As2_7p}(a) show a clear cusp at 150~K that is slightly higher in temperature than $T_{\rm N} = 146$~K in ${\rm Eu_{0.96}K_{0.04}Mn_2As_2}$ and 142~K in ${\rm EuMn_2As_2}$ and thus show a systematic increase in the Mn $T_{\rm N}$ with increasing K~concentration. 

\subsection{\label{Sec:EuKMn2As2_7p_Rho} Electrical Resistivity}

The $ab$-plane $\rho(T)$ of an ${\rm Eu_{0.93}K_{0.07}Mn_2As_2}$ crystal is shown in Fig.~\ref{fig:rho_EuKMn2As2_7p}. The overall behavior of $\rho(T)$ is similar to that of ${\rm Eu_{0.96}K_{0.04}Mn_2As_2}$ in Fig.~\ref{fig:rho_EuKMn2As2}. The $\rho(T)$ of ${\rm Eu_{0.93}K_{0.07}Mn_2As_2}$ is metallic, arising from the doped holes induced by K~substitution.  The $\rho$ shows an S-shaped behavior on cooling through the Mn $T_{\rm N} = 150$~K with an inflection point at $T_{\rm N}$, similar to the shape and magnitude of $\rho$ exhibited by ${\rm Eu_{0.96}K_{0.04}Mn_2As_2}$. There is almost no thermal hysteresis in the data as shown in Fig.~\ref{fig:rho_EuKMn2As2_7p}. A double peak in $\rho(T)$ is observed at $T_{\rm N1} = 14.5$~K and $T_{\rm N2} = 11.5$~K as shown in the inset of Fig.~\ref{fig:rho_EuKMn2As2_7p}.  The appearance of these two peaks may be associated with the formation of superzone gaps as inferred for ${\rm Eu_{0.96}K_{0.04}Mn_2As_2}$. However, compared to ${\rm Eu_{0.96}K_{0.04}Mn_2As_2}$, we see two transitions in the low-$T$ $\rho(T)$ data for ${\rm Eu_{0.93}K_{0.07}Mn_2As_2}$  compared to one in former compound.  This suggests a possible change in the AFM structure of the Eu moments at the higher K~concentration.

\section{\label{Conclusion} Summary and Discussion}

\subsection{EuMn$_2$As$_2$ and Related Compounds}

\begin{table*}
\caption{\label{Tab:OrdTemps} Trigonal lattice parameters and unit cell volume (hexagonal setting) at room temperature and magnetic ordering temperatures $T_{\rm m}$ (FM or AFM) for $AM_2X_2$ compounds with the ${\rm CaAl_2Si_2}$-type structure.  Also included are the maximum observed  ordered or saturation moments $\mu_{\rm sat}$. PW means the present work and NA means not applicable.}
\begin{ruledtabular}
\begin{tabular}{lccccccccc}
Compound 				& $a$ 		& $c$ 		& $c/a$		& $V_{\rm cell}$ 	& Eu $T_{\rm m}$ 	& Eu $\mu_{\rm sat}$	&  Mn $T_{\rm m}$	& 	Mn $\mu_{\rm sat}$	& Ref.				\\
& (\AA)		& (\AA)		&	& (\AA$^3$)	& (K)	& $\left(\frac{\mu_{\rm B}}{\rm Eu~atom}\right)$ &	(K)	&	$\left(\frac{\mu_{\rm B}}{\rm Mn~atom}\right)$	\\
\hline
${\rm SrMn_2P_2}$		& 4.1680(5)	& 7.137(1)	& 1.7123		& 107.38	& NA		&	& AFM~53(1)\footnotemark[1] 	& 	&\onlinecite{Brock1994}		\\
${\rm EuMn_2P_2}$		& 4.1294(3)	& 6.9936(8)	& 1.6936		& 103.28	& 	AFM 16.5(3)\footnotemark[1]$^,$\footnotemark[3]	& 6.98(35)\footnotemark[1]	&  	& 	&\onlinecite{Payne2002}		\\
\hline
${\rm CaMn_2As_2}$		& 4.2377(1)	& 7.0333(2)	& 1.6597		& 109.39	& NA		&		& AFM~62(3)\footnotemark[2]$^,$\footnotemark[3] 	&	&\onlinecite{Sangeetha2016}	\\
${\rm SrMn_2As_2}$		&			&			&			&				& NA		&			& AFM~120(2)\footnotemark[2]$^,$\footnotemark[3]		&		&\onlinecite{Sangeetha2016}	\\
${\rm SrMn_2As_2}$		&			&			&			&		& NA		&		& AFM~118(2)\footnotemark[1]		&	3.6(2)\footnotemark[1]	&\onlinecite{Das2016}		\\
${\rm SrMn_2As_2}$		& 4.2966(1)	& 7.2996(2)	& 1.6989		& 116.70			& NA				&					& AFM~125\footnotemark[3]			&						&\onlinecite{Wang2011}		\\
${\rm EuMn_2As_2}$		& 4.2846(1)	& 7.2217(3)	& 1.6855		& 114.81	&  AFM 15.0, 5.0\footnotemark[2]$^,$\footnotemark[3]	& 7.0\footnotemark[3]	& AFM~142\footnotemark[2]	&	& PW		\\
\hline
${\rm CaMn_2Sb_2}$		& 4.5328(2)	& 7.4870(4)	& 1.6517		& 133.22	& NA				&	& AFM 85\footnotemark[3] 	&	$\approx 3.3$\footnotemark[1] 		&\onlinecite{Bridges2009}	\\
${\rm CaMn_2Sb_2}$		&			&			&			&		& NA	&	& AFM 88(1)\footnotemark[1]$^,$\footnotemark[2] & 2.8(1)\footnotemark[1]	&\onlinecite{Ratcliff2009}	\\
${\rm CaMn_2Sb_2}$		&	&	&	&	& NA	&	& FM 220\footnotemark[2]$^,$\footnotemark[3], AFM 85\footnotemark[2]$^,$\footnotemark[3] 	&0.007\footnotemark[3] (FM)	&\onlinecite{Simonson2012}	\\
${\rm EuMn_2Sb_2}$		& 4.526(2)	& 7.445(3)	& 1.645		& 132.1	& AFM~9.4(1)\footnotemark[3]		& $\approx 7$\footnotemark[3]	& 	AFM~128\footnotemark[2]	&	&\onlinecite{Schellenberg2010}		\\
${\rm EuZn_2Sb_2}$		& 4.4932(7)	& 7.617(1)	& 1.6952		& 133.18	&  AFM 13.1\footnotemark[2]$^,$\footnotemark[3]	&6.87(2)\footnotemark[3]	& NA	&		&	\onlinecite{Zhang2008}		\\
${\rm EuZn_2Sb_2}$		& 4.4852(4)	& 7.593(1)	& 1.6929		& 132.28	&  AFM 13\footnotemark[2]$^,$\footnotemark[3]	&7.0\footnotemark[3]	& NA	&		&	\onlinecite{May2012}		\\
${\rm YbZn_2Sb_2}$		& 4.4366(3)	& 7.401(1)	& 1.6682		& 126.15	& NA (Yb$^{+2}$)\footnotemark[4]	&	& NA	&	&\onlinecite{May2012}		\\
${\rm YbMn_2Sb_2}$		& 4.5289(4)	& 7.4503(8)	& 1.6451		& 132.34	& NA (Yb$^{+2}$)\footnotemark[4]	&	& AFM 120\footnotemark[1]	&	3.6(1)\footnotemark[1]	&\onlinecite{Morozkin2006}	\\
\hline
${\rm CaMn_2Bi_2}$		&	&	&	&		& NA		&	& AFM 154\footnotemark[1]$^,$\footnotemark[2] 	&	3.85\footnotemark[1]	&\onlinecite{Gibson2015}		\\
${\rm CaMn_2Bi_2}$		& 4.63(1)	& 7.64(1)	& 1.650	& 141.8		&&&&& \onlinecite{Cordier1976}		\\
\end{tabular}
\end{ruledtabular}
\footnotetext[1]{From neutron diffraction measurements}
\footnotetext[2]{From heat capacity measurements}
\footnotetext[3]{From magnetization and/or magnetic susceptibility measurements}
\footnotetext[4]{Yb$^{+2}$ has a $4f^{14}$ electronic configuration and hence does not carry a local moment}
\end{table*}

We have presented the results of a  comprehensive study of the physical properties of Eu$_{1-x}$K$_x$Mn$_2$As$_2$ ($x=0$, 0.04, 0.07) single crystals based on $\chi(T)$, $M(H)$, $C_{\rm p}(T)$  and $\rho(T)$ measurements.

For undoped \ema, the $\rho(T)$ data indicate a small band-gap semiconducting behavior (activation energy $\approx 52$~meV)  with an insulating ground state.  The $\chi(T)$ and $C_{\rm p}(T)$ data together reveal long-range AFM ordering of Eu$^{+2}$ moments with $S = 7/2$ below $T_{\rm N1} = 15.0$~K with a spin reorientation transition at $T_{\rm N2} = 5.0$~K\@. The ordered moment of $7~\mu_{\rm B}$/Eu observed from the $M(H)$ isotherms agrees with the calculated saturation moment $\mu_{\rm sat}=gS\mu_{\rm B}$/Eu with $S=7/2$ and $g=2$. This spin value is also consistent with the magnetic entropy $S_{\rm mag}(T=50$~K) derived from $C_{\rm mag}(T)$. The $\chi(T)$ and $M(H)$ data suggest that the AFM structure  of the Eu spins is a noncollinear, noncoplanar structure.  AFM ordering of the Mn spins in \ema\ is inferred from the high-$T$ $\chi(T)$ and $C_{\rm p}(T)$ data which indicate that the Mn N\'eel temperature is $T_{\rm N} =  142$~K\@.  Although our measurements yield no information on the AFM structure of the Mn spins below their $T_{\rm N}$, we note that the Mn spins in the very similar ${\rm SrMn_2As_2}$ compound have a collinear AFM structure with the easy axis aligned within the $ab$~plane \cite{Das2016}.  We see no clear evidence from $M(H)$ isotherms between~2 and~20~K of metamagnetic transitions in \ema\ in fields up to 14~T\@.

Some representative crystallographic and magnetic ordering data for $AM_2X_2$ compounds with the trigonal ${\rm CaAl_2Si_2}$-type structure are given in Table~\ref{Tab:OrdTemps} \cite{May2012, Payne2002, Schellenberg2010, Simonson2012, Gibson2015, Sangeetha2016, Das2016, Wang2011, Brock1994, Bridges2009, Ratcliff2009, Morozkin2006, Cordier1976}, including data for \ema\ crystals from the present work.  The Eu AFM ordering temperature of $T_{\rm N} = 9$--16~K does not vary much between the different compounds.  Similarly, for Mn the $T_{\rm N}$ values are all in the relatively narrow range 53 to 154~K\@.  In contrast, ${\rm BaMn_2As_2}$ \cite{Singh2009a, An2009, Singh2009b, Johnston2011} and ${\rm BaMn_2Bi_2}$ \cite{Saparov2013} with stacked Mn square lattices in the body-centered tetragonal ${\rm ThCr_2Si_2}$-type structure have higher $T_{\rm N} = 625$~K and 400~K, respectively.  

\begin{table}
\caption{\label{Tab:Js} Exchange interactions between Mn spins in ${\rm ThCr_2Si_2}$-type ${\rm BaMn_2As_2}$ (Ref.~\onlinecite{Johnston2011}) and ${\rm BaMn_2Bi_2}$ (Ref.~\onlinecite{Calder2014}) and ${\rm CaAl_2Si_2}$-type ${\rm CaMn_2Sb_2}$ (Ref.~\onlinecite{McNally2015}). All Mn--Mn exchange interactions are positive (antiferromagnetic) for each compound.  Also listed are the measured $T_{\rm N}$ values and the $T_{\rm N}$ and Weiss temperature~$\theta_{\rm p}$ in Eq.~(\ref{eq:C-W}) calculated from the exchange interactions using the MFT predictions in Eqs.~(\ref{Eqs:TNQp}) assuming Mn spin $S=2$. The $f$ ratios for the compounds are also given.  A number in parentheses is the estimated error in the last digit of the listed quantity.}
\begin{ruledtabular}
\begin{tabular}{lccc}
  						& ${\rm BaMn_2As_2}$ & ${\rm BaMn_2Bi_2}$ & ${\rm CaMn_2Sb_2}$\\
\hline
$SJ_1$ (meV)					&	33(3)		&	21.7(15)		&	7.9(6)		\\
$z_1$						&	4			&	4			&	3			\\
$SJ_2$ (meV)					&	9.5(13)		&	7.85(140)		&	1.3(2)		\\
$z_2$ 						&	4			&	4			&	6			\\
$SJ_c$ (meV)					&	1.5			&	1.26(2)		&	0.51(5)		\\
$z_c$						&	2			&	2			&	2			\\
$T_{\rm N}$ (K, observed)		&	625			&	400			&	88			\\
$T_{\rm N}$ (K, MFT, $S=2$)		&	1125(200)		&	670(140)		&	200(40)		\\
$\theta_{\rm p}$ (K, MFT, $S=2$)	&	$-2010(200)$	&	$-1400(140)$	& 	$-380(40)$	\\
$f\equiv\theta_{\rm p}/T_{\rm N}$ (MFT)	&	$-1.8(6)$	&	$-2.1(8)$		&	$-1.9(7)$		\\
\end{tabular}
\end{ruledtabular}
\end{table}

We now discuss the variation in $T_{\rm N}$ of these materials with reference to the Mn--Mn exchange interactions determined from inelastic neutron scattering measurements on ${\rm ThCr_2Si_2}$-type ${\rm BaMn_2As_2}$ \cite{Johnston2011} and ${\rm BaMn_2Bi_2}$ \cite{Calder2014} and ${\rm CaAl_2Si_2}$-type ${\rm CaMn_2Sb_2}$ \cite{McNally2015}.  The spin-wave spectra were all modeled in terms of the $J_1$-$J_2$-$J_c$ model, where $J_1$ and $J_2$ are the nearest- and next-nearest-neighbor exchange interactions within the Mn $ab$-plane square-lattice layer in ${\rm BaMn_2As_2}$ or within the $ab$-plane corrugated Mn honeycomb lattice layer in ${\rm CaMn_2Sb_2}$ and $J_c$ is the nearest-neighbor interaction along the $c$~axis of the respective layers.  Here the Heisenberg exchange interaction between a distinct pair of spins is written $J_{ij}{\bf S}_i\cdot{\bf S}_j$.  These exchange interactions are listed in Table~\ref{Tab:Js} along with the number $z_j$ of respective neighbors~$j$ of a given spin~$i$.

Within MFT, $T_{\rm N}$ and the Weiss temperature $\theta_{\rm p}$ in the Curie-Weiss law for a system containing identical crystallographically-equivalent spins are given by \cite{Johnston2012, Johnston2015}
\bse
\label{Eqs:TNQp}
\bea
T_{\rm N} &=& -\frac{S(S+1)}{3k_{\rm B}}\sum_j J_{ij}\cos\phi_{ji},\label{Eq:TNMFT}\\
\theta_{\rm p} &=& -\frac{S(S+1)}{3k_{\rm B}}\sum_j J_{ij},
\eea
\ese
where the sums are over neighbors $j$ of a given spin $i$ that interact with spin~$i$ with Heisenberg exchange constant $J_{ij}$ and $\phi_{ji}$ is the angle between ordered moments $\vec{\mu}_j$ and $\vec{\mu}_i$ in the AFM-ordered state.  Positive interactions $J_1$, $J_2$ and $J_c$ are AFM, negative ones are FM, and the angles are $\phi_{ji}=\pi$ for $J_1$ and $J_c$ and 0 for $J_2$ for both AFM structures discussed here.  The values of $T_{\rm N}$ and $\theta_{\rm p}$ predicted by Eqs.~(\ref{Eqs:TNQp}) are given in Table~\ref{Tab:Js} (see also \cite{McNally2015Error}).  The $T_{\rm N}$ predicted by MFT from the exchange constants is seen to overestimate the measured $T_{\rm N}$ by roughly a factor of two for the three compounds.  On the other hand, classical Monte Carlo simulations of $T_{\rm N}$ using the exchange constants in Table~\ref{Tab:Js} for ${\rm BaMn_2As_2}$ predicted a $T_{\rm N}$ in close agreement with the observed value \cite{Johnston2011}. This implies that classical thermal fluctuations associated with the quasi-low dimensionality of the spin lattice and bond frustration associated with $J_2$ are responsible for the suppresion of $T_{\rm N}$ from the MFT value in all three compounds.  A different interpretation for the suppression of the observed $T_{\rm N}$ in ${\rm CaMn_2Sb_2}$ from its MFT $T_{\rm N}$ value (but see Ref.~\onlinecite{McNally2015Error}) is given in Ref.~\onlinecite{McNally2015}, where the authors infer that the suppression is due to classical fluctuations of the system between the observed collinear N\'eel AFM structure and two competing noncollinear (``spiral'') AFM structures.  

\subsection{\ekxma, $x$ = 0.04 and 0.07}

We find that a 4\% substitution of K for Eu in ${\rm Eu_{0.96}K_{0.04}Mn_2As_2}$ is sufficient to change the ground state of ${\rm EuMn_2As_2}$ from insulating to metallic as revealed from $\rho(T)$ measurements.  This insulator-to-metal transition occurs due to doping of itinerant holes into the system arising from the partial substitution of Eu$^{+2}$ by K$^{+1}$.  A similar insulator to metal transition arises from a small 1.6\% substitution of K for Ba in tetragonal ${\rm BaMn_2As_2}$ \cite{Pandey2012}.  ${\rm BaMn_2As_2}$ is a so-called Mott-Hund insulator \cite{McNally2015a}, and K-doping that compound leads at small concentrations to a weakly-correlated metallic Fermi liquid of doped holes with a Fermi surface \cite{Pandey2012, McNally2015a, Yeninas2013}.

We also see evidence for a change in magnetic structure of the Eu moments in ${\rm Eu_{0.96}K_{0.04}Mn_2As_2}$ from that in undoped ${\rm EuMn_2As_2}$. The low-$T$ $\chi(T)$ exhibits AFM ordering of Eu$^{+2}$ moments below $T_{\rm N1} = 13.5$~K and a spin reorientation at $T_{\rm N2} = 9.0$~K\@. The high-$T$ $\chi(T)$ exhibits an anomaly at $\sim 150$~K that we interpret as the Mn AFM transition temperature. The high-$T$ $C_{\rm p}(T)$ confirms the Mn $T_{\rm N}= 146$~K\@. The $\chi^{-1}(T)$ data also suggest a transition at 275~K of unknown origin.  Like ${\rm EuMn_2As_2}$, the low-$T$ $C_{\rm p}(T)$ of ${\rm Eu_{0.96}K_{0.04}Mn_2As_2}$ shows a single anomaly at $T_{\rm N1}$ (no anomaly at $T_{\rm N2}$) related to AFM ordering of Eu$^{+2}$. 

With the higher K~concentration in ${\rm Eu_{0.93}K_{0.07}Mn_2As_2}$ we see further changes in the physical properties. The anisotropy at low~$T$ between $\chi_{ab}$ and $\chi_c$  cannot be explained by the conventional shape, magnetic-dipole and single-ion anisotropies.  The low-$T$ $\chi(T)$ of ${\rm Eu_{0.93}K_{0.07}Mn_2As_2}$ shows two transition temperatures at $T_{\rm N1} = 14.5$~K and $T_{\rm N2} = 12.5$~K\@. The high-$T$ $\chi(T)$ anomaly at the Mn $T_{\rm N}$ which appears as a result of 4\% K-substitution in ${\rm EuMn_2As_2}$ disappears at 7\% K concentration.  On the other hand, both $C_{\rm p}(T)$ and $\rho(T)$ of ${\rm Eu_{0.93}K_{0.07}Mn_2As_2}$ show clear anomalies near 150~K similar to the case of ${\rm Eu_{0.96}K_{0.04}Mn_2As_2}$ that are ascribed to AFM ordering of the Mn spins. Both $C_{\rm p}(T)$ and $\rho(T)$ of ${\rm Eu_{0.93}K_{0.07}Mn_2As_2}$ exhibit two well-defined anomalies related to AFM ordering of Eu$^{+2}$ whereas only one anomaly is seen for $x = 0$ and~0.04. Further investigations are required to understand the complex magnetic structures of Eu$_{1-x}$K$_x$Mn$_2$As$_2$ ($x=0,\ 0.04$ and 0.07). 

Perhaps the most interesting feature of both ${\rm Eu_{0.96}K_{0.04}Mn_2As_2}$ and ${\rm Eu_{0.93}K_{0.07}Mn_2As_2}$ is an anomalous $T$ dependence of $\rho$.  Both compounds show a strongly S-shaped metallic behavior of $\rho(T)$ where $d\rho/dT>0$ for $T>50$~K, but where an inflection point occurs at the Mn $T_{\rm N}$ of 146~K and 150~K, respectively.  Thus upon cooling, $\rho$ exhibits a strong downturn below $\sim 200$~K, reaches maximum positive slope at the Mn $T_{\rm N}$, and then continues to decrease but more slowly below $T_{\rm N}$.  Normally, one only sees such effects below magnetic ordering temperatures.  Here, the resistivity seems to be probing the dynamic short-range AFM order of the Mn spins above their $T_{\rm N}$ where there is no static ordered moment.  Perhaps the reason it can be observed is the large magnitude of $\rho$ in these doped compounds.  

\acknowledgments

Helpful discussions with Pinaki Das, Alan Goldman, Andreas Kreyssig, Abhishek Pandey and N.~S.~Sangeetha are gratefully acknowledged.  This research was supported by the U.S. Department of Energy, Office of Basic Energy Sciences, Division of Materials Sciences and Engineering.  Ames Laboratory is operated for the U.S. Department of Energy by Iowa State University under Contract No.~DE-AC02-07CH11358.


\end{document}